\documentclass[useAMS,usenatbib]{mn2e}
\usepackage{times,amsmath,amssymb,units,graphicx,aas_macros}
\usepackage{multirow,txfonts,natbib,xspace}
\usepackage{ulem}

\renewcommand\emph[1]{\textit{#1}}

\usepackage{color}

\newcommand\V{\mathbf v}

\newcommand{\dd}{\operatorname{d}}

\newcommand{\simgt}%
           {\,\hbox{\lower0.35ex\hbox{$\sim$}\llap{\raise0.35ex\hbox{$>$}}}\,}
\newcommand{\simlt}%
           {\,\hbox{\lower0.35ex\hbox{$\sim$}\llap{\raise0.35ex\hbox{$<$}}}\,}

\newcommand\NIRVANA{\textsc{nirvana}\xspace}
\newcommand\NIII{\textsc{nirvana-iii}\xspace}

\def\la{\hbox{\raise.35ex\rlap{$<$}\lower.6ex\hbox{$\sim$}\ }}
\def\ga{\hbox{\raise.35ex\rlap{$>$}\lower.6ex\hbox{$\sim$}\ }}

\def\beqa{\begin{eqnarray}}
\def\eeqa{\end{eqnarray}}
\def\beq{\begin{equation}}
\def\eeq{\end{equation}}

\def\order#1{{\cal O}\left({#1}\right)}

 \title[Vertical shear instability in discs]%
    {Linear and nonlinear evolution of the vertical shear instability
        in accretion discs}
            \author[Nelson, Gressel  \& Umurhan]%
             {Richard~P.~Nelson$^1$\thanks{E-mail: r.p.nelson@qmul.ac.uk (RPN); o.gressel@qmul.ac.uk (OG); oumurhan@ucmerced.edu (OMU)},
              Oliver~Gressel$^{1,2}$$^\star$ and  %
              Orkan~M.~Umurhan$^{1,3}$$^\star$ \\
              $^1\,$Astronomy Unit, Queen Mary University of London, Mile End Road, London E1 4NS\\
              $^2\,$NORDITA, KTH Royal Institute of Technology and Stockholm University,
Roslagstullsbacken 23, 106 91 Stockholm, Sweden \\
              $^3\,$School of Natural Sciences, University of California, Merced, 5200 North Lake Rd, Merced, CA 95343, USA}

\begin{document}
\pdfimageresolution=100

\date{Accepted 1988 December 15. %
      Received 1988 December 14; %
      in original form 1988 October 11}

\pagerange{\pageref{firstpage}--\pageref{lastpage}} \pubyear{2002}

\maketitle

\label{firstpage}

%

\begin{abstract}
We analyse the stability and nonlinear dynamical evolution of power-law accretion 
disc models. 
These have midplane densities that follow radial power-laws,
and have either temperature or entropy distributions that are strict power-law 
functions of cylindrical radius, $R$. 
We employ two different hydrodynamic codes to perform high resolution 2D-axisymmetric and 3D 
simulations that examine the long-term evolution of the disc models as a function of the
power-law indices of the temperature or entropy, the thermal relaxation time of the fluid,
and the disc viscosity. We present an accompanying stability analysis of the problem,
based on asymptotic methods, that we use to interpret the results of the simulations.
We find that axisymmetric disc models whose temperature or entropy profiles cause the 
equilibrium angular velocity to vary with height are unstable to the growth of modes with wavenumber ratios $|k_R/k_Z| \gg 1$ when the thermodynamic response of the fluid
is isothermal, or the thermal evolution time is comparable to or shorter than the local dynamical time scale. These discs are subject to the Goldreich-Schubert-Fricke (GSF) or `vertical shear' linear instability. Development of the instability
involves excitation of vertical breathing and corrugation modes in the disc, with the corrugation modes in particular being a feature of the nonlinear saturated state.
Instability is found to operate when the dimensionless disc kinematic viscosity 
$\nu < 10^{-6}$, corresponding to Reynolds numbers ${\rm Re}=H c_{\rm s}/\nu > 2500$. In three dimensions the instability generates a quasi-turbulent flow, and the 
associated Reynolds stress produces a fluctuating effective viscosity coefficient whose 
mean value reaches $\alpha \sim 6 \times 10^{-4}$ by the end of the simulation. 
The evolution and saturation of the vertical shear instability in astrophysical disc models which include realistic treatments of the thermal physics has yet to be examined. Should it occur on either global or local scales, however, our results suggest that it will have significant consequences for their internal dynamics, transport properties, and
observational appearance.

\end{abstract}


%

\section{Introduction}
\label{sec:intro}
Accretion discs play important roles in a broad range of astrophysical phenomena.
Protostellar discs orbiting young stars provide conduits through which
most of the mass accretes during the star formation process, and they are the nascent
environments for planetary system formation. Mass transfer through Roche lobe overflow 
in close binary systems leads to highly energetic and time variable phenomena because 
of accretion through a disc onto compact objects such as white dwarfs in cataclysmic 
variables, and neutron stars or black holes in low mass X-ray binaries. Quasars 
and active galactic nuclei are powered by disc accretion onto supermassive black
holes. Understanding the dynamics and evolution of astrophysical discs is key to understanding
these and related phenomena.

Since the early work of \cite{1973A&A....24..337S} and \cite{1974MNRAS.168..603L},
significant efforts have been made to understand the internal dynamics of
discs, including mechanism(s) that allow them to transport angular momentum and accrete
at their observed rates. Several ideas based on dynamical instabilities in
non-magnetised flows have been explored, including the Papaloizou-Pringle 
instability for thick accretion tori driven by unstable non-axisymmetric wave 
modes \citep{1984MNRAS.208..721P}, convective instability 
\citep{1973Icar...18..377C, 1980MNRAS.191...37L, 1988ApJ...329..739R, 1992ApJ...388..438R}, gravitational instability \citep{1964ApJ...139.1217T, 1987MNRAS.225..607L, 1991MNRAS.248..353P}, the global and subcritical baroclinic
instabilities \citep{2003ApJ...582..869K, 2006ApJ...636...63J, 
2007ApJ...658.1252P, 2010A&A...513A..60L}, 
and the vertical shear instability 
\citep{1998MNRAS.294..399U, 2003A&A...404..397U, 2004A&A...426..755A} 
which is closely related to the Goldreich-Schubert-Fricke (GSF) instability \citep{1967ApJ...150..571G, 1968ZA.....68..317F} developed in the context of
differentially rotating stars. 
The presence of a weak magnetic field in the disc, however, leads to the development of 
magnetohydrodynamic turbulence driven by the magnetorotational instability
\citep{1991ApJ...376..214B,1991ApJ...376..223H}, and it is now generally accepted that this
is likely to be the source of anomalous viscosity in most accretion discs
in which the magnetic field is well-coupled to the gas.

In this paper we present an extensive analysis of the hydrodynamic
stability and nonlinear dynamics of disc models with power-law midplane density
distributions, and either temperature or entropy profiles that are
power-law function of $R$ only, where $R$ is the cylindrical radius.
Our investigation employs high resolution 2D-axisymmetric and 3D hydrodynamic
simulations and a linear stability analysis based on asymptotic
methods. Models with power-law temperature profiles, adopting
locally isothermal equations of state, have been used extensively
in the study of protostellar disc dynamics and disc planet interactions.
These studies normally assume the disc is viscous
\citep[e.g.][]{2001ApJ...547..457K, 2007A&A...473..329C, 2010A&A...511A..77F,
2010ApJ...724..730D}, or magnetised \citep[e.g.][]{2006A&A...457..343F, 2011MNRAS.416..361B},
but in this study we include neither viscosity nor magnetic fields.
Adoption of radial variations in temperature or entropy in the models causes them
to have angular velocity profiles that are a function of both radius and height,
$\Omega(R,Z)$. The height-variation of $\Omega$ is often referred to as the
thermal wind in studies of atmospheric dynamics.

We find that disc models for which $d\Omega/dZ \neq 0$, and which experience
thermal relaxation on $\sim$ dynamical time scales or shorter, are unstable to
the growth of modes with $|k_R/k_Z| \gg 1$, where ($k_R$, $k_Z$) are the radial
and vertical wavenumbers. This instability appears to be closely related
to the GSF instability, as studied by \citet{1998MNRAS.294..399U}, \citet{2003A&A...404..397U}
and \citet{2004A&A...426..755A} in the context of accretion discs, and confirmed
by our own linear stability analysis. Growth of the instability is favoured
when the thermodynamic response of the gas is isothermal, of near-isothermal,
although the strength of this dependence varies with model parameters
with steeper thermal gradients displaying a greater tendency toward instability.
The one 3D simulation we presents suggests that nonlinear development
of the instability leads to a turbulent flow whose associated Reynolds stress
leads to an effective alpha parameter $\alpha \sim 10^{-3}$, causing
non-negligible outward angular momentum transport.

This paper is organised as follows. In Sect.~\ref{sec:eqns} we present the basic 
equations of the problem, and the disc models we examine. 
In Sect.~\ref{sec:stability} we discuss the hydrodynamic stability of rotating
shear flows, and previous work in the literature relevant to the present study.
In Sect.~\ref{sec:methods} we describe the numerical methods employed,
and in Sect.~\ref{sec:Results} we present the results of the
nonlinear simulations. A stability analysis of the problem using asymptotic
methods is presented in Sect.~\ref{sec:analysis}, and a reworking of the
analysis of \citet{1967ApJ...150..571G} for a fully compressible fluid is given 
in the appendix. We draw our conclusions and
discuss ideas for future work in Sect.~\ref{sec:conclusion}.


\section{Basic equations}
\label{sec:eqns}

In this paper we make use of both cylindrical polar coordinates ($R$,
$\phi$, $Z$) and spherical polar coordinates ($r$, $\theta$, $\phi$).
We solve the continuity, momentum and internal energy equations of 
hydrodynamics
\begin{eqnarray}
      \partial_t\rho +\nabla\cdot(\rho \V) & = & 0\,, \nonumber\\
      \partial_t(\rho\V) +\nabla\cdot
      \left[ \rho\mathbf{vv} \right]
            & = & \!\! -\nabla P - \nabla \Phi, \nonumber \\
      \partial_t (e) + \nabla \cdot (e \V) 
            &=& -P \nabla \cdot \V + {\cal S} - {\cal Q}
\label{eqn:motion}
\end{eqnarray}
where $\rho$ is the density, $\V$ is the velocity,
$P$ is is the pressure, $e$ is the internal energy per unit volume,
${\cal S}$ and ${\cal Q}$ are energy source and sink terms,
and $\Phi=-GM/r$ is the gravitational potential due to the central star.
Here $G$ is the gravitational constant and $M$ is the mass of the star.

\subsection{Disc models}
\label{sec:disc-model}
We are concerned primarily with two basic equilibrium disc models in 
this paper. In the first, the temperature, $T$, and midplane density, 
$\rho_{\rm mid}$, are simple power-law functions of cylindrical 
radius:
\begin{eqnarray}
  T(R) & = & T_0  \left(\frac{R}{R_0}\right)^q \label{eqn:TR} \\
  \rho_{\rm mid}(R) & = & \rho_0 \left( \frac{R}{R_0} \right)^p,
\label{eqn:rhoR}
\end{eqnarray}
where $T_0$ is the temperature at the fiducial radius $R_0$,
and $\rho_0$ is the midplane gas density there. Adopting an 
ideal gas equation of state
\begin{equation}
P= \frac{\cal R}{\mu} T \rho,
\label{eqn:eos1}
\end{equation}
where ${\cal R}$ is the gas constant and $\mu$ is the 
mean molecular weight, we note that the isothermal sound 
speed is related to the temperature through the expression 
$c_s^2={\cal R} T/{\mu}$, such that $q$ also represents the
radial power-law dependence of $c_s^2(R)$:
\begin{equation}
c_s^2(R) = c_0^2 \left(\frac{R}{R_0}\right)^q.
\label{eqn:csR}
\end{equation}

In the second disc model, we adopt a power-law function 
for the midplane density, as in eqn.~(\ref{eqn:rhoR}), 
and specify the entropy function, $K_s$, as a strict power-law 
function of $R$ in the initial model:
\begin{equation}
 K_s(R) = K_0 \left(\frac{R}{R_0}\right)^s,
\label{eqn:ent-func}
\end{equation}
where the entropy function is defined through the expression
\begin{equation}
P=K_s \rho^{\gamma},
\label{eqn:eos2}
\end{equation}
and $\gamma$ is assumed to be constant. The entropy per unit mass is given
by
\begin{equation}
S= c_{\rm v} \log{(T \rho^{1-\gamma})}
\label{eqn:entropy}
\end{equation}
where $c_{\rm v}$ is the specific heat at constant volume, and the entropy
function is given in terms of the entropy by the expression
\begin{equation}
K_s=c_{\rm v} (\gamma-1) \exp{\left(\frac{S}{c_{\rm v}} \right)}.
\label{eqn:Ks-S}
\end{equation}

\subsubsection{Equilibrium solutions}
\label{sec:equilibria}

In order to construct initial conditions for our simulations
we need to obtain equilibrium disc models.
The equations of force balance in the radial and vertical
directions are given by
\begin{eqnarray} 
 R \Omega^2  & - & \frac{GM R}{(R^2 + Z^2)^{3/2}}
-\frac{1}{\rho} \frac{\partial P}{\partial R} = 0 
\label{eqn:force-R} \\
 & - & \frac{GM Z}{(R^2+Z^2)^{3/2}} - \frac{1}{\rho}
 \frac{\partial P}{\partial Z} = 0.
\label{eqn:force-Z}
\end{eqnarray}
Combining eqns.~(\ref{eqn:force-R}), (\ref{eqn:force-Z}), (\ref{eqn:TR})
and (\ref{eqn:rhoR}) leads to expressions for the equilibrium density and
angular velocity, $\Omega$, as functions of ($R$, $Z$) for the disc with
a power-law temperature profile: 
\begin{eqnarray}
\rho(R, Z) &=& \rho_0 \left( \frac{R}{R_0} \right)^p 
\exp{\left(\frac{G M}{c^2_s} \left[\frac{1}{\sqrt{R^2 + Z^2}} 
- \frac{1}{R} \right] \right)}, \label{eqn:density1} \\
\Omega(R,Z) &=& \Omega_K \left[(p+q) \left(\frac{H}{R}\right)^2
+ (1+q) - \frac{q R}{\sqrt{R^2 + Z^2}} \right]^{1/2}
\label{eqn:Omega1}
\end{eqnarray}
where $\Omega_K=\sqrt{G M / R^3}$ is the keplerian angular velocity
at radius $R$, and $H = c_s / \Omega_K$ is the local disc scale height \citep[also see][]{2002ApJ...581.1344T, 2011A&A...534A.107F}.
The definition of $c_s$ given in eqn.~(\ref{eqn:csR}) implies that
\begin{equation}
H = H_0 \left(\frac{R}{R_0} \right)^{(q+3)/2} 
\label{eqn:H}
\end{equation}
where $H_0 = c_0/\sqrt{G M / R_0^3}$ is the disc scale height at
radius $R_0$. 

Similarly, the equilibrium density and angular velocity
for the disc model with a power-law entropy function profile are
given by:
\begin{eqnarray}
\rho(R,Z) &=& \left( \rho_{\rm mid}^{(\gamma - 1)} + \frac{(\gamma-1)}{\gamma}
\frac{G M}{K_s} \left[\frac{1}{\sqrt{R^2 + Z^2}} - \frac{1}{R} \right]
\right)^{1/(\gamma - 1)} \label{eqn:density2} \\
\Omega(R,Z) &=& \Omega_K \left[ \frac{p}{{\cal M}^2_{\rm mid}} + 
\frac{s}{\gamma {\cal M}^2} + (1+s) - \frac{s R}{\sqrt{R^2 + Z^2}} \right]^{1/2},
\label{eqn:Omega2}
\end{eqnarray}
where ${\cal M_{\rm mid}}=v_K/a_{\rm mid}$ is the Mach number at the disc 
midplane and ${\cal M}= v_K/a_s$ is the Mach number at each disc location.
$v_K= \sqrt{GM/R}$ is the keplerian velocity at radius $R$,
and the adiabatic sound speed $a_s=\sqrt{\gamma P/\rho}$, which takes the
value $a_{\rm mid}$ at the disc midplane. We note that combining
equations ~(\ref{eqn:eos1}), (\ref{eqn:ent-func}), (\ref{eqn:eos2}) and 
(\ref{eqn:density2}) demonstrates that the temperature in this model
is a function of both $R$ and $Z$ in general:
\begin{equation}
T(R,Z)= K_{\rm s} \rho^{\gamma -1} \frac{\mu}{\cal R}.
\label{eqn:T_RZ}
\end{equation}
As such, this model provides a useful contrast to the one with temperature 
constant on cylinders, and is convenient to implement numerically 
because of the existence of analytic solutions for the equilibrium
$\rho(R,Z)$ and $\Omega(R,Z)$ profiles.

Equations (\ref{eqn:TR}), (\ref{eqn:rhoR}), (\ref{eqn:density1}) and 
(\ref{eqn:Omega1}) fully specify the initial disc models with power-law
temperature profiles that we examine in this paper, subject to appropriate
choices for $p$ and $q$.
The expressions (\ref{eqn:rhoR}) and (\ref{eqn:ent-func}), along with
(\ref{eqn:density2}) and (\ref{eqn:Omega2}), specify the initial models
for which the entropy function is a power-law function of cylindrical 
radius, subject again to appropriate choices for the power-law exponents 
$p$ and $s$. We note that for all models in which the initial temperature,
$T$, or entropy function, $K_s$, are strict power-law functions of $R$,
the equilibirum angular velocities are explicit functions of both $R$ and 
$Z$, a fact that appears to play a key role in the disc instability
that we examine in this paper. The dependence of $\Omega$ on $Z$ is often 
referred to as the `thermal wind' in studies of planetary atmosphere dynamics. 

\subsection{Thermodynamic evolution}
\label{sec:thermodynamics}

The thermodynamic evolution of both disc models described above in Sect.~\ref{sec:disc-model} 
is assumed to be one of three types in the simulations presented here:
{\it locally isothermal}, for which the local temperature at each 
($R$, $Z$) position in the disc is kept strictly equal to its original value;
{\it isentropic}, where the entropy of the fluid is kept constant (equivalent
to there being no source/sink term in the energy eqn.~[\ref{eqn:motion}]);
{\it thermally relaxing}, where we relax the temperature at each location
in the disc toward its initial value on some time scale, $\tau_{\rm relax}$.
The thermal relaxation model we adopt is
\begin{equation}
\frac{\dd T}{\dd t} = - \frac{(T - T_0)}{\tau_{\rm relax}}
\label{eqn:t-relax}
\end{equation}
where $T_0$ is the initial temperature. For simplicity, we assume that
$\tau_{\rm relax}$ is a function of $R$, being a fixed multiple or 
fraction of the local keplerian orbital period. Equation~(\ref{eqn:t-relax})
has a simple analytic solution of the form
\begin{equation}
T(t+\Delta t) = T_0 + (T(t) - T_0) \exp{\left(-\frac{\Delta t}{\tau_{\rm relax}}\right)},
\end{equation}
where $T(t)$ is the temperature at time $t$, and $T(t + \Delta t)$
is the temperature at some later time $t+\Delta t$.

In the locally isothermal models we use an isothermal equation
of state $P=c_s^2 \rho$, and do not evolve the energy equation
in eqn.~(\ref{eqn:motion}). In the isentropic models we use
the equation of state $P=(\gamma -1) e$, solve the energy
equation in eqn.~(\ref{eqn:motion}), and neglect the source and
sink terms. The energy equation is also solved in the thermally
relaxing models, along with eqn.~(\ref{eqn:t-relax}) which 
plays the role of the source and sink terms in the energy equation
(\ref{eqn:motion}).

\section{Hydrodynamic stability of disc models}
\label{sec:stability}
\subsection{The Rayleigh and Solberg-H{\o}iland criteria}
The Rayleigh criterion indicates that accretion discs with strictly
keplerian angular velocity profiles are hydrodynamically stable since
$$\frac{d j^2}{dr} > 0,$$
where $j=R^2 \Omega$ and $\Omega=\sqrt{G M/R^3}$. More generally, a
differentially rotating, compressible fluid with angular velocity varying with
height and radius, $\Omega(R,Z)$, subject to axisymmetric isentropic 
perturbations (i.e. $DS/Dt=0$, where $D/Dt$ is the total time derivative for
fluid elements) is stable if both of the Solberg-H{\o}iland 
criteria are satisfied \citep[e.g.][]{1978trs..book.....T}:
\begin{equation}
\frac{1}{R^3} \frac{\partial j^2}{\partial R} + \frac{1}{\rho C_p} 
\left(-\nabla P \right) \cdot \nabla S > 0
\label{eqn:hoiland1}
\end{equation}
\begin{equation}
-\frac{\partial P}{\partial Z} \left(
\frac{\partial j^2}{\partial R} \frac{\partial S}{\partial Z} - 
\frac{\partial j^2}{\partial Z} \frac{\partial S}{\partial R} \right) > 0.
\label{eqn:hoiland2}
\end{equation}
Accretion disc models generally possess negative radial pressure gradients, 
and a negative vertical pressure gradient in the disc hemisphere above the midplane, leading to stability criteria~(\ref{eqn:hoiland1}) and (\ref{eqn:hoiland2}) in the form:
\begin{equation}
\frac{1}{R^3} \frac{\partial j^2}{\partial R} + \frac{1}{\rho C_p}
\left( \left| \frac{\partial P}{\partial R} \right| \frac{\partial S}{\partial R}
+\left| \frac{\partial P}{\partial Z} \right| \frac{\partial S}{\partial Z} \right) > 0
\label{eqn:hoiland3}
\end{equation}
\begin{equation}
\frac{\partial j^2}{\partial R} \frac{\partial S}{\partial Z} -
\frac{\partial j^2}{\partial Z} \frac{\partial S}{\partial R} > 0.
\label{eqn:hoiland4}
\end{equation}
Considering a disc with a strictly keplerian $j$ profile we see that
stability according to eqn.~(\ref{eqn:hoiland4}) requires 
$\partial S/\partial Z >0$,
in agreement with the Schwarzschild condition for convective stability.
Condition~(\ref{eqn:hoiland3}) shows that a large amplitude negative 
radial entropy gradient $\partial S/\partial R <0$ can also drive instability
in principle, provided the gradient is strong enough to overcome the positive
angular momentum gradient. This is distinct from the global baroclinic and/or 
subcritical baroclinic instability discussed in the introduction that does not require 
violation of the Solberg-H{\o}iland criteria to operate, but does require thermal evolution
of the fluid on time scales similar to the dynamical time to re-establish the initial 
radial entropy gradient, in addition to the presence of finite amplitude perturbations
\citep{2010A&A...513A..60L}.

Considering a quasi-keplerian disc whose angular velocity depends on height and radius,
$\Omega(R,Z)$, eqn.~(\ref{eqn:hoiland4}) shows that such a disc
with $\partial S/\partial Z >0$ can be destabilised through the combination
$(\partial j^2/\partial Z)(\partial S/ \partial R) >0$. 

We consider two basic disc models in this paper that are the subject of
nonlinear simulations in which the fluid evolution is isentropic and for 
which the Solberg-H{\o}iland criteria determine hydrodynamic stability. One model
assumes the temperature profile is a strict power law function of radius, $R$,
such that the density and angular velocity are given by 
eqns.~(\ref{eqn:density1}) 
and (\ref{eqn:Omega1}). With values $p=-1.5$ and $q=-1$, -0.5, -0.25 and 0 in eqns.~(\ref{eqn:TR}) 
and (\ref{eqn:rhoR}), respectively, these discs are stable according to criterion (\ref{eqn:hoiland3}). 
The term involving $\partial S/ \partial R$ provides the only destabilising contribution 
(because $\partial S/\partial Z >0$), but is always far too small to overcome the positive radial 
gradient in $j^2$. This disc is also stable according to criterion (\ref{eqn:hoiland4}) as 
$(\partial j^2/\partial R)(\partial S/\partial Z) > 
(\partial j^2/\partial Z <0)(\partial S/\partial R)$.

The second set of disc models we consider assume that the entropy is a strict power law
function of radius $R$. The density and angular velocity are given by the expressions
(\ref{eqn:density2}) and (\ref{eqn:Omega2}), and the values ($p=0$, $s=-1$) and
($p=-1.5$, $s=0$) are adopted in eqns.~(\ref{eqn:rhoR}) and (\ref{eqn:ent-func}). 
Both of these models are stable according to criterion (\ref{eqn:hoiland3}) as the 
destabilising term proportional to $\partial S/ \partial R$ is either zero or too small 
(note that $\partial S/ \partial Z=0$ in these models). Criterion (\ref{eqn:hoiland4}), 
however, predicts instability for the ($p=0$, $s=-1$) model because only the second term on 
the left-hand side is non zero, with $\partial j^2/\partial Z <0$ and $\partial S/\partial R <0$. 
The isentropic simulation that we present in the results section later in this paper will provide an example of nonlinear evolution of this model that is unstable according to one of the 
Solberg-H{\o}iland criteria. The model with ($p=-1.5$, $s=0$) is stable according to criterion
(\ref{eqn:hoiland4}).

\subsection{The Goldreich-Schubert-Fricke instability}
\label{sec:GSF}
When thermal and viscous diffusion play a role the stability of rotating flows
are no longer controlled by the Solberg-H{\o}iland criteria, but instead are
determined by stability criteria obtained originally by \cite{1967ApJ...150..571G} and
\cite{1968ZA.....68..317F} in application to the radiative zones of differentially
rotating stars. Axisymmetric rotating flows are susceptible to the Goldreich-Schubert-Fricke
(GSF) instability when viscous diffusion is much less efficient than thermal
diffusion, such that a fluid element retains its initial angular momentum
but quickly attains the entropy of the surrounding fluid when perturbed from
its equilibrium location. Under these circumstances the stabilising influence
of entropy gradients provided by the Solberg-H{\o}iland criteria diminish
and instability ensues for wave modes satisfying the instability criterion
\begin{equation}
\frac{\partial j^2}{\partial R} - \frac{k_R}{k_Z}
\frac{\partial j^2}{\partial Z} < 0.
\label{eqn:GSF}
\end{equation}
For a rotating flow where the angular momentum is a function of both $R$ and $Z$,
and the appropriate conditions on thermal and viscous diffusion are satisfied,
unstable modes are guaranteed to exist since wavevectors with ratios $k_R/k_Z$
that satisfy eqn.~(\ref{eqn:GSF}) can always be found. In general, we expect
$|\partial j/\partial R | \gg |\partial j/ \partial Z |$ in a quasi-keplerian
accretion disc, such that unstable modes will have $|k_R/k_Z| \gg 1$
(i.e. unstable modes will have radial wavelengths that are very much shorter
than vertical ones).

Application of the GSF instability to accretion discs has not received a
great deal of attention in the literature (but see the discussion below). 
Indeed, the study presented in this paper has arisen from unrelated attempts to 
generate 3D models of protoplanetary disc models within which turbulence is generated 
by the MRI in active layers near the disc surface, and in which extensive dead zones that
are magnetically inactive (or at least stable against the MRI) exist
near the midplane. The adoption of a locally isothermal equation of state
in these models, combined with the absence of a physical viscosity, rendered them unstable to the growth of vertical corrugation
oscillations that in the nonlinear regime became quite violent. This behaviour appears
to be a nonlinear manifestation of the GSF instability, which we study here in more detail. 
The two classes of models we consider (temperature constant on cylinders and entropy constant 
on cylinders) both have angular velocity profiles that depend on both $R$ and $Z$, and we expect 
that models in which perturbations evolve quasi-isothermally will display the GSF instability.

Previous analysis of the GSF instability in the context of accretion
discs was initiated by \cite{1998MNRAS.294..399U} in which they presented
a local linear analysis utilising the short-wavelength approximation. A more extensive
analysis was presented in \cite{2003A&A...404..397U} where it was suggested that the GSF
may provide a source of hydrodynamic turbulence in accretion discs, and nonlinear
simulations were presented by \cite{2004A&A...426..755A}. These nonlinear simulations
adopted a basic disc model with a strictly isothermal equation of state.
As such, the underlying equilibrium disc model has an angular velocity profile
that is independent of height, $Z$, (see eqn.~\ref{eqn:Omega1}) and should not
be susceptible to the GSF instability. \cite{2004A&A...426..755A}, however, adopted
initial conditions that allowed relaxation around the equilibrium state leading to 
variations of $\Omega$ with height, and
when initial perturbations with $|k_R/k_Z| \gg 1$ were applied growth of the GSF was observed. It is also worth noting the related linear and nonlinear study of isentropic accretion disc models presented by \cite{2002A&A...391..781R}
at this point. This study examined the stability of discs whose angular momentum and entropy profiles rendered them stable according to the previously discussed Solberg-H{\o}iland criteria. As expected, applied perturbations were found to always decay.
We present a linear analysis of the instability we study in Sect.~\ref{sec:analysis}
for a fully compressible fluid under the assumption that the equation of state 
is locally isothermal, and in the appendix we present an analysis of the problem that closely follows the derivation of the GSF stability in \citet{1967ApJ...150..571G}.

\begin{table}
\begin{center}
  \caption{Simulation parameters and results. Labels beginning with a
   `T' denote runs with $T$ being a function of $R$. Those
    with a `K' denotes runs where $K_s=K_s(R)$. The letter R denotes
     reflecting boundary conditions, and a letter O denotes outflow b.c's.
     Digits after the hyphens denote thermal relaxation times. Note that all runs have
        $H_0/R_0=0.05$.
  \label{table-NIRVANA}}
  \begin{tabular}{@{}lccccc@{}}
  \hline Simulation & $p$ & $q$ or $s$ & $\tau_{\rm Relax}$
  &$N_r \times N_{\theta} \times N_{\phi}$ & 
  Unstable\\ 
                    &     &            &     &
  &  ? \\
  \hline

T1R/O--0 & -1.5 & -1    & 0.00 & $1328 \times 1000 \times 1$ & Y \\
T2R/O--0 & -1.5 & -0.5  & 0.00 & $1328 \times 1000 \times 1$ & Y \\
T3R/O--0 & -1.5 & -0.25 & 0.00 & $1328 \times 1000 \times 1$ & Y \\
T4R/O--0 & -1.5 & 0     & 0.00 & $1328 \times 1000 \times 1$ & N \\
\hline
T5R--0.01 & -1.5 & -1   & 0.01 & $1328 \times 1000 \times 1$ & Y \\
T6R--0.1  & -1.5 & -1   & 0.10 & $1328 \times 1000 \times 1$ & N \\
T7R--1.0  & -1.5 & -1   & 1.00 & $1328 \times 1000 \times 1$ & N \\
T8R--10.0 & -1.5 & -1   & 10.0 & $1328 \times 1000 \times 1$ & N \\
T9R--$\infty$ & -1.5 & -1   & $\infty$ & $1328 \times 1000 \times 1$ & N \\
\hline
K1R--0    & 0 & -1      & 0.00 & $1328 \times 1000 \times 1$ & Y \\
K5R--0.01 & 0 & -1   & 0.01 & $1328 \times 1000 \times 1$ & Y \\
K6R--0.1  & 0 & -1   & 0.10 & $1328 \times 1000 \times 1$ & Y \\
K7R--1.0  & 0 & -1   & 1.00 & $1328 \times 1000 \times 1$ & Y \\
K8R--10.0 & 0 & -1   & 10.0 & $1328 \times 1000 \times 1$ & Y \\
K9R--$\infty$ & 0 & -1   & $\infty$ & $1328 \times 1000 \times 1$ & N \\
K10R--0.01 & -1.5 & 0 & 0.01 & $1328 \times 1000 \times 1$ & N \\
\hline
T1R--0--3D & -1.5  & -1     & 0.00 & $1328 \times 1000 \times 300$ & Y \\
\hline
  \end{tabular}
\end{center}
\end{table}

\section{Numerical methods}
\label{sec:methods}


The simulations presented in this paper were performed using two
different codes that utilise very different numerical schemes.  We use
an older version of \NIRVANA, which uses an algorithm very similar to
the \textsc{zeus} code to solve the equations of ideal MHD
\citep{1997CoPhC.101...54Z, 1992ApJS...80..753S}.  This scheme uses
operator splitting, dividing the governing equations into source
and transport terms. Advection is performed using the second-order
monotonic transport scheme \citep{1977JCoPh..23..276V}. We also use
the more modern \NIII code, which is a second order Godunov-type MHD code
\citep{2004JCoPh.196..393Z}, which has recently been extended to
orthogonal-curvilinear coordinate systems
\citep{2011JCoPh.230.1035Z}. All presented simulations were performed using a
standard spherical coordinate system ($r$, $\theta$, $\phi$).

\subsection{Initial and boundary conditions}

The numerical study presented here applies to the two general classes of
disc model discussed in Sect.~\ref{sec:disc-model}, and as such we adopt 
a numerical set up that is not specific to any particular physical system 
(although our motivation for undertaking this study arose from earlier attempts 
to establish stationary equilibrium solutions for protostellar disc models).
Based on numerous test calculations performed with a wide range of resolutions
during an early stage of this project, we know that the instability we study here 
is characterised by having a radial wavelength much shorter than the vertical 
wavelength (i.e. $k_R\gg k_Z$) during its early growth phase.
Consequently we consider disc models with fairly narrow radial domains
to facilitate high resolution simulations. The spherical polar grid we adopt has 
inner radius $r_{in}=R_0=1$ and outer radius $r_{out}=2$, and most simulations we 
present are axisymmetric. The one non-axisymmetric simulation we present covers
a restricted azimuthal domain of $\pi/4$, again to facilitate a high resolution study.

Disc models in which the initial temperature is a strict function
of cylindrical radius, $R$, have meridional domains
$\pi/2 - 5 H_0/R_0 \le \theta \le \pi/2 + 5 H_0/R_0$.
For a disc with radial temperature profile $q=-1$, corresponding to
a disc with constant $H/R$, the meridional domain covers $\pm 5$ scale 
heights above and below the midplane. For larger (less negative) values
of $q$, the disc covers $\pm 5$ scale heights at the inner radius, 
but a reduced number of scale heights as one moves out in radius. Prior
to initiating these simulations, the disc models are specified using 
eqns.~(\ref{eqn:rhoR}), (\ref{eqn:TR}), (\ref{eqn:density1}) and 
(\ref{eqn:Omega1}).
In all models the initial velocity field was seeded with random noise
distributed uniformly in each component with a peak amplitude equal to 
1 \% of the local sound speed. 

Discs for which the initial entropy function, $K_s$, depends only on $R$
have physical surfaces where the density goes to zero. In these models
the meridional boundaries are placed at a location that is 5\% smaller
than the angular distance from the midplane where the density vanishes.
The initial models are specified using eqns.~(\ref{eqn:rhoR}),
(\ref{eqn:ent-func}), (\ref{eqn:density2}) and (\ref{eqn:Omega2}).
In order to determine the value of $K_0$ in eqn.~(\ref{eqn:ent-func}),
we specify the midplane Mach number, ${\cal M}_{\rm mid}$, at radius $R_0$.
We adopt ${\cal M}_{\rm mid}=20$ to be consistent with the models
for which $H/R=0.05$. Seed noise with amplitude equal to 1\% of the sound speed was again
added to all velocity components.

For most simulations we adopt standard outflow or reflecting boundary
conditions at the inner and outer radial boundaries.
Periodic boundaries are applied in the azimuthal direction, and either
standard outflow or reflecting conditions are applied at the meridional
boundaries (all simulations performed using \NIII adopted outflow
boundary conditions at the radial and merdional boundaries).
The density is obtained in ghost zones by means of linear extrapolation.
A variety of different boundary conditions were used in test
simulations at an early stage of this project, and the results were found
to be insensitive to the choice adopted. These tests included the adoption 
of damping boundary conditions that absorb incoming waves, indicating
that reflecting boundaries are not required to drive the instability 
discussed in this paper.

Owing to the unsplit character of finite volume schemes, such as used in \NIII, it is
difficult numerically to preserve static equilibria \citep{2002ApJS..143..539Z}. In particular, 
problems arise with the constant extrapolation of the density profile in the vertical direction. 
This is because the weight of the gas is not balanced by the now vanishing pressure gradient in 
the case of an isothermal equation of state, leading to a standing accretion shock in the first grid 
cell of the meridional domain. In the case of an adiabatic evolution equation, the same problem arises 
in the boundary condition for the thermal energy (which enters the boundary condition for the total 
energy). To alleviate this situation, we obtain the density (and in the case of solving an energy 
equation, the pressure) by integrating the hydrostatic equilibrium in each cell adjacent to the domain 
boundary. This is done with a second-order Runge-Kutta shooting method.

Unless indicated otherwise, we use a fixed resolution of 
$N_r\times N_\theta$ of $1328\times 1000$ grid cells for our
\NIRVANA runs with $H_0/R_0=0.05$ or ${\cal M}_{\rm mid}=20$ at $r=R_0$.
The \NIII runs used a resolution of $1344 \times 1024$

We adopt a system of units in which $M=1$, $G=1$ and $R_0=1$.
When presenting our results, the unit of time is the orbital
period at the disc inner edge, $P_{in}=2 \pi$.

%
\begin{figure}
  \center\includegraphics[width=0.9\columnwidth]{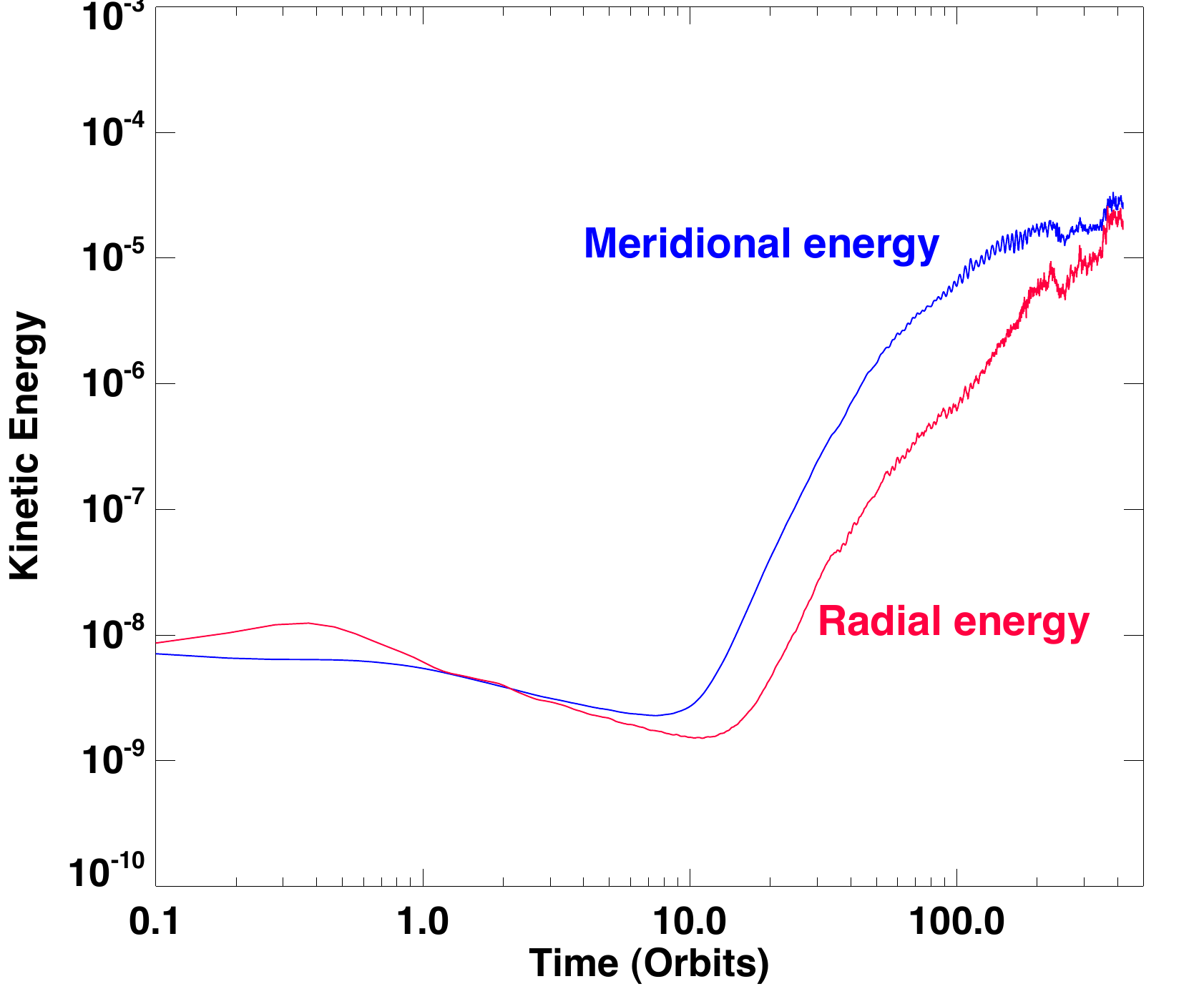}
  \caption{Time evolution of the normalised perturbed kinetic energy in the
   meridional and radial coordinate directions for model T1R-0 with
   $p=-1.5$, $q=-1$ and reflecting boundary conditions at the
   meridional boundaries.}\label{corr-breath}
\end{figure}

\section{Results}
\label{sec:Results}

The main aims of the following simulations are to
delineate conditions under which the disc models
outlined in Sect.~\ref{sec:disc-model} are unstable
to the growth of the vertical shear instability,
 and to examine the effect that different  physical and numerical 
set-ups have on the growth and evolution of this instability. We 
also aim to characterise the final saturated state of these unstable
discs, although our adoption of mainly axisymmetric simulations provides 
some restriction in achieving this final aim.


We define the volume integrated meridional and radial kinetic energies through 
the expressions
\begin{equation}
e_{\theta} = \frac{1}{2} \int_V \rho v^2_{\theta} dV,  \;\;\; e_{r} = \frac{1}{2} \int_V \rho v^2_{r} dV.
\label{eqn:ecorr}
\end{equation}
When presenting our results we normalise these energies by the
total kinetic energy contained in keplerian motion in the disc initially.

We begin discussion of our results below by describing simulations
of locally isothermal discs for which $T(R) \sim R^{-q}$ and 
$\rho(R) \sim R^{-p}$.
We describe one fiducial model in detail, before discussing
briefly the influence of the temperature profile in
controlling the instability. We present a comparision between
results obtained using the two codes described in Sect.~\ref{sec:methods},
and also demonstrate how the instability evolves as a function of disc 
viscosity.

The next set of results we present are for disc models
in which the initial temperature is a strict function of
$R$, but we set $\gamma=1.4$, solve the energy in 
eqn.~(\ref{eqn:motion}),
and allow the temperature to relax toward its initial value
using eqn.~(\ref{eqn:t-relax}). In this section, we examine how the
thermal relaxation rate contols the instability,
covering the full range of thermodynamic behaviour from
locally isothermal through to isentropic.

In the penultimate part of our numerical study, we consider disc models for which 
the entropy function, $K_s$, is a strict function of $R$,
and again employ thermal relaxation to examine the
conditions under which accretion discs display the vertical shear instability.
The final numerical experiment we present examines the instability in
a full 3D model, and provides an estimate of the Reynolds stress induced by
the instability.

We present an analytic model in the discussion section which 
illustrates the basic mechanism of the instability and 
delineates the conditions under which it operates.


\subsection{A fiducial model}
\label{sec:fiducial}
We begin presentation of the simulation results by
discussing one particular model in detail to illustrate 
the nature of the instability that is the focus of this paper.
The fiducial model is TR1--0 listed in Table \ref{table-NIRVANA},
with temperature, $T$, defined as a function of $R$ only, and
$p=-1.5$ and $q=-1$ in eqns.~(\ref{eqn:rhoR}) and (\ref{eqn:TR}).
A locally isothermal equation of state is adopted.
As such, this disc model has parameters
very similar to those used in numerous previous studies 
of disc related phenomena \citep[e.g.][]{2001ApJ...547..457K, 
2006A&A...450..833C, 2011arXiv1102.0671F, 2010A&A...520A..14P}, 
although we focus primarily on the inviscid non-magnetised evolution here.

\begin{figure*}
 \includegraphics[width=0.6\columnwidth]{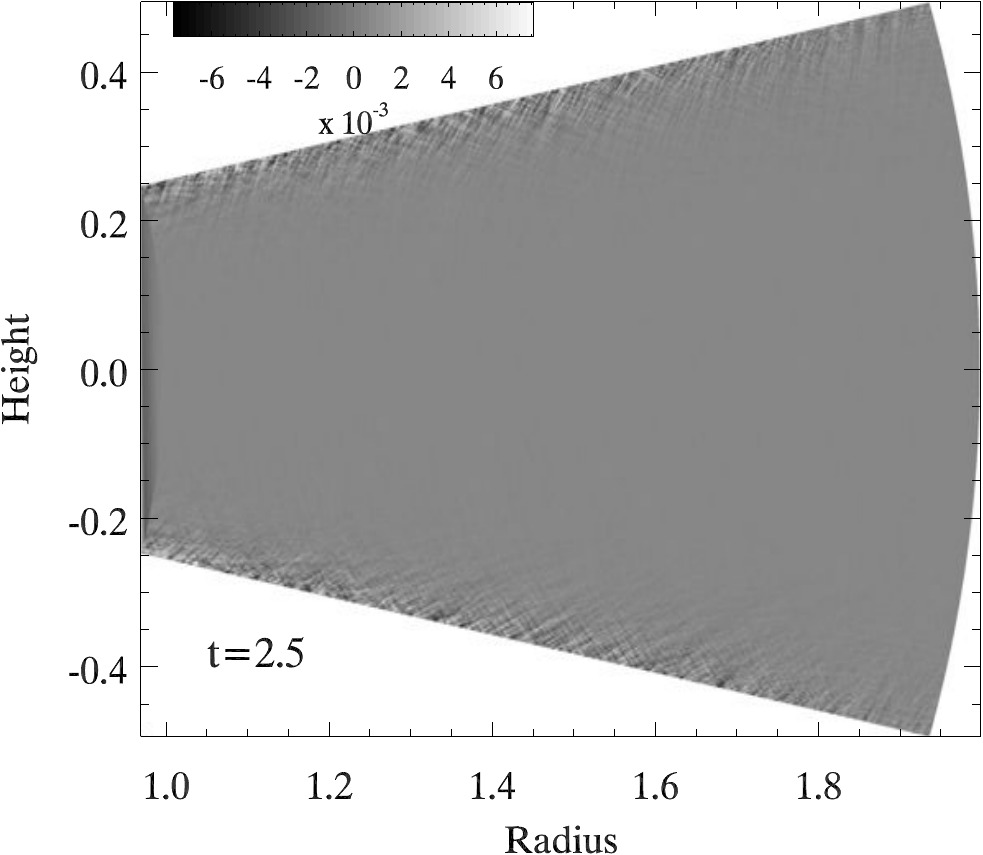}\hfill%
 \includegraphics[width=0.6\columnwidth]{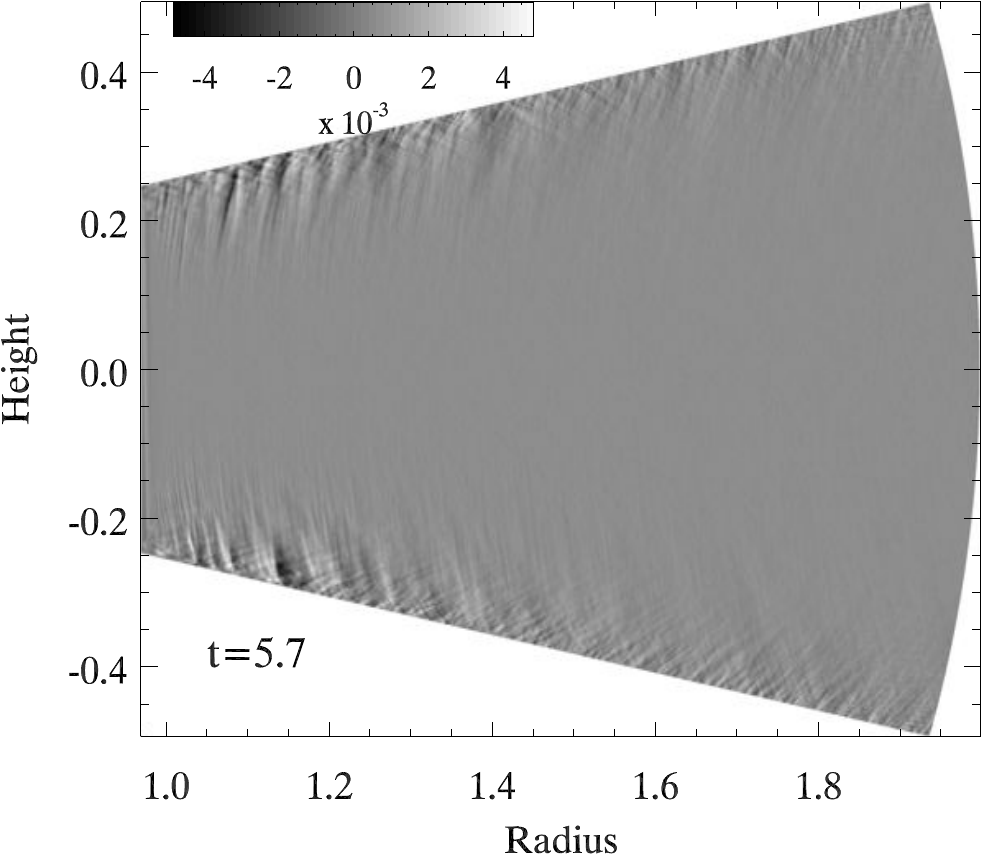}\hfill%
 \includegraphics[width=0.6\columnwidth]{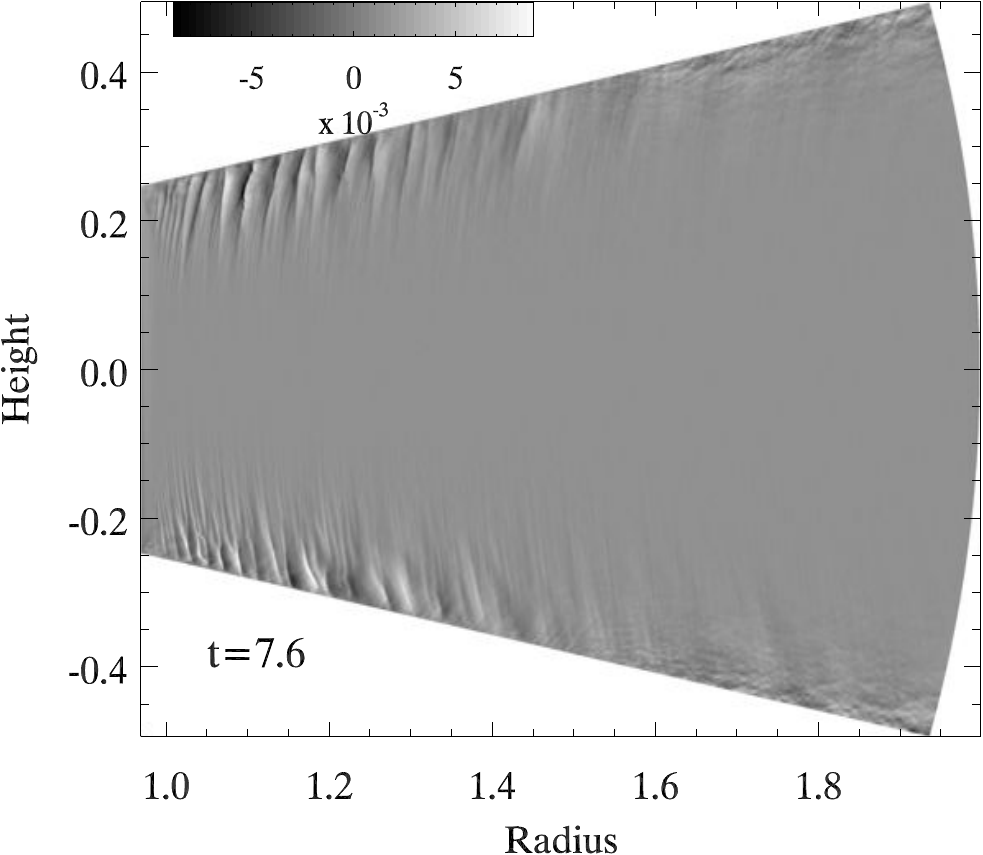}\hfill
 \includegraphics[width=0.6\columnwidth]{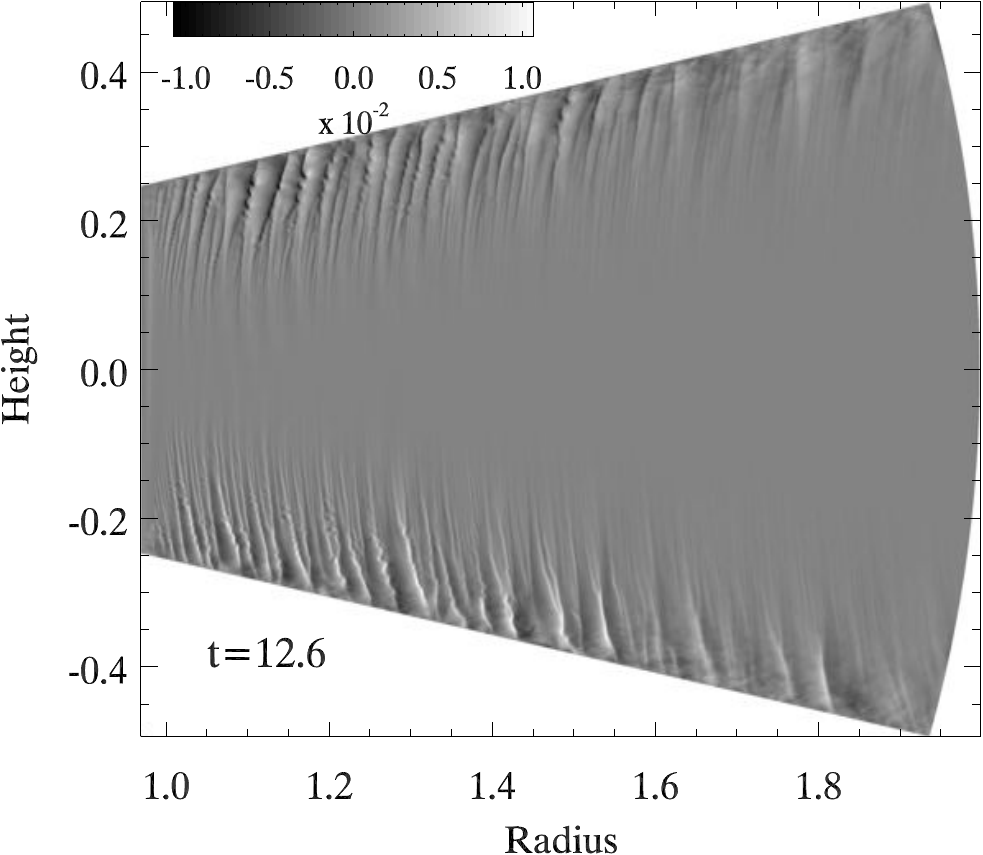}\hfill%
 \includegraphics[width=0.6\columnwidth]{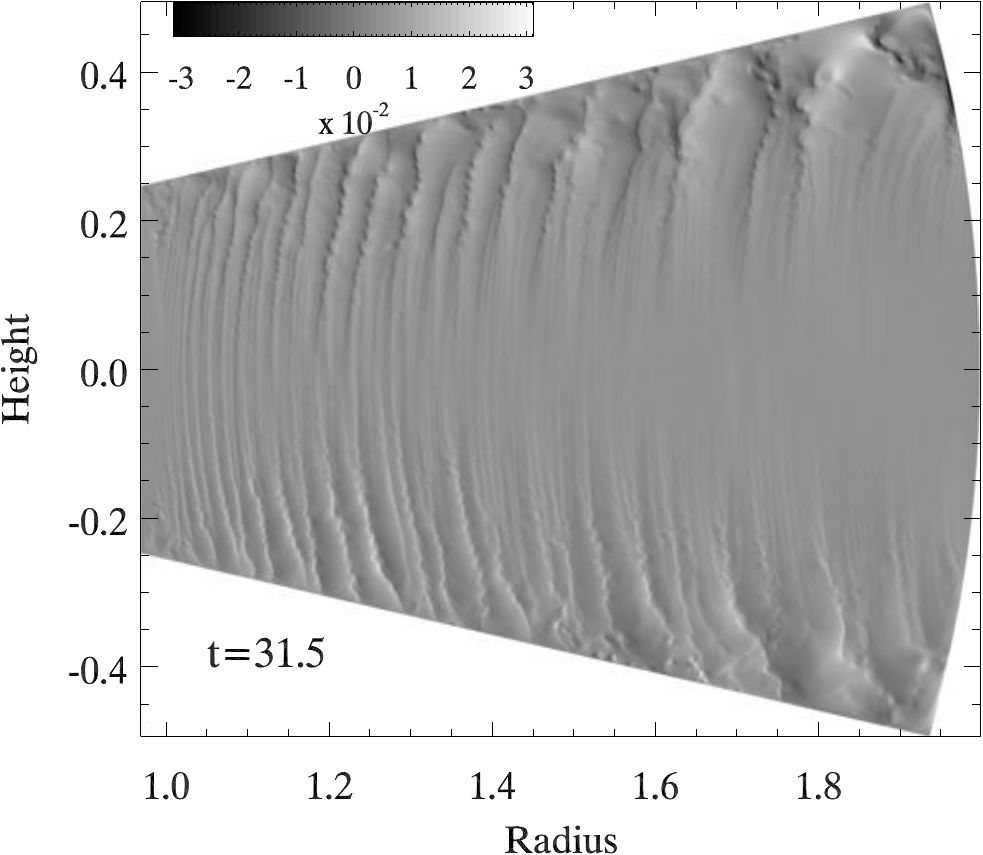}\hfill%
 \includegraphics[width=0.6\columnwidth]{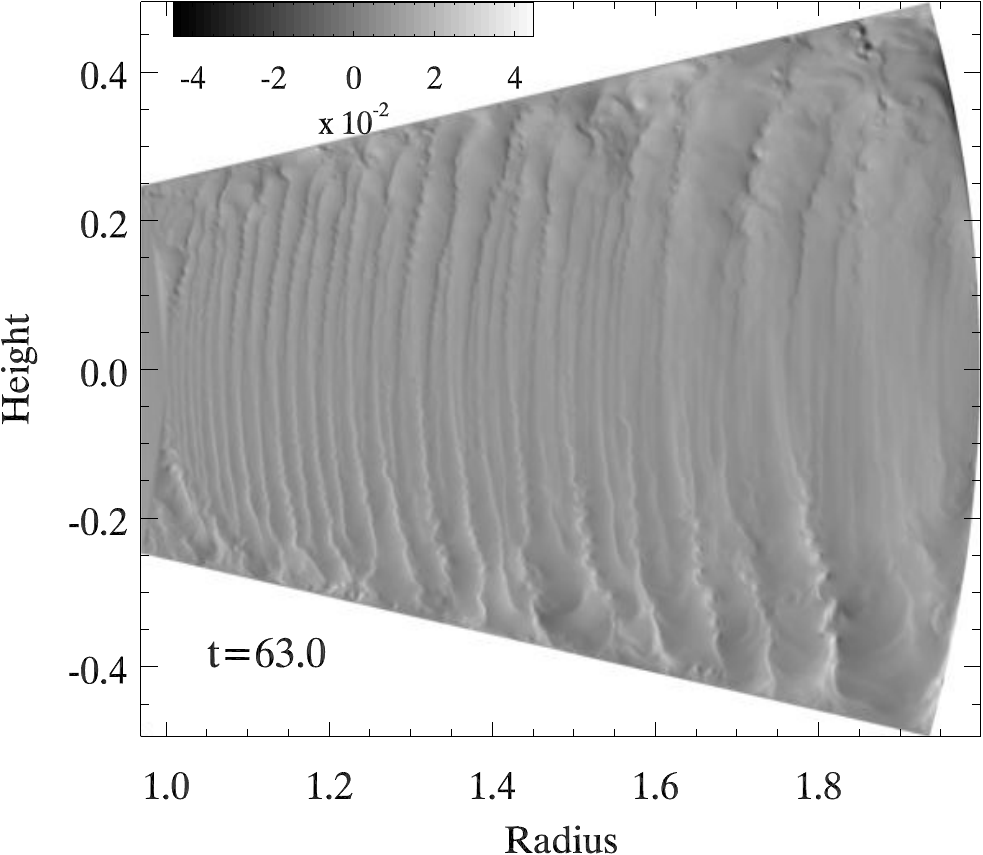}
 \caption{Edge-on contours of the perturbed vertical velocity
 as a function of $R$, $Z$ and time for model T1R--0.} \label{fig:vel-image-AO}
\end{figure*}

\begin{figure*}
 \includegraphics[width=0.6\columnwidth]{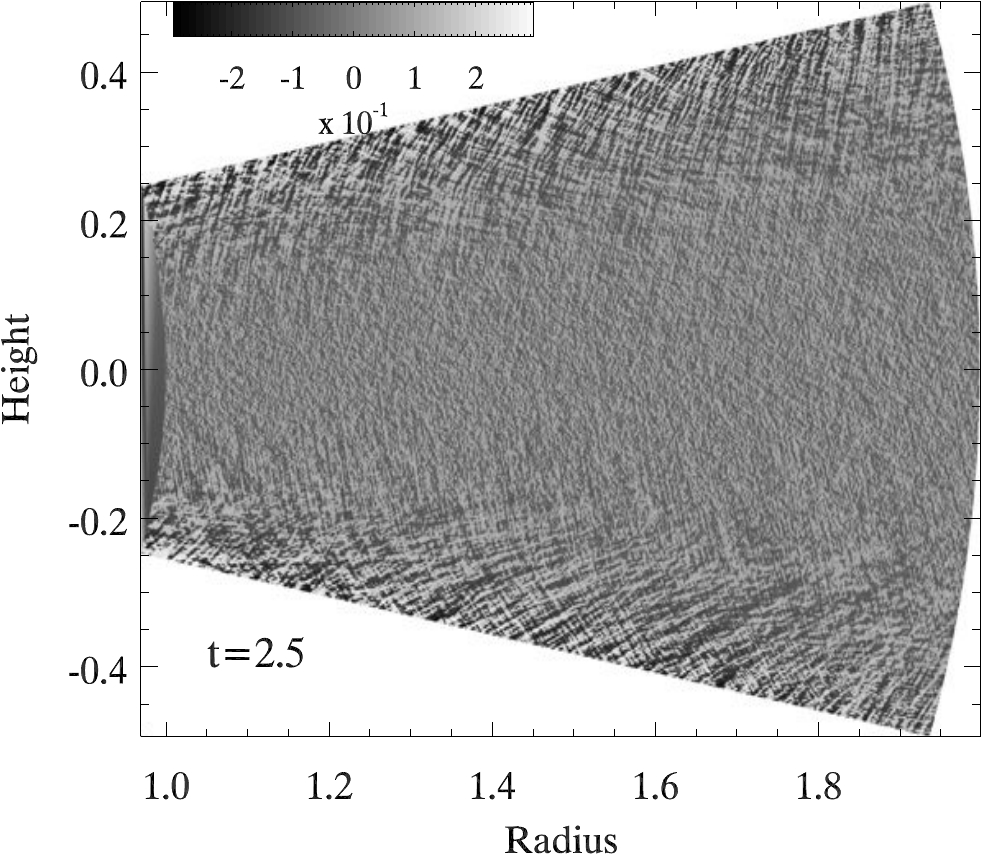}\hfill%
 \includegraphics[width=0.6\columnwidth]{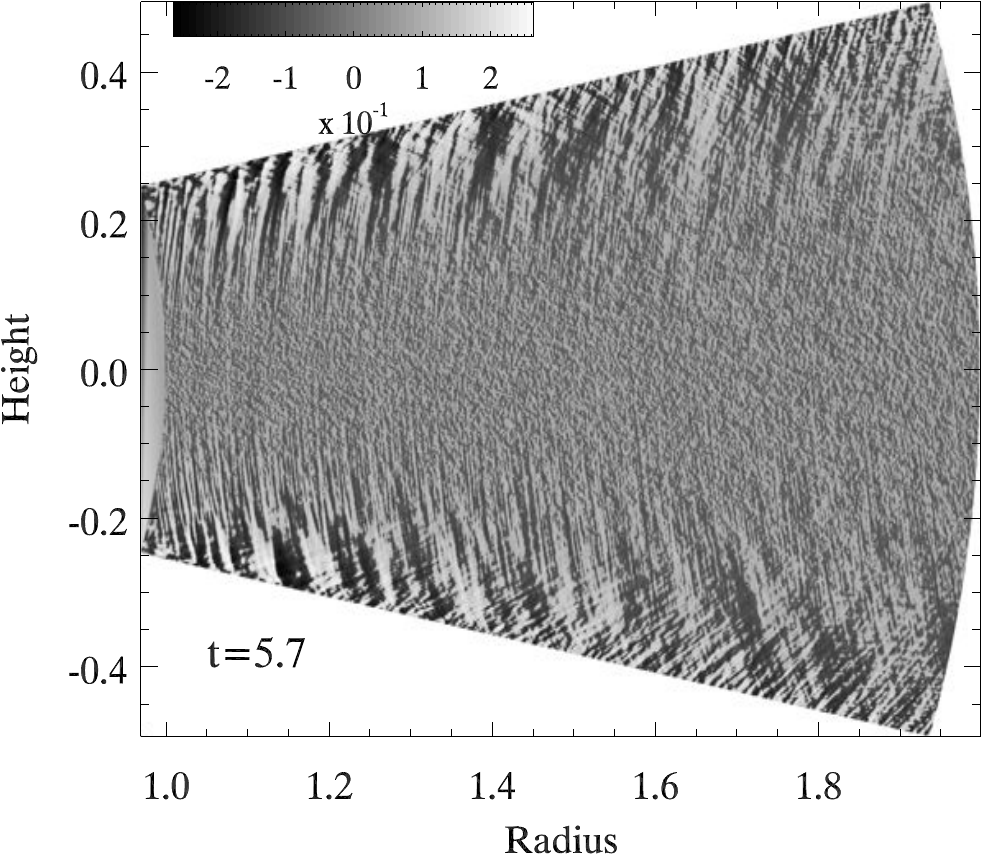}\hfill%
 \includegraphics[width=0.6\columnwidth]{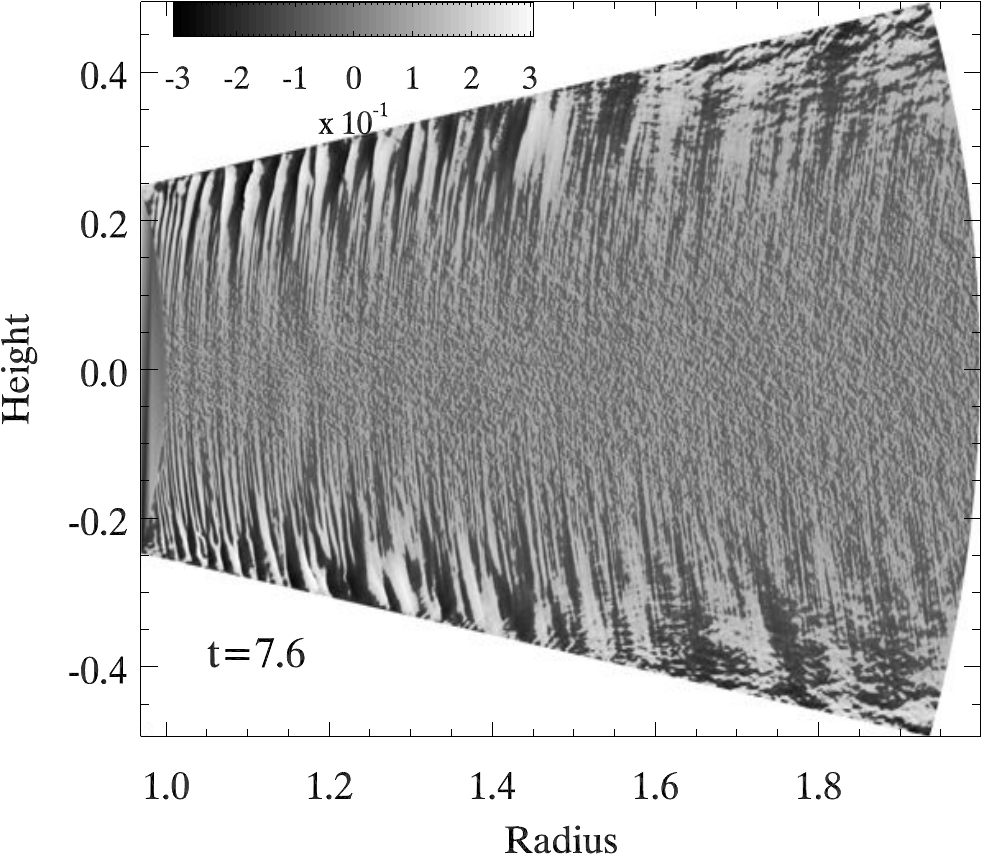}\hfill
 \includegraphics[width=0.6\columnwidth]{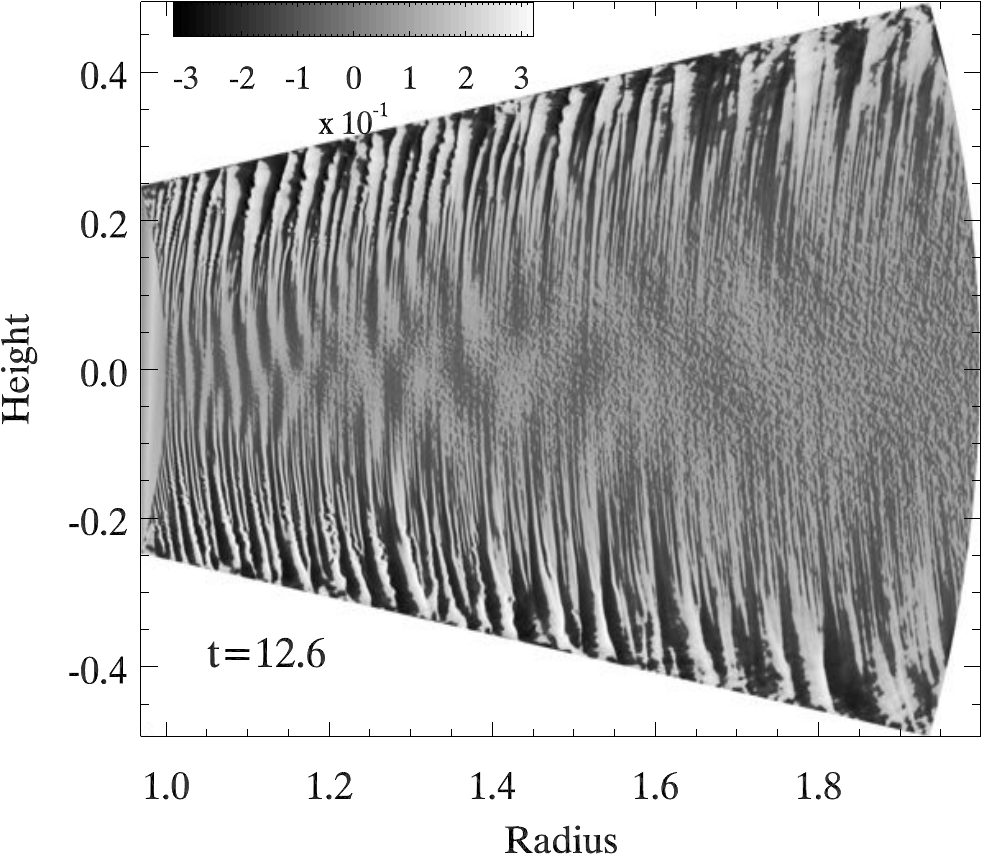}\hfill%
 \includegraphics[width=0.6\columnwidth]{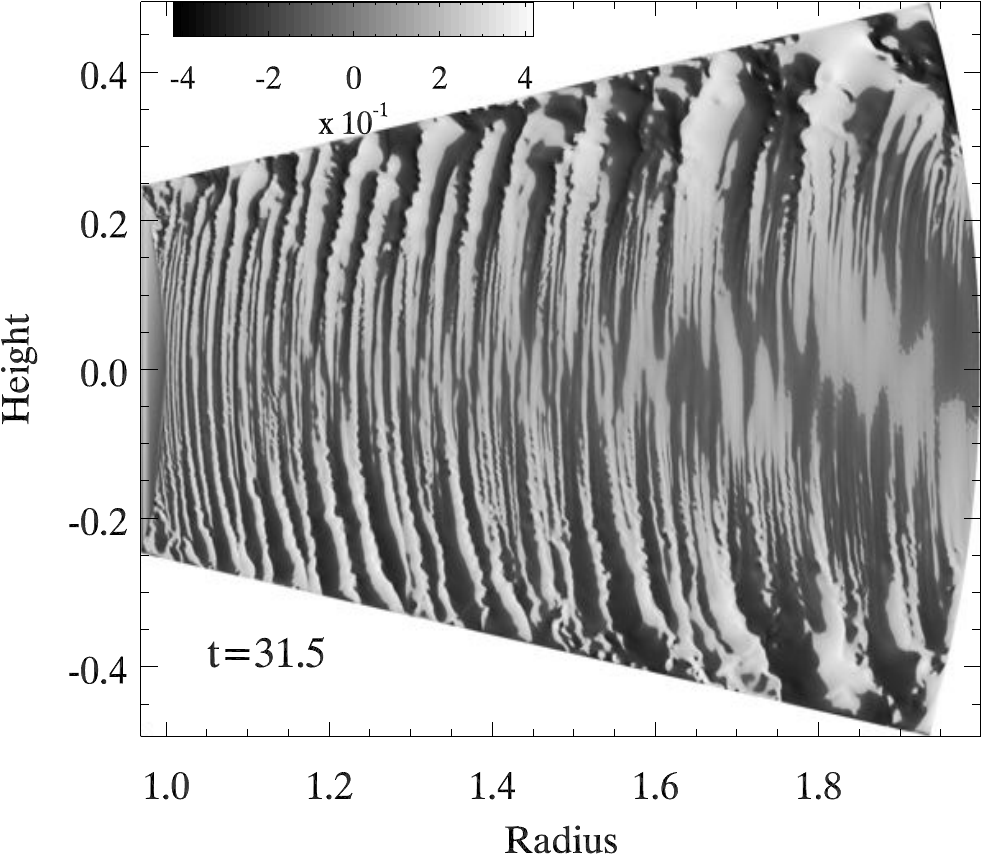}\hfill%
 \includegraphics[width=0.6\columnwidth]{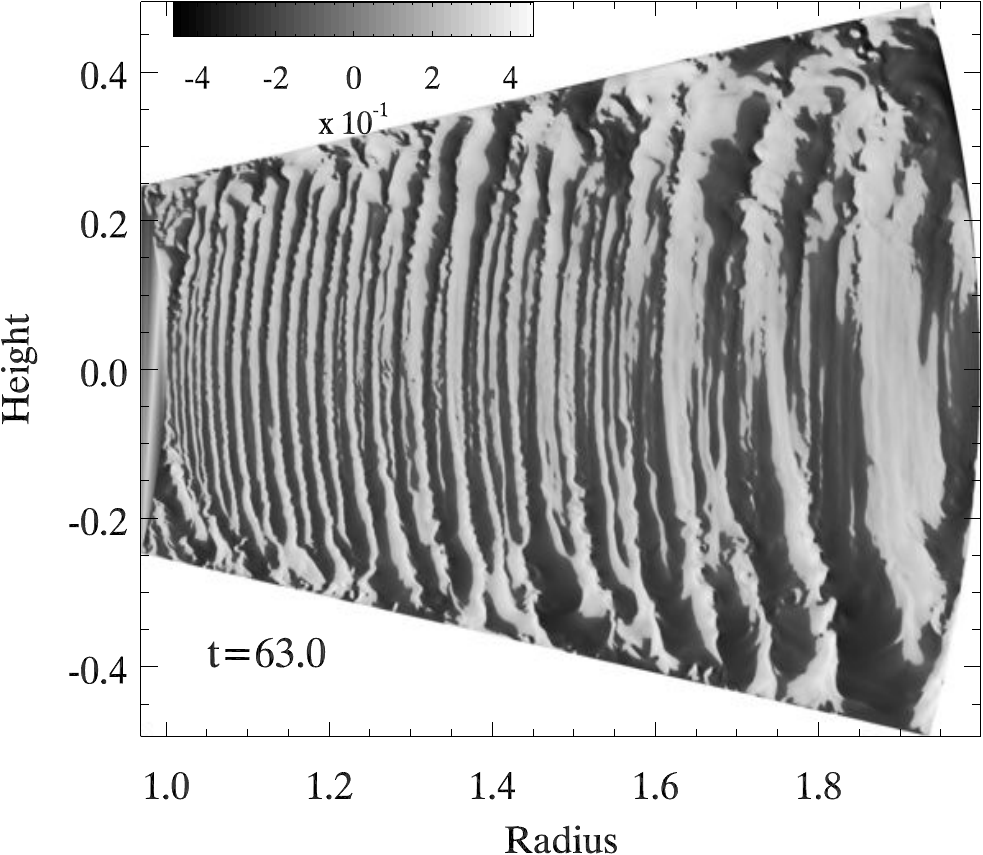}
 \caption{Edge-on contours of the perturbed vertical velocity
 as a function of $R$, $Z$ and time for model TR1--0. Note that for clarity,
  the grey-scale of the
 image has been streteched by plotting the quantity ${\rm sign}(v_Z) \times |v_Z|^{1/4}$.
 Note that the spectrum bar shows values of $v_Z^{1/4}$.} \label{fig:vel-image-AO-stretch}
\end{figure*}

The time evolution of the normalised meridional and radial kinetic
energies defined in eqn.~(\ref{eqn:ecorr})
are shown in Fig.~\ref{corr-breath}. The intial values at $t=0$
originate from the seed noise, and we observe that after 
$\sim 10$ orbits, during which the perturbed kinetic
energies damp slightly, rapid growth of the perturbation energies
arises. The normalised energies reach non linear saturation
after $\sim 400$ orbits having reached values of a few $\times 10^{-5}$.
Inspection of the evolution of the sum of the meridional plus radial kinetic
on a log-linear plot indicates that the linear growth rate of the perturbed energy
in Fig.~\ref{corr-breath} is $\simeq 0.24$ orbit$^{-1}$.

\begin{figure}
  \center\includegraphics[width=0.9\columnwidth]{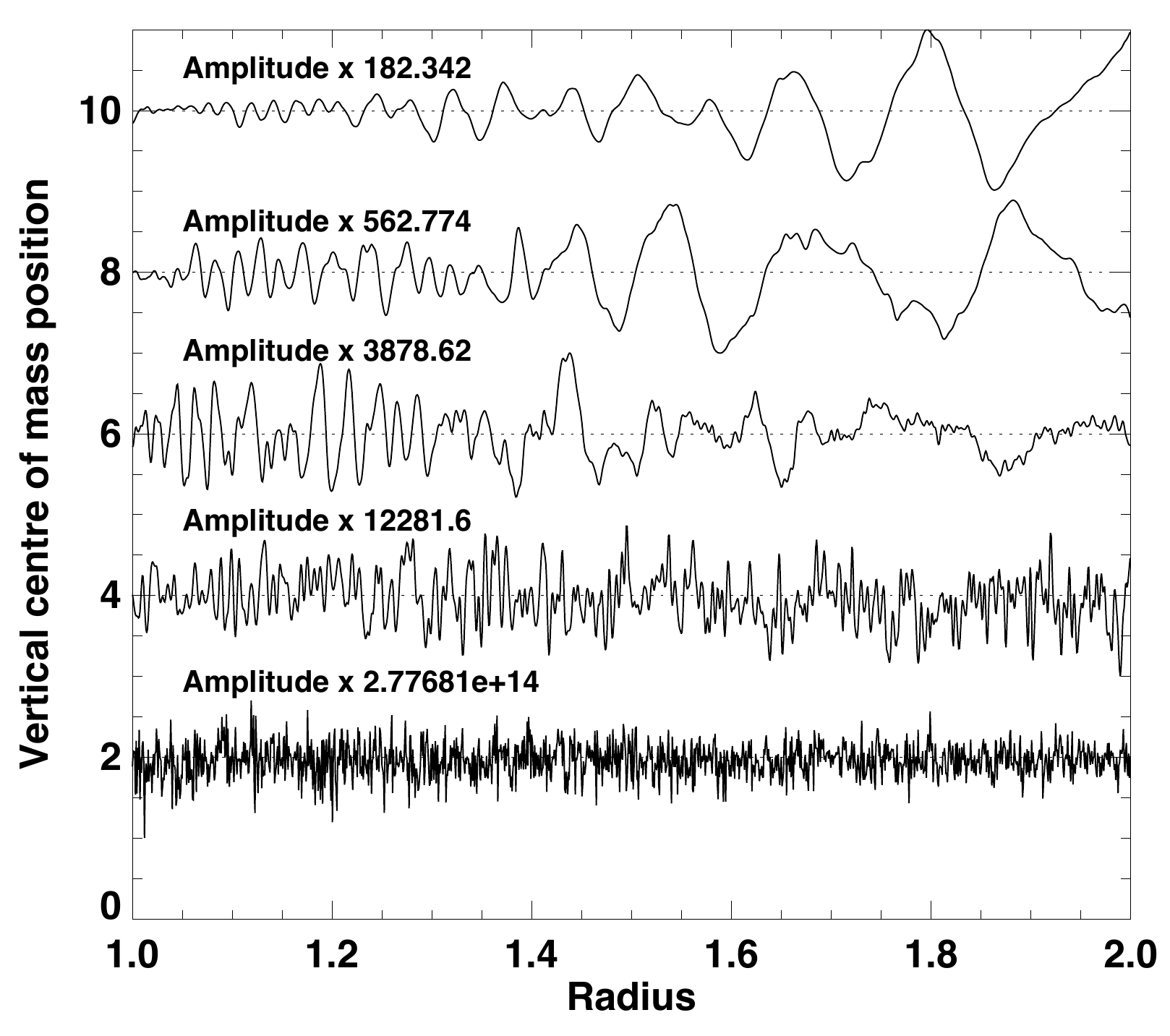}
  \caption{Time evolution of the vertical centre of mass position for each
   radial location in the disc for the fiducial model T1R--0 with $H/R=0.05$.
   Note that the vertical centre of mass position has been normalised by the
   local scale height at each radius. Starting from bottom to top the
   plots correspond to times (in orbits): $9 \times 10^{-4}$, 9.18, 27.86, 65.00,
   92.66. The multiplicative factor indicated in each legend causes the
   maximum amplitude of the normalised c.o.m. position in each graph
   to equal unity.
   }\label{fig:corrugation}
\end{figure}

\begin{figure*}
 \includegraphics[width=0.6\columnwidth]{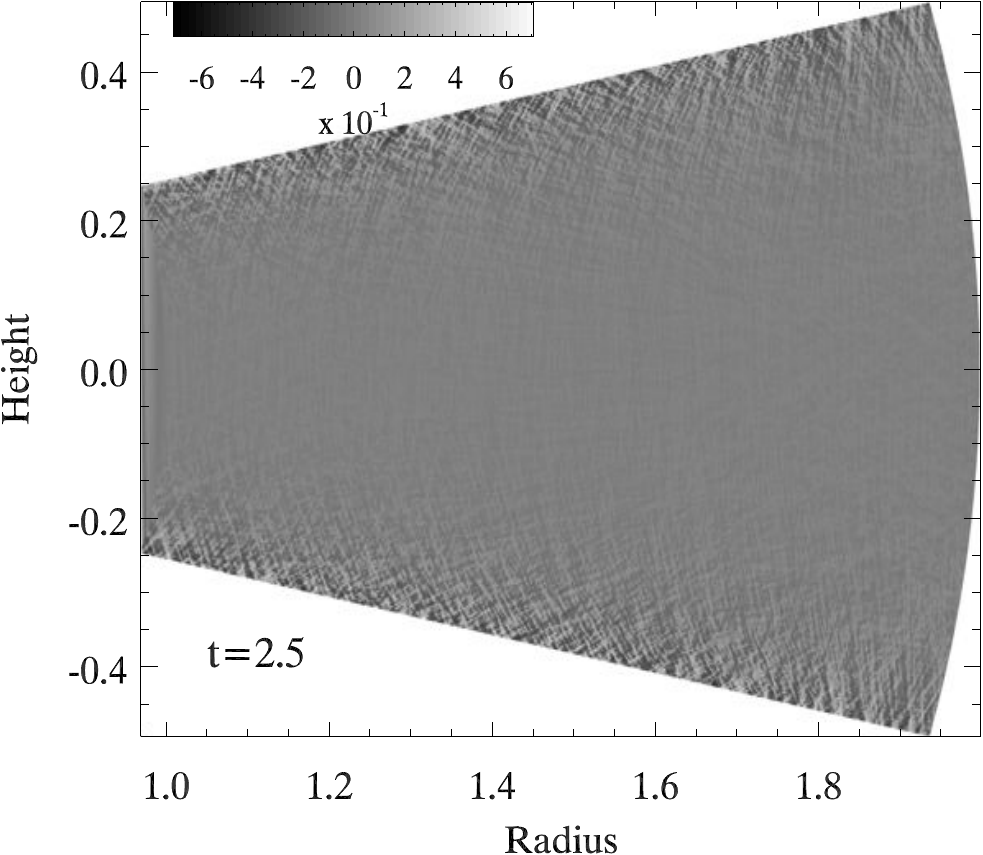}\hfill%
 \includegraphics[width=0.6\columnwidth]{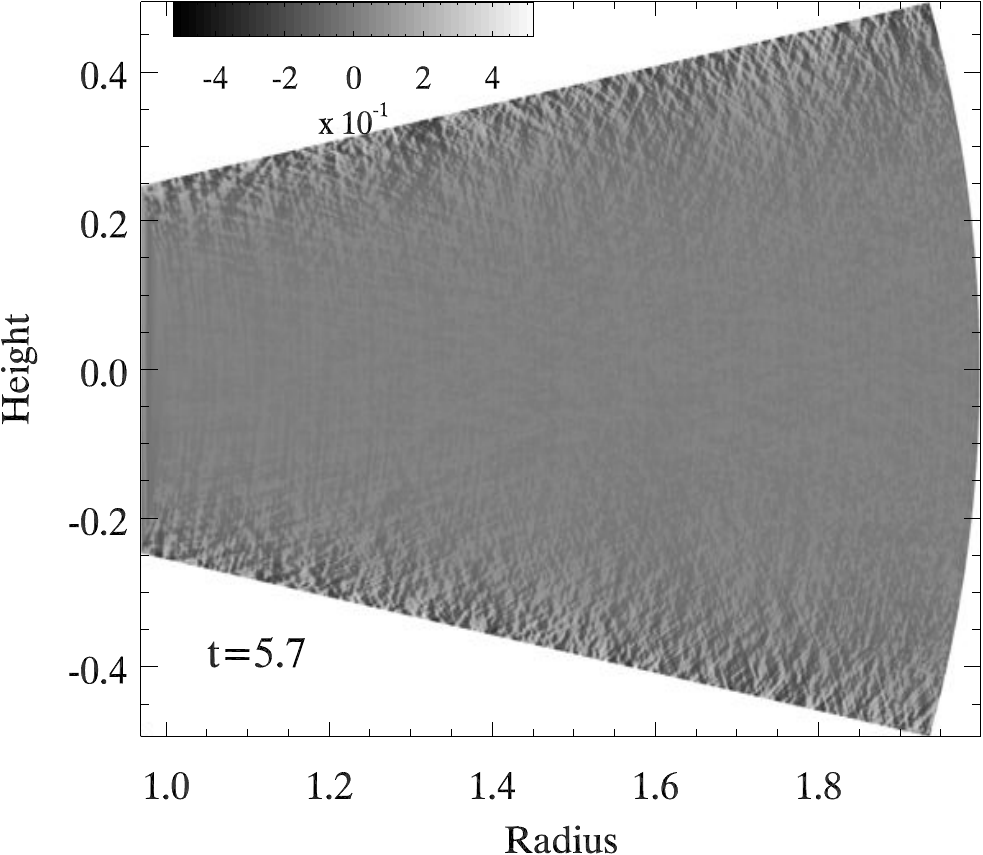}\hfill%
 \includegraphics[width=0.6\columnwidth]{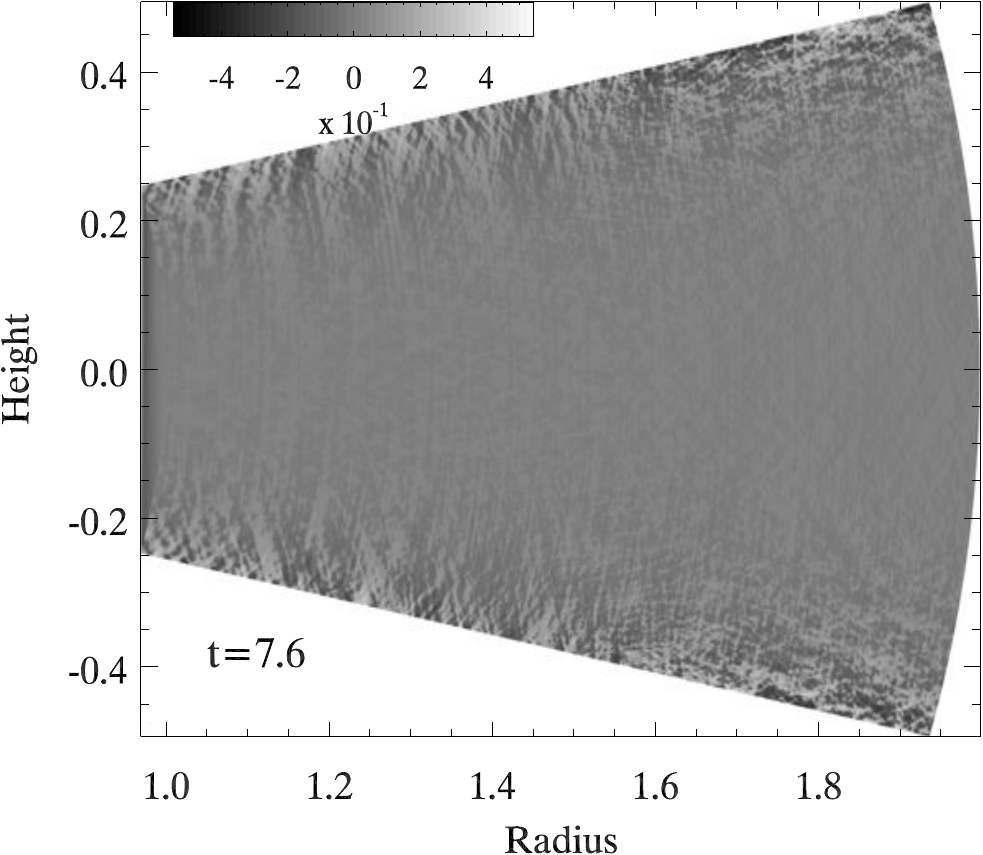}\hfill
 \includegraphics[width=0.6\columnwidth]{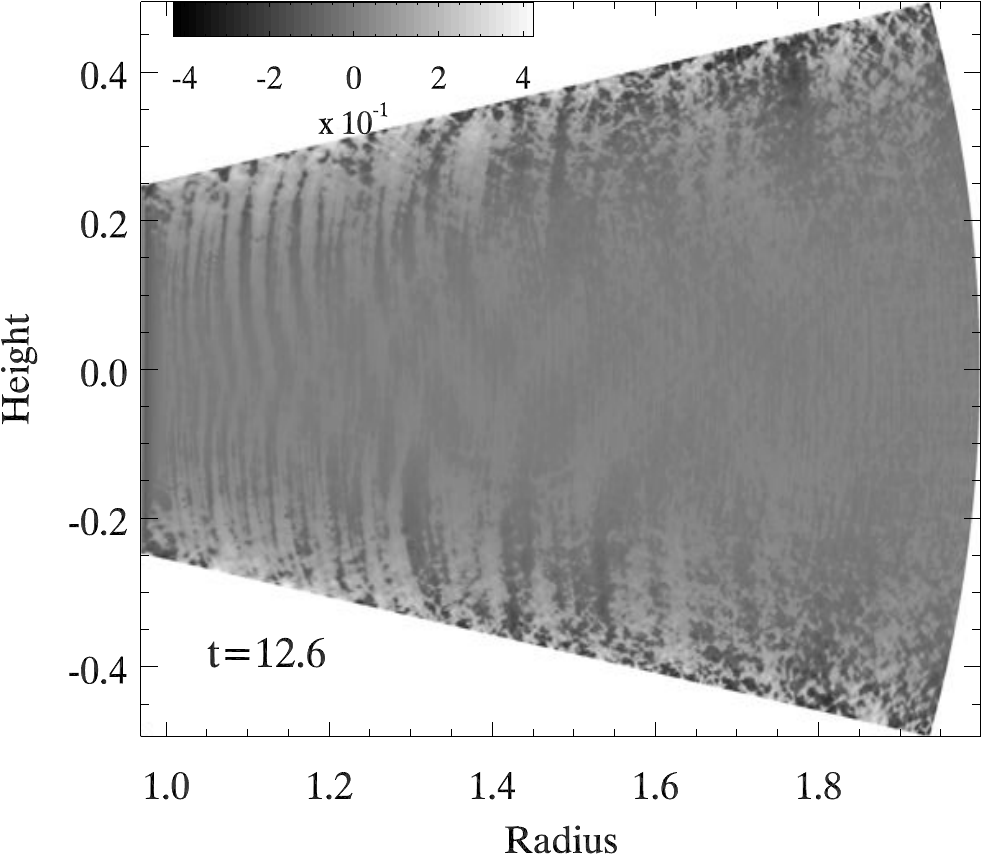}\hfill%
 \includegraphics[width=0.6\columnwidth]{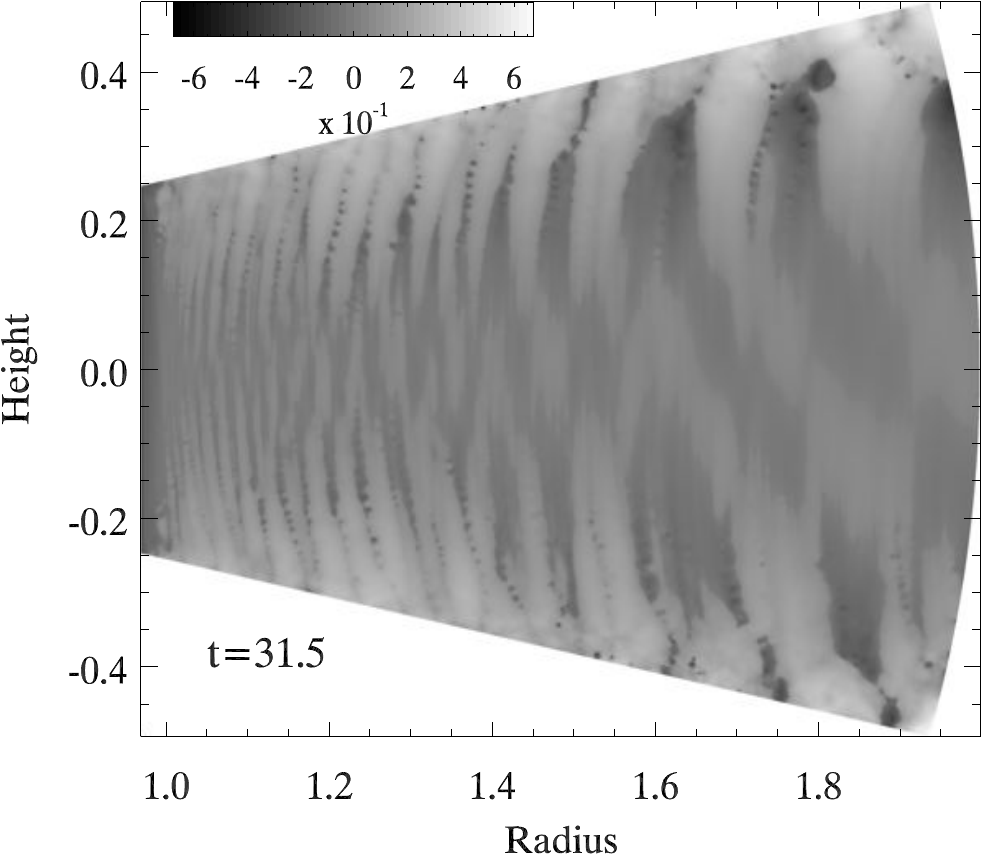}\hfill%
 \includegraphics[width=0.6\columnwidth]{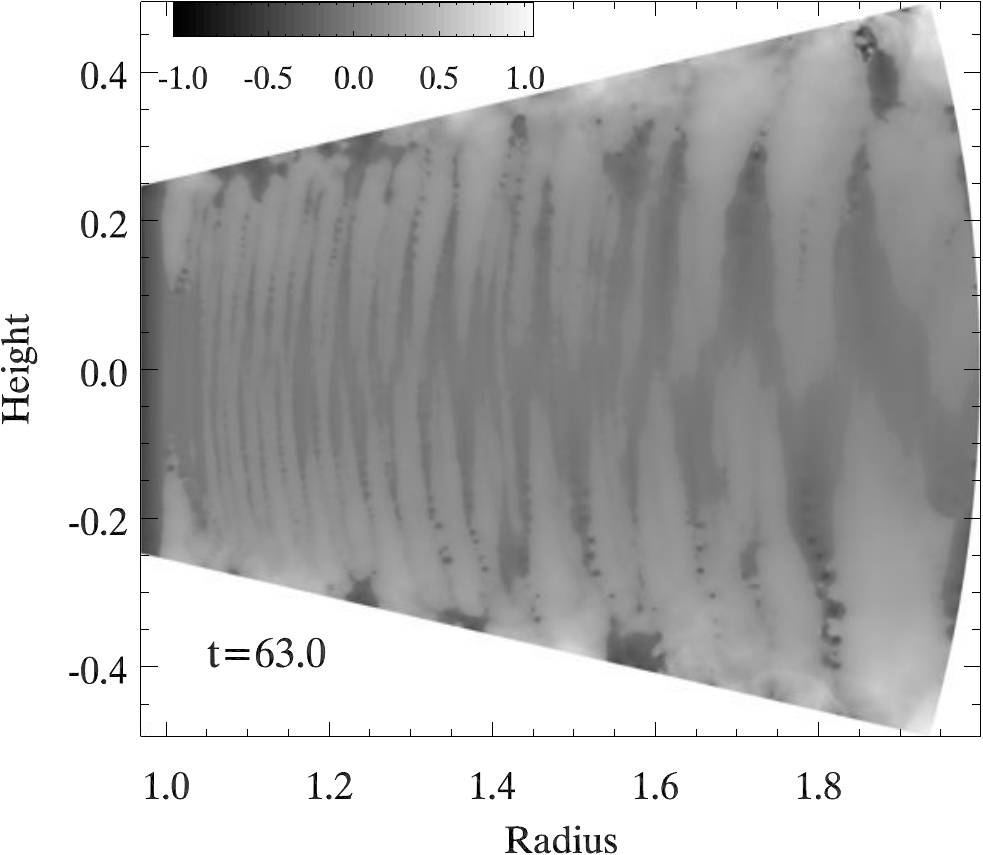}\hfill%
 \caption{Edge-on contours of the disc relative density perturbations
  $\delta \rho/\rho_0$ as a function of $R$, $Z$ and time. Note that we have
  effectively stretched the grey-scale by plotting the quantity
   ${\rm sign} (\delta \rho) \times \sqrt{ \delta \rho / \rho_0 }$}.
 \label{fig:density-image-p1-5q1}
\end{figure*}

Contour plots of vertical velocity perturbations, $v_Z$, that 
arise at different stages of the disc evolution are shown in 
Figs.~\ref{fig:vel-image-AO} and \ref{fig:vel-image-AO-stretch}. 
These two figures show the perturbed velocity field at identical
times, but whereas Fig.~\ref{fig:vel-image-AO} maps linearly
between the velocity values and the grey-scale, 
Fig.~\ref{fig:vel-image-AO-stretch} plots the values
$\rm{sign}(v_Z) \times |v_z|^{1/4}$ so that the grey-scale is
stretched to enable the morphology of the perturbations to be
more clearly discerned. Both figures demonstrate that perturbations
start to grow near the upper and lower disc surfaces, where
$|{\rm d} \Omega /{\rm d} Z|$ is largest, and toward the inner edge of the disc. 
The perturbations are characterised by having short radial and long 
vertical wavelengths, as expected for the vertical shear instability 
described in Sects.~\ref{sec:GSF} and \ref{sec:analysis}. The short 
radial wavelength gives rise
to significant radial shear in the vertical velocity ${\rm d} v_Z/{\rm d} R$, and 
this apparently causes small scale eddies to form at the 
shearing interfaces. As time proceeds the instability extends toward 
the disc midplane and out to larger radii, until the entire disc 
participates in the instability (although it should be noted that the
midplane where ${\rm d} \Omega /{\rm d} Z = 0$ is formally stable to local
growth of the vertical shear instability).

Close inspection of the lower panels in Fig.~\ref{fig:vel-image-AO-stretch}
show that the velocity perturbations have odd symmetry about the midplane
initially (this is particularly apparent in the fourth panel between
radii $1.1 \le r \le 1.4$). In other words the disc exhibits a breathing
motion about the midplane as the instability first becomes apparent
at lower disc latitudes. As time proceeds, however, the velocity
perturbations become symmetric about the midplane as demonstrated by
the fifth and sixth panels. These perturbations correspond to corrugation
of the disc characterised by coherent oscillations of the 
vertical centre of mass position whose phase depends on radius in a time
dependent manner. The development of this disc corrugation is illustrated by
Fig.~\ref{fig:corrugation} which shows the vertical centre of mass position
of the disc at each radius for five different times (note that each plot is
off-set in the vertical direction to aid clarity, and each curve
has been multiplied by a unique factor so that the corrugation may be
observed). The vertical centre of mass has been normalised by the local disc
scale height at each radius. Moving from the lower curve to the upper curve, 
we note that the vertical centre of mass position has a very small variation 
with radius after one time step, but this becomes progressively larger in 
amplitude and more spatially coherent as time progresses (times corresponding 
to each curve are given in the figure caption). The final curve, corresponding
to an evolution time of $\sim 92$ orbits, has a maximum vertical displacement 
approximately equal to $0.006 H$. It is interesting 
to note that the initial disc instability begins with $|k_R/k_Z| \gg 1$ due to the
short radial wavelengths of the fastest growing modes of the vertical
shear instability. As the disc approaches the nonlinear state,
however, the development of coherent corrugation waves causes the
radial wavelengths of the most apparent perturbations to approach
or modestly exceed the local scale height. At the end of the simulation
($\sim 420$ orbits) the maximum vertical displacement of the disc centre
of mass reaches $\sim 0.01 H$, but we are cautious about interpreting the results
at this late stage of evolution as the reflecting boundary conditions may play a role.

Contour plots of the density perturbations $\delta \rho/\rho_0$ corresponding to 
the previously discussed velocity contours are displayed in 
Fig.~\ref{fig:density-image-p1-5q1}. As with the velocity contours, we see 
perturbations first arise near the disc surface. Coherent density structures 
first become visible in the third panel, but are substantially more apparent in 
the lower panels.

The final saturated state consists of locally unstable disc annuli that oscillate
vertically at close to the local frequency, $\Omega_{\rm k}(R)$, superposed on 
which are a spectrum of oscillations with different frequencies caused by 
travelling waves excited by vertical oscillations at 
neighbouring disc radii. A region of the disc lying at intermediate radii will thus 
experience a locally generated corrugation, in addition to inward travelling corrugation 
waves that propagate as inertial modes (or $r$-modes) launched from exterior disc locations,
and outward propagating corrugation waves propagating as
acoustic or fundamental modes launched from interior disc locations
\citep{1993ApJ...409..360L, 1998ApJ...504..983L}.

Although we only present simulations with finite amplitude initial perturbations
to the velocity fields in this paper, we have conducted numerous experiments in
which the peak amplitude of the imposed perturbations varies, including cases
where perturbations just grow from numerical round-off errors. Although this requires
a larger time interval for the instability to become apparent, we nonetheless observe
instability for all perturbation amplitudes, demonstrating that the instability
is linear.

\subsection{Evolution as a function of the radial temperature profile}
\label{sec:t-profile}

In this section we discuss simulations T1R--0 to T4R--0 (which
utilise reflecting boundary conditions in the meridional direction) and
T1O--0 to T4O--0 (which use open boundary conditions). These simulations
have $p=-1.5$ and a range of values for $q$ running from $q=-1$ (a 
constant $H/R$ disc) up to $q=0$ (a purely isothermal disc in
which $H/R \propto R^{1/2}$). 

\begin{figure}
  \includegraphics[width=0.5\columnwidth]{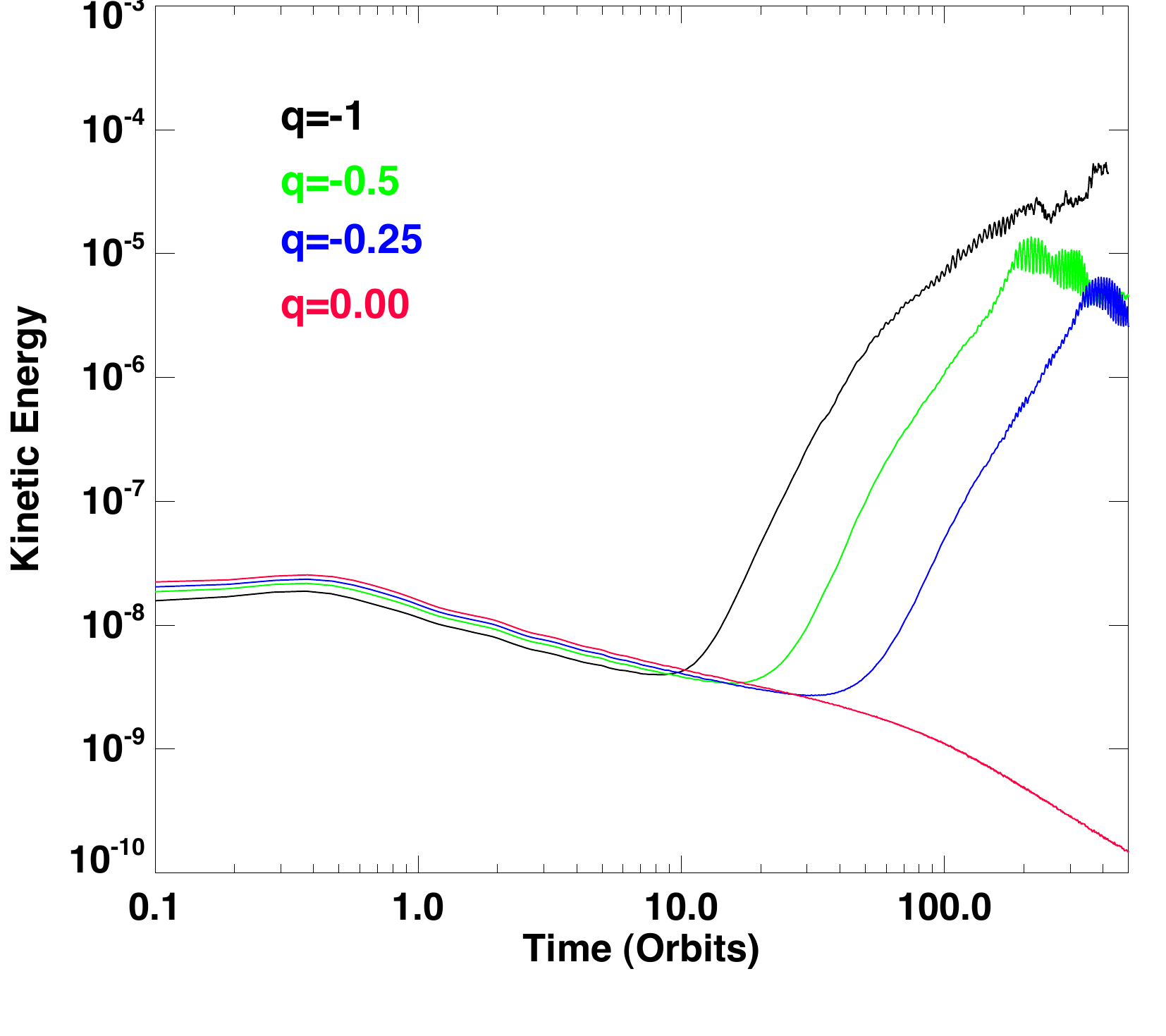}
  \includegraphics[width=0.48\columnwidth]{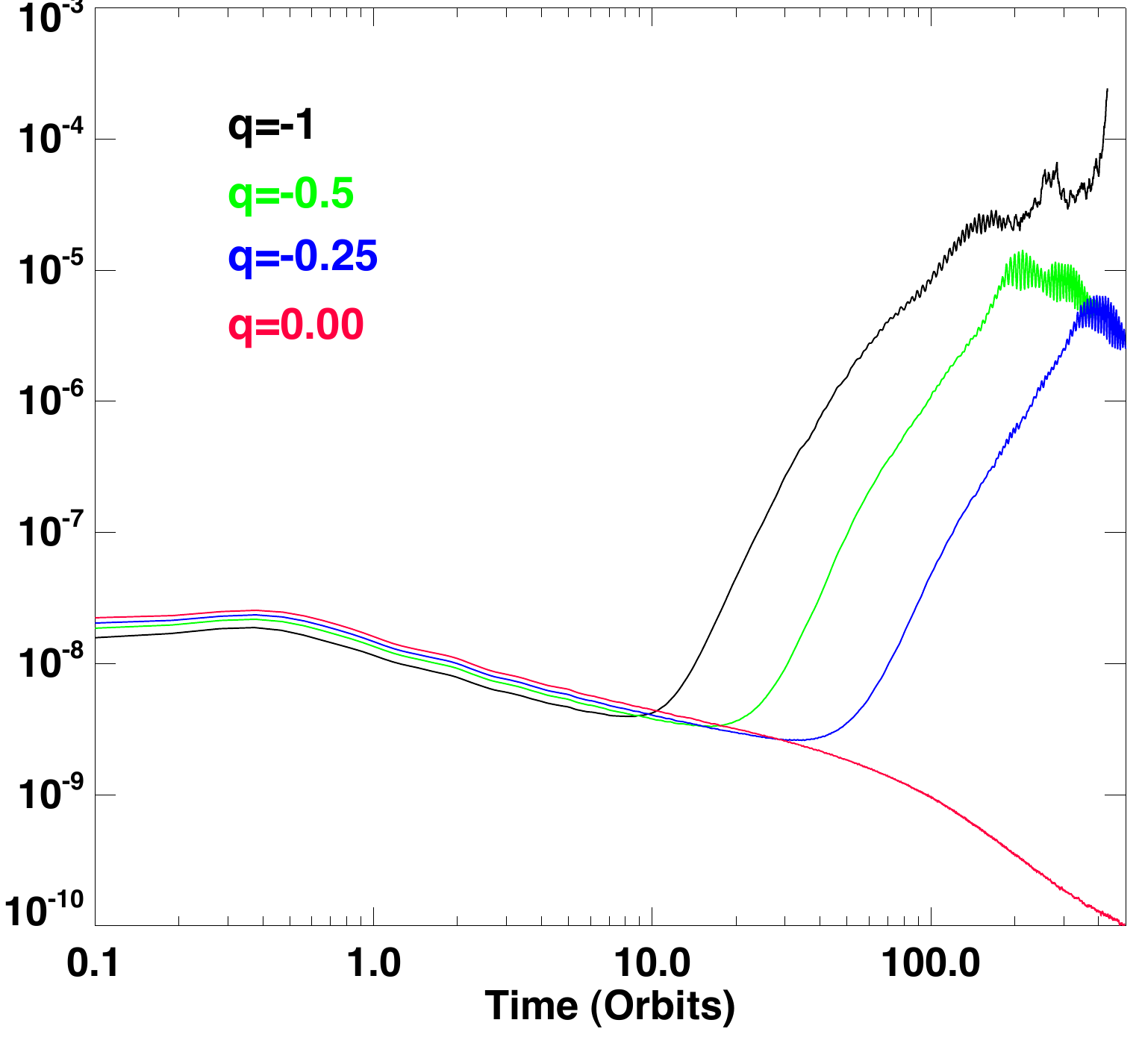}
  \caption{Time evolution of the perturbed meridional plus radial kinetic
   energies (normalised) as a function of the radial temperature profile.
   The left/right panel shows results for simulations which adopted
   reflecting/outflow boundary conditions at the upper and lower disc
   surfaces.} \label{energy-AO-AR}
\end{figure}

The left panel of Fig.~\ref{energy-AO-AR} shows the time evolution
of the normalised perturbed kinetic energy summed over the radial 
and merdional directions in the disc models T1R--0 to T4R--0.
We see that as the value of $q$ increases from $q=-1$ through
to $q=-0.25$, the growth rate of the instability decreases,
and for $q=0$ it switches off altogether. 
Inspection of the evolution of the sum of the meridional and radial kinetic
energies on a log-linear plot indicates that the linear growth rate for the
$q=-0.5$ case is $\simeq 0.12$ orbit$^{-1}$, and for the $q=-0.25$ run is
$\simeq 0.052$ orbit$^{-1}$ (to be contrasted with the
growth rate $\simeq 0.24$ orbit$^{-1}$ obtained for the $q=-1$ run).
The azimuthal velocity in the $q=0$ model is independent of $Z$, 
as indicated by eqn.~(\ref{eqn:Omega1}), so the observed stability of 
this disc model is in agreement with the expectations discussed in 
Sect.~\ref{sec:stability}. 
It is also noteworthy that the saturated values of the perturbation energies 
in each unstable disc, normalised to the total energy in keplerian motion,
differ substantially from one another in accord with the trend in
the growth rates.

The right panel of Fig.~\ref{energy-AO-AR} shows the corrugation
energies for runs T1O--0 to T4O--0, which are identical to runs T1R--0 to T4R--0
except that the boundary conditions applied at the meridional boundaries
are outflow rather than reflecting. During the growth phase of the
instability the results of the T1O--0 -- T4O--0 runs are essentially identical 
to the corresponding T1R--0 -- T4R--0 calculations, and the saturated energies
are very similar. We also observe the important result that the
transition between stable and unstable disc models is independent of
the boundary conditions, with both $q=0$ models showing decay of the
perturbed meridional and radial kinetic energies during the simulations. 
It is clear that the existence of a radial temperature profile plays a 
fundamental role in determining whether or not a disc becomes unstable.

\subsection{Code comparison}
\label{sec:code}

Numerical modelling of the Navier-Stokes equations can pose arcane
pitfalls, particularly if hydrodynamic instabilities are involved. 
In the absence of rigorous analytical reference solutions, 
it has become customary to substantiate the physical reliability of the
solutions obtained by means of code comparisons. While we have already
shown the generality of a GSF-like instability occurring under various 
physical settings, we demonstrate here its comparative development 
using two numerical schemes. Although very similar in their names, the two 
codes we used are fundamentally different in the numerical schemes they apply.  
\NIRVANA utilises the same finite-difference scheme as the \textsc{zeus} 
code, whereas \NIII applies a finite-volume Godunov scheme very similar to the 
ones used in, e.g., \textsc{ramses} \citep{2002A&A...385..337T} or 
\textsc{athena} \citep{2008JCoPh.227.4123G}.

\begin{figure}
  \center\includegraphics[width=0.9\columnwidth]{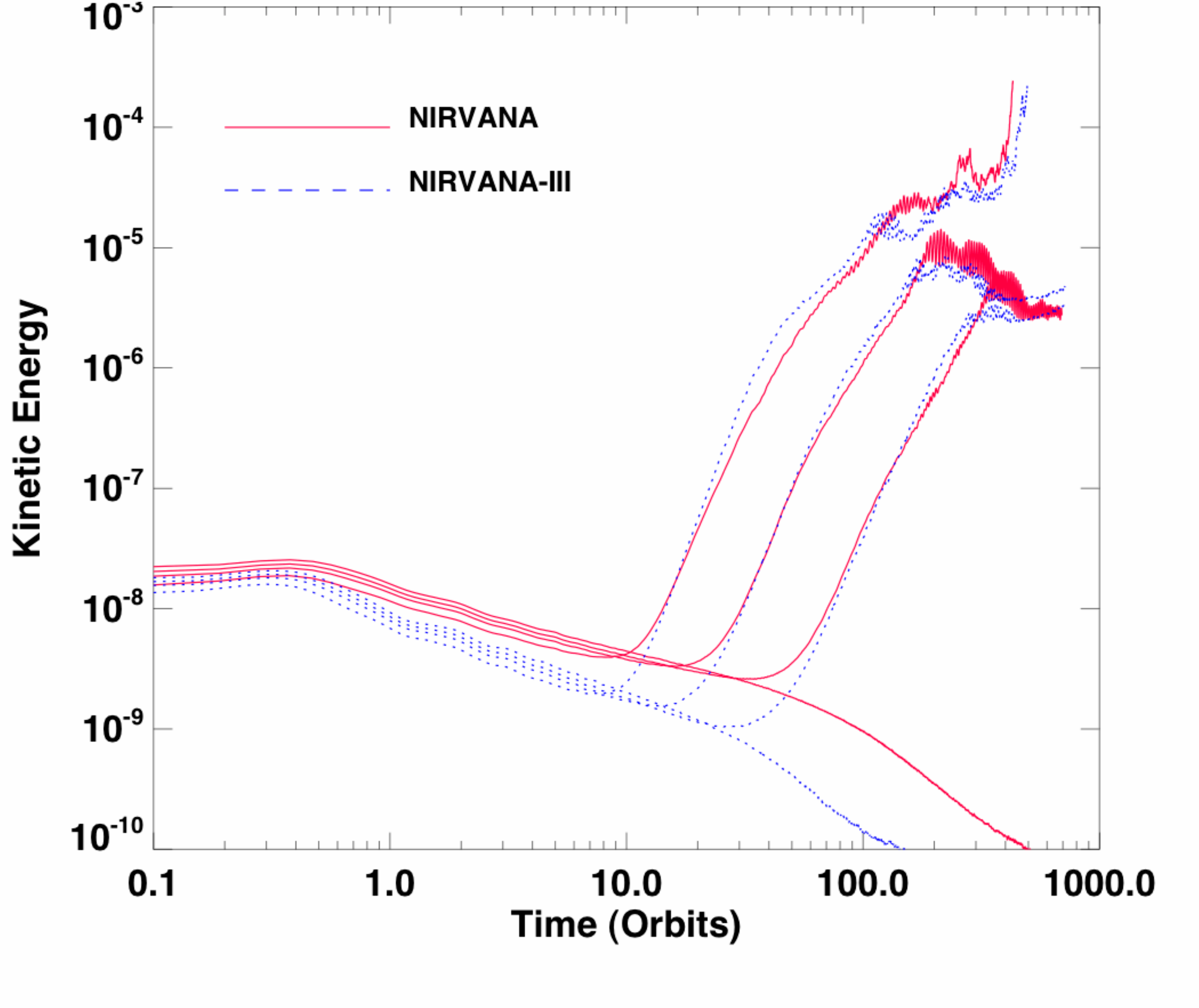}
  \caption{Comparison of the time evolution of the normalised 
   (meridional + radial) kinetic energies for models T1O--0 to T4O--0 
   that were run with the two codes.}
    \label{fig:code_cmp}
\end{figure}

\begin{figure}
  \center\includegraphics[width=0.9\columnwidth]{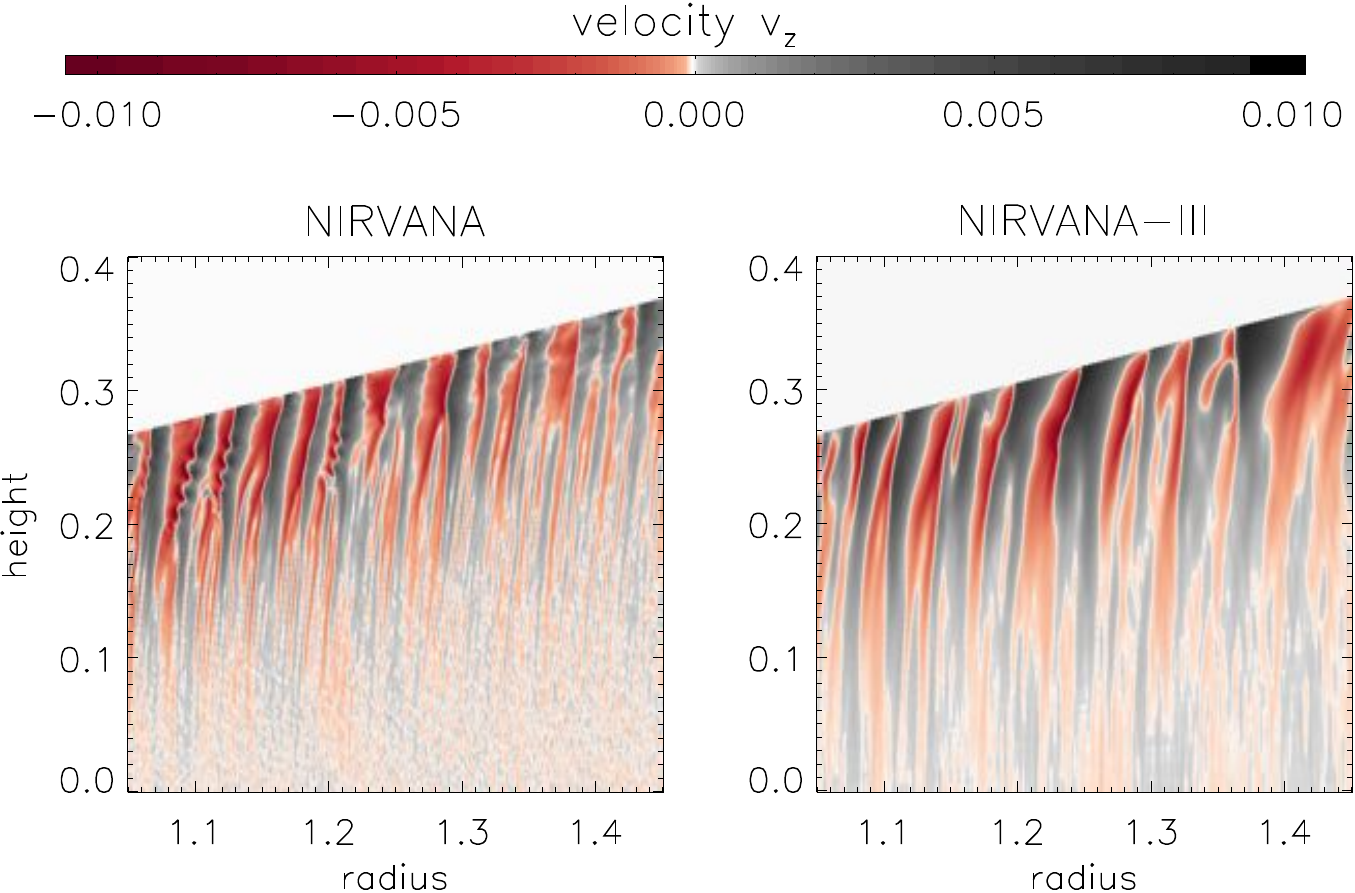}
  \caption{Comparison between the vertical velocity perturbations
  at time $t=12.8$ orbits for the \NIRVANA and \NIII runs. The results
  are for simulation T1O--0.}
    \label{fig:code_cmp2}
\end{figure}

We have run the simulations labelled as T1O--0 to T4O--0 in Table~1 using
both codes. The codes generally agree well in the development of
the instability and its saturation level as demonstrated
by Fig.~\ref{fig:code_cmp}, although it should be noted that the realisation
of the initial seed noise in the two runs is slightly different. The decay of 
these initial seed perturbations occurs slightly faster in the NIRVANA-III simulations, 
and the saturation amplitude is slightly smaller. A detailed look at the vertical velocity
perturbations at time $t \simeq 12.8$ orbits is shown in 
Fig.~\ref{fig:code_cmp2}, and both codes show the characteristic
$|k_R/k_Z| \gg 1$ perturbations associated with the early development of the
vertical shear instability. The codes are in decent agreement about the
magnitude of the velocity perturbations and also in the dominant wavelength
of instability. The codes, however, also show some differences
in their solution at this time. The continued presence of the initial seed noise
is more apparent in the \NIRVANA run than in the \NIII run, in agreement with
the evolution of the kinetic energies in Fig.~\ref{fig:code_cmp}, and the
\NIRVANA run shows a greater degree of structure in the velocity perturbations, 
perhaps indicative of higher-order modes being present at this time.
Overall the comparision is very satisfactory and demonstrates that the 
conditions for the vertical shear instability to occur are predicted accurately
by both numerical schemes, which also show reasonable agreement for the
growth rates under different radial temperature profiles.

\subsection{Evolution as a function of viscosity}
\label{sec:viscosity}
\begin{figure}
  \center\includegraphics[width=0.9\columnwidth]{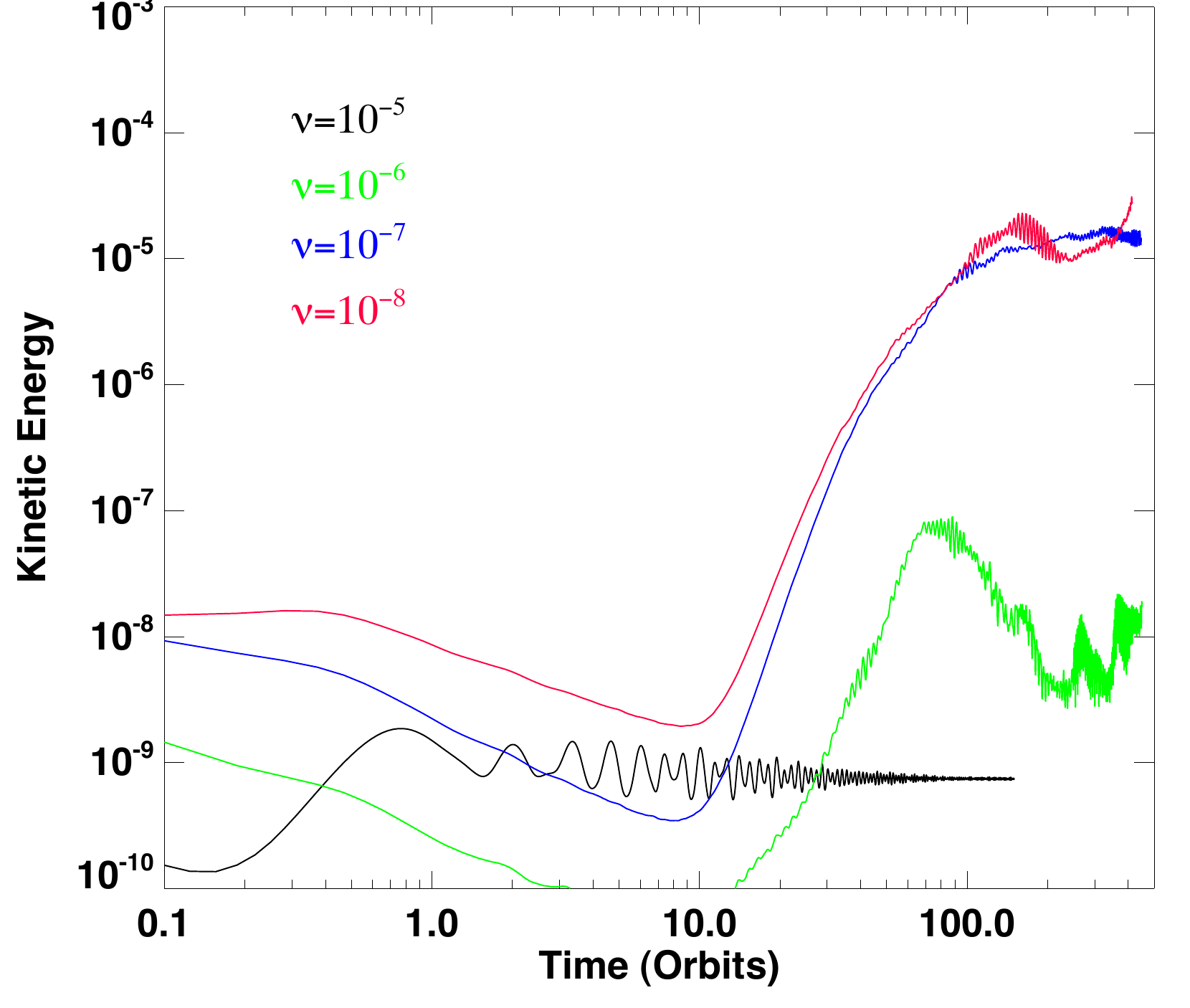}
  \caption{Time evolution of the normalised perturbed kinetic energy in the
   meridional and radial coordinate directions for model T1R--0 with
   $p=-1.5$, $q=-1$ and reflecting boundary conditions at the
   meridional boundaries. Each curve corresponds to a different value
   of the imposed kinematic viscosity, $\nu$, as indicated.}\label{fig:viscous}
\end{figure}

In this section we present results from simulations that examine the amplitude 
of the saturated state as a function of imposed viscosity.
We apply a constant kinematic viscosity, $\nu$, to the disc model T1R--0
and vary its value between $10^{-8} \le \nu \le 10^{-5}$
(a value of $\nu =10^{-6}$ corresponds to the Shakura-Sunyaev
viscous stress parameter $\alpha=4 \times 10^{-4}$ at $R=1$
\citep{1973A&A....24..337S}, and to a Reynolds number ${\rm Re}=H c_s / \nu=2500$).
The results are shown in Fig.~\ref{fig:viscous}, which shows the time evolution
of the perturbed meridional plus radial kinetic energies. As expected,
the results have a strong dependence on viscosity. For $\nu=10^{-5}$ the instability
is damped completely, which explains why previous 3D simulations of locally
isothermal discs have not reported seeing the vertical shear instability 
\citep[e.g.][]{2001ApJ...547..457K, 2006A&A...450..833C, 2011arXiv1102.0671F, 2010A&A...520A..14P}.
For decreasing values of $\nu$ the amplitude of instability increases, until at a value
of $\nu=10^{-8}$ there is little difference between the result in Fig.~\ref{fig:viscous}
and the inviscid result shown in the left panel of Fig.\ref{energy-AO-AR}.

Interestingly, it is found that in fully turbulent models where the 
MRI is active throughout the disc and $\alpha \simeq 0.01$, the
corrugation instability is not observed \citep{2006A&A...457..343F}.
We have computed models similar to those presented in 
\citet{2006A&A...457..343F}
and find that corrugation of the disc does not develop. Although
these MHD simulations adopt a significantly lower resolution than the
pure hydrodynamic runs we have presented here, we note that hydrodynamic
runs performed at low resolution still show the development of the
instability even when the short radial wavelength perturbations of the initial
growth phase are not resolved. Instead, we find that the disc displays
longer wavelength breathing and corrugation modes that become unstable
and cause the disc to oscillate vertically in quite a violent manner.
In magnetised global disc models with dead zones whose vertical
height covers $\simeq 2.5$ scale heights, which support Reynolds stresses
in the dead zone with an effective value of $\alpha \simeq 10^{-4}$,
the development of these corrugation oscillations {\emph is} observed in models
that adopt a locally isothermal equation of state with $q=-1$.

\subsection{Thermal relaxation in models with $T(R)$}

\label{sec:thermal-relax-TR}
We now consider the evolution of models where we relax the
locally isothermal assumption associated with the response of
the fluid to perturbations. We evolve the energy equation in 
(\ref{eqn:motion}), and introduce thermal relaxation by 
integrating eqn.~(\ref{eqn:t-relax}). We adopt the equation of state
$P=(\gamma-1) e$, and set $\gamma=1.4$. The gas is assumed to be
inviscid. Power-law profiles for the initial temperatures, $T(R)$,
and midplane density, $\rho_{\rm mid} (R)$, are adopted with
$q=-1$ and $p=-1.5$ in eqns.~(\ref{eqn:TR}) and (\ref{eqn:rhoR}).
The aim of these models is to examine the robustness of the
vertical shear instability as a function of the thermal
relaxation time, $\tau_{\rm relax}$, defined in eqn.~(\ref{eqn:t-relax})
and expressed as a fixed multiple or fraction of the local 
orbital period. These runs are labelled T5R--0.01 -- T9R--$\infty$ in
Table 1.

\begin{figure}
 \center\includegraphics[width=0.9\columnwidth]{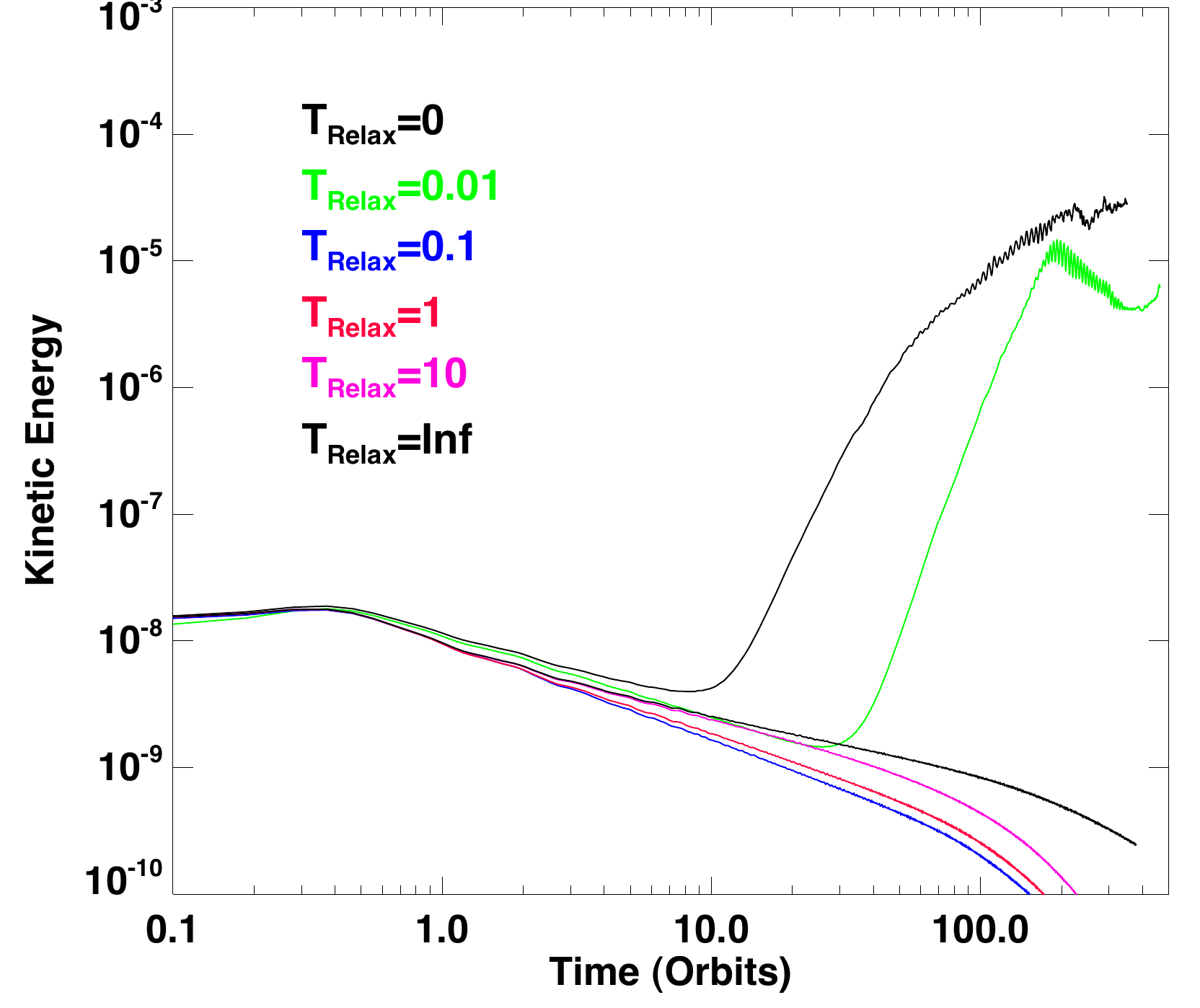}
  \caption{Time evolution of the sum of the (normalised) perturbed radial and 
  meridional kinetic energy in discs where the temperature was initially constant 
  on cylinders, as a function of the thermal relaxation time. Note that only the
  ${\tau_{\rm Relax}}=0$ and 0.01 cases show growth.} \label{TR-trelax}
\end{figure}

The evolution of the normalised perturbed kinetic energies for a number 
of models with relaxation times in the range 
$0 \le \tau_{\rm relax} \le \infty$ are plotted in Fig.~\ref{TR-trelax}.
It is immediately obvious that instability only occurs in either the
locally isothermal case ($\tau_{\rm relax}=0$) and when
$\tau_{\rm relax}=0.01$ orbits. All other simulations
result in the perturbed kinetic energy contained in the initial 
seed noise decaying with time. We note that the case of $\tau_{\rm relax}=\infty$ is 
directly comparable to a previous study on the adiabatic evolution of a stratified disc 
by \citet{2002A&A...391..781R}. The authors considered the hydrodynamic stability under the 
Solberg-H{\o}iland criterion and also find stability in this case.

Our results indicate that the vertical shear instability requires that the initial 
temperature profile of the fluid is re-established rather rapidly during dynamical 
evolution, at least for the equilibrium temperature and density profiles adopted in 
these particular models. 

The requirement for near-isothermal evolution suggests that the 
vertical shear instability is most likely to operate in the optically 
thin regions of astrophysical discs whose global properties are similar 
to those considered here. For example, the outer regions of protoplanetary 
discs lying beyond $\sim$ 50--100 AU may be prone to this instability,
provided that MHD turbulence is present at low enough levels that the
instability is not damped by the turbulent viscosity. 
This seems to be a likely prospect given that low density regions may be stabilised 
by ambipolar diffusion \citep{2011ARA&A..49..195A}.
It should also be noted,
however, that the simple thermal relaxation model we employ does not capture
the fact that the thermal evolution time of a mode with radial wavelength $\lambda_R$
scales as $\sim \lambda_R^2/{\cal D}$ (where ${\cal D}$ is the thermal diffusion
coefficient), such that very short wavelength modes may remain unstable in optically
thick discs. A reduction in spatial scales on which the instability operates, however, will
presumably affect the resulting turbulent flow and reduce the associated Reynolds stress
and transport coefficients in the nonlinear saturated state.
\begin{figure}
 \center\includegraphics[width=0.9\columnwidth]{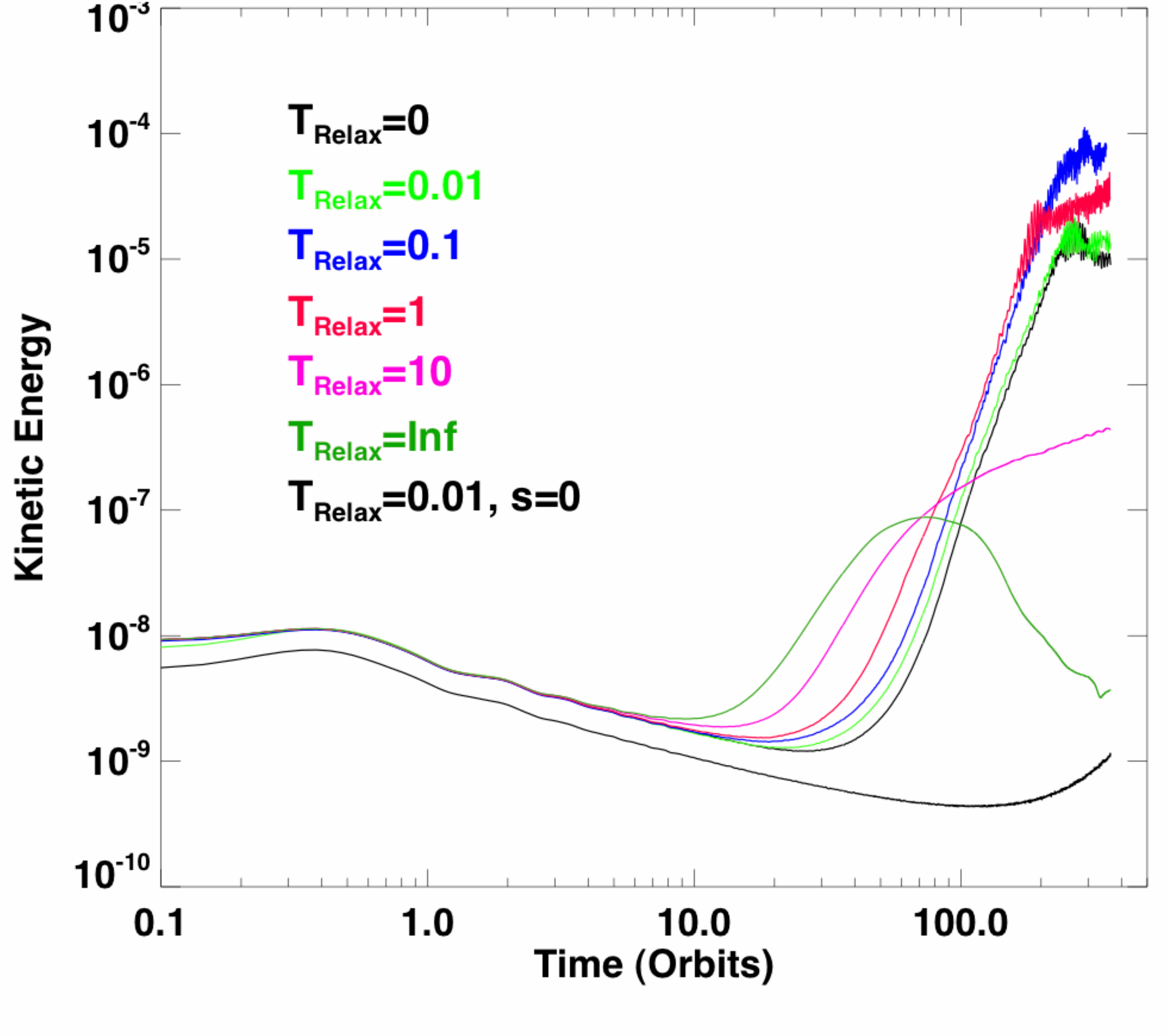}
  \caption{Time evolution of the sum of the normalised radial and meridional
  kinetic energies in discs where the entropy function, $K_s(R)$, was initially
  constant on cylinders.}
  \label{KR-trelax}
\end{figure}

\subsection{Thermal relaxation in models with $K_s(R)$}
\label{sec:thermal-relax-KR}
We now consider models in which the initial entropy function,
$K_s(R)$, follows a strict power-law function of radius 
given by eqn.~(\ref{eqn:ent-func}), and the midplane density, 
$\rho_{\rm mid}(R)$, follows a radial power-law given by eqn.~(\ref{eqn:rhoR}).
For all models except one we adopt the values $s=-1$ and $p=0$, 
leading to the midplane Mach number, ${\cal M}_{\rm mid}$, 
being constant at all radii. Our normalisation of $K_s(R_0)$ sets 
${\cal M}_{\rm mid}=20$. We also consider a single model with $p=-1.5$ 
and $s=0$, so that there is no initial radial entropy gradient.
The entropy in this case is normalised so that ${\cal M}_{\rm mid}=20$
at $R=R_0=1$. These runs are listed in Table 1 as K1R--0 to K10R--0.01.

We impose reflecting conditions at the meridional boundaries, and consider inviscid 
evolution. As described in Sect.~\ref{sec:equilibria}, these models are convenient to 
implement numerically because analytic solutions can be obtained for the equilibrium 
density and velocity field  through eqns.~(\ref{eqn:rhoR}) and (\ref{eqn:Omega2}). As such, 
these models allow us to explore the vertical shear instability as a function of the thermal
relaxation time, $\tau_{\rm relax}$, in discs where the initial
distribution of temperature no longer follows a power-law function of
cylindrical radius, but instead varies with both $R$ and $Z$. 
We note that eqn.~(\ref{eqn:Omega2}) also demonstrates that a radial 
power-law in $K_s(R)$ with $s=-1$ gives rise to an equilibrium 
$v_{\phi}$ that varies with height $Z$ at each radius $R$, but adopting
$s=0$ implies that no vertical gradient in $v_{\phi}$ exists.

The normalised sum of the radial and meridional kinetic energies is plotted in 
Fig.~\ref{KR-trelax} for thermal relaxation times in the range $0 \le \tau_{\rm relax} \le \infty$ 
local orbits. The model with $s=0$ and $p=-1.5$ employed a relaxation time $\tau_{\rm relax}=0.01$ 
orbits, and is labelled as `$T_{\rm relax}=0.01$, $s=0$'. 

\begin{figure}
  \center\includegraphics[width=0.9\columnwidth]{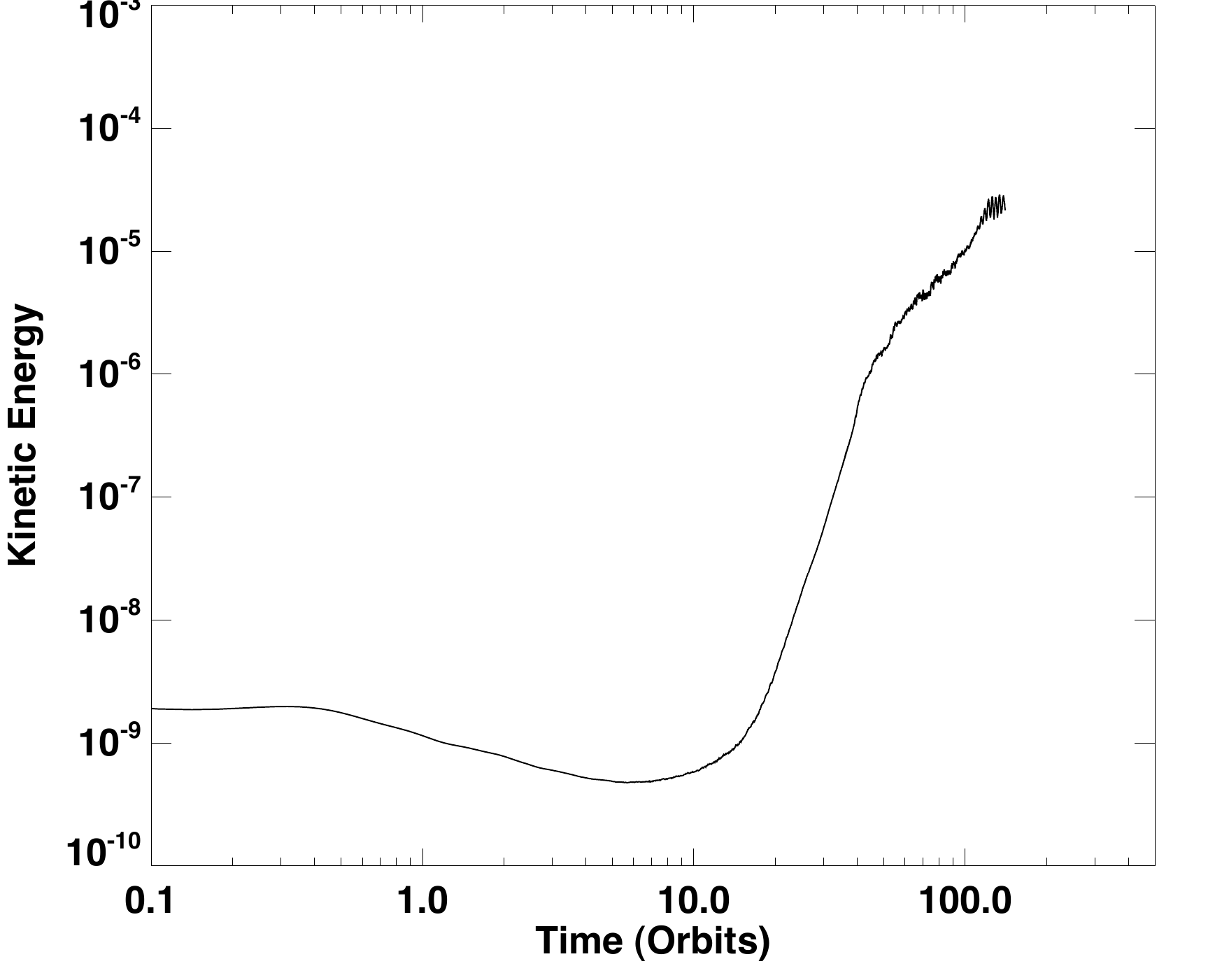}
  \caption{Time evolution of the perturbed meridional + radial
   kinetic energy (normalised by the total energy in keplerian motion)
  for the full 3D simulation T1R--0--3D.} \label{energy-NA}
\end{figure}

Interestingly, for the model with $s=-1$ and $p=0$, all values of 
$\tau_{\rm relax}$ result in growth of the perturbed meridional
and radial energies, and only the strictly isentropic simulation 
with $\tau_{\rm relax}=\infty$ shows eventual decay over long time 
scales of $\sim 100$ orbits. It is noteworthy that this isentropic
model is unstable according to one of the Solberg-H{\o}iland criteria discussed
in Sect.~\ref{sec:stability}, and this is probably the cause of the initial growth
in perturbed energy.
The subsequent decay may arise because the instability causes the equilibrium disc
to move to a stable state. Inspection of the perturbed velocity profiles in
contour plots similar to Figs.~\ref{fig:vel-image-AO} and \ref{fig:vel-image-AO-stretch} 
(not shown here) for all of the runs 
with $p=0$ and $s=-1$ shows that the previously discussed
characteristic perturbations with $|k_R/k_Z| \gg 1$ grow initially, even
for the isentropic disc model with no thermal relaxation. All models with 
finite thermal relaxtion 
time show long term growth in perturbed energy, although in the case of 
$\tau_{\rm relax}=10$ orbits the growth time is very long indeed.
The instability displayed by the remaining models with $\tau_{\rm relax} \le 10$ orbits 
shows that the requirement of very rapid thermal relaxation observed in the 
models presented in Sect.~\ref{sec:thermal-relax-TR} actually depends on the 
detailed temperature and density structure of the disc. It is clear that disc  
models exist for which thermal relaxation times in the range 
$0 \le \tau_{\rm relax} \le 10$ orbits lead to the growth of the instability,
and as such its range of applicability in the study of the dynamics of 
astrophysical discs is probably broader than suggested by the results presented in
Sect.~\ref{sec:thermal-relax-TR}. A full exploration and understanding
of the range of applicability, however, is beyond this paper, and will
require a dedicated and detailed future study that accounts for 
the thermal evolution of the disc with greater realism.

Turning to the run with $p=-1.5$ and $s=0$, we see that the initial perturbation 
energy decreases for the first $\sim 200$ orbits before increasing again.
The expectation is that this model will not display the vertical shear instability,
and inspection of velocity contour plots (not shown here) confirms that the characteristic
perturbations with $|k_R/k_Z| \gg 1$ do not appear in this case. These velocity plots, however,
indicate that over secular time scales sound waves are generated close to the meridional
boundaries, and this appears to be the reason for the up-turn in the perturbed kinetic 
energies seen in Fig.~\ref{KR-trelax} after 200 orbits.

\subsection{Non-axisymmetric model}
\label{sec:nonaxi}

We now consider briefly the evolution of a non-axisymmetric model T1R--0--3D, in which the 
azimuthal domain covered $\pi/4$ radians. This model is the 3D equivalent of model
T1R--0. The simulation was performed using the \NIRVANA code, and the
model was set up using equations~(\ref{eqn:rhoR}), 
(\ref{eqn:TR}) and (\ref{eqn:Omega1}), with values $p=-1.5$ and $q=-1$. 
The velocity field was seeded with noise (amplitude $0.01 c_s$).
Details of the model are given in Table~\ref{table-NIRVANA}. 

\begin{figure}
  \center\includegraphics[width=0.9\columnwidth]{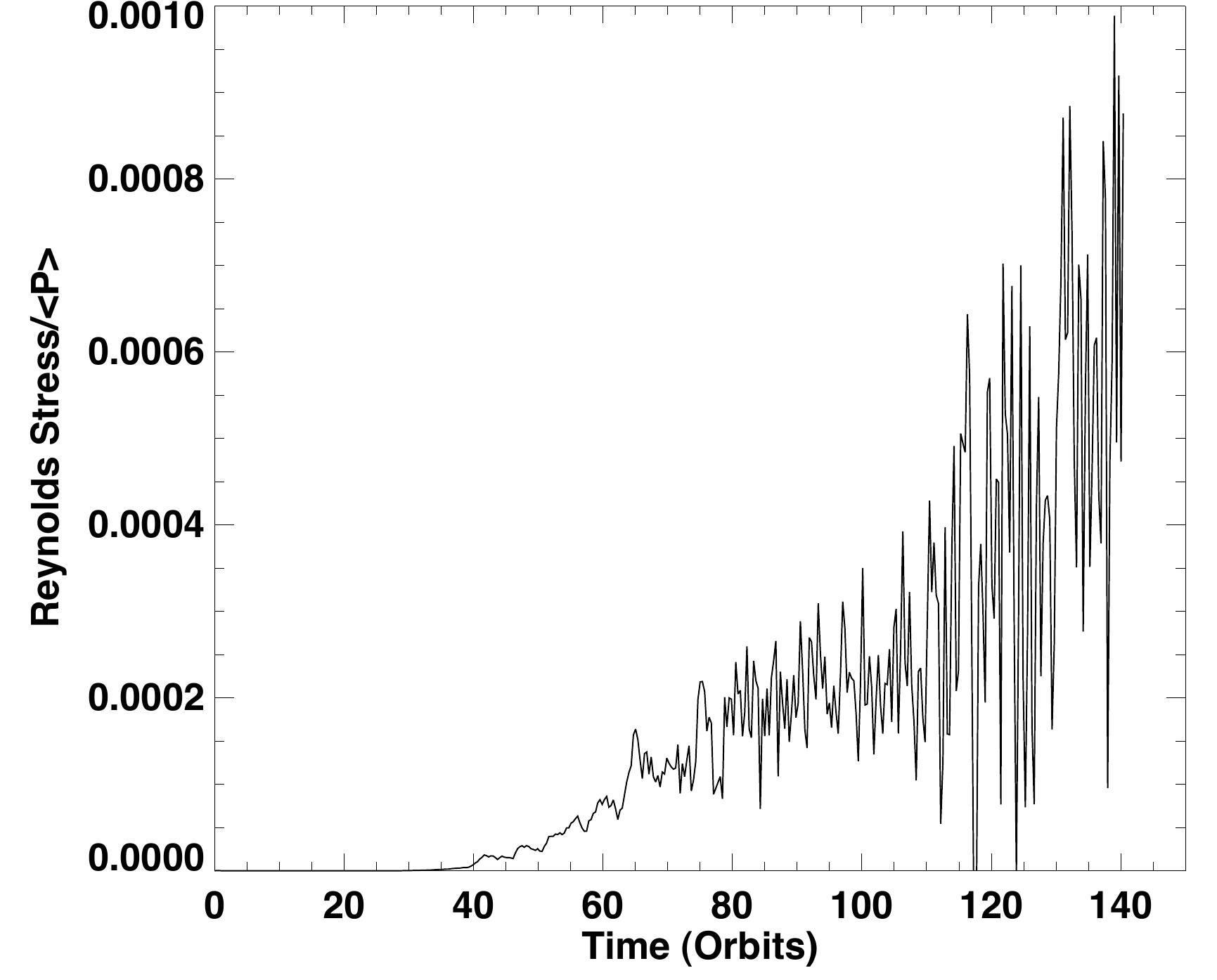}
  \caption{Time evolution of the volume averaged Reynolds stress
   (normalised by the mean pressure)
  for the full 3D simulations T1R--0--3D} \label{alpha-v-time}
\end{figure}

The total normalised kinetic energy (meridional + radial) versus time is displayed in Fig.~\ref{energy-NA}. 
Comparing with the equivalent axisymmetric plot (Fig.\ref{energy-AO-AR}) 
we see that the evolution is similar, 
with growth in the perturbed energy occurring after $\sim 10$ orbits, and saturation at 
a value of a few $\times 10^{-5}$ beginning to occur after $\sim 100$ orbits.

In Fig.~\ref{alpha-v-time} we plot the time evolution of the volume averaged
Reynolds stress normalised by the mean pressure in the disc. This quantity is computed as
follows. We define an azimuthally averaged Reynolds stress 
$T_{\rm R}(r, \theta)$ obtained by averaging the quantity $\rho \delta v_{\rm R} \delta v_{\phi}$
over azimuth. Here $\delta v_{\rm R}$ and $\delta v_{\phi}$ are the local radial
and azimuthal velocity fluctuations. 
We also define a density-weighted mean pressure as a function of $r$, ${\overline P}(r)$,
obtained by averging over $\theta$ and $\phi$. We define a local value of
the Shakura-Sunyaev stress parameter $\alpha(r,\theta)=T_{\rm R}(r,\theta)/{\overline P}(r)$.
The simple average of $\alpha(r,\theta)$ over $r$ and $\theta$ is the quantity plotted
in Fig.~\ref{alpha-v-time}.

\begin{figure}
  \center\includegraphics[width=0.9\columnwidth]{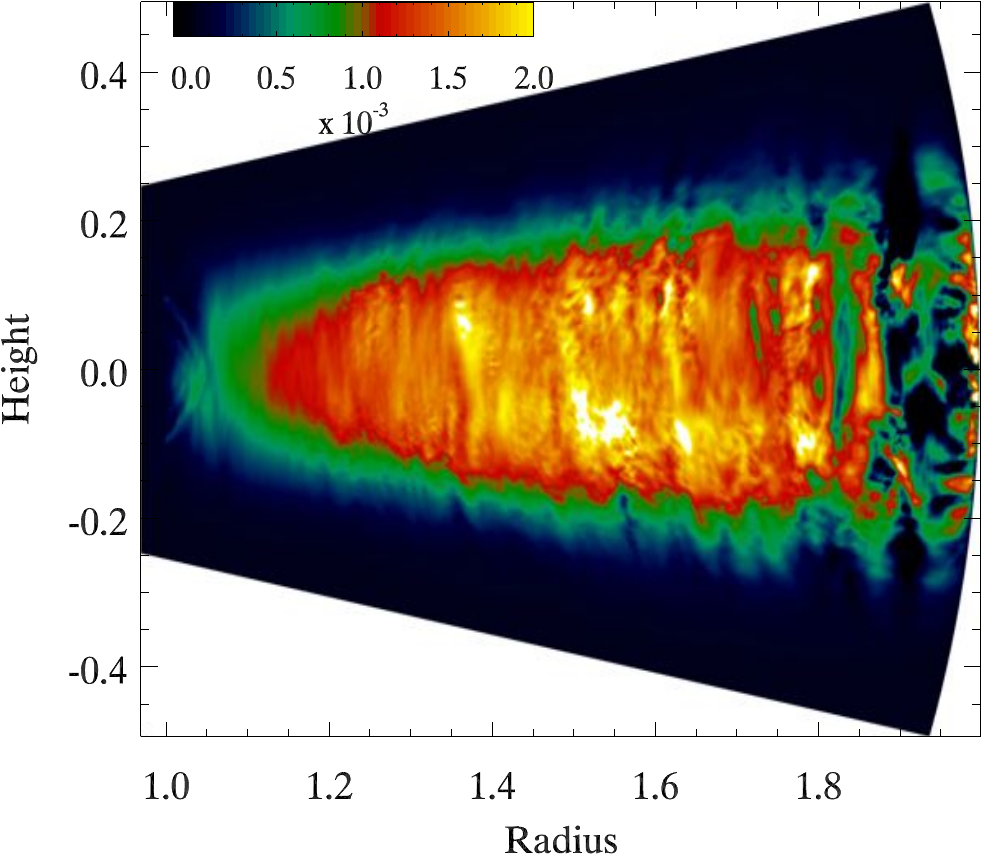}
  \caption{Spatial distribution of the time and horizontally
           averaged Reynolds stress (normalised by the mean pressure at each radius)
           for the model T1R--0--3D.}
      \label{alpha-v-space}
\end{figure}

Although rather noisy, we see that the normalised stress approaches average values 
$\sim 6 \times 10^{-4}$ by the end of the simulation (and appears to be still
growing at this point). The spatial distribution of $\alpha(r,\theta)$, time averaged
during the last 10 orbits of the run, is shown in Fig.~\ref{alpha-v-space}. Here we see that local values of the stress reach $\sim 2 \times 10^{-3}$, indicating that the vertical shear instability generates a quasi-turbulent flow capable of supporting significant outward angular momentum transport in astrophysical discs, given favourable conditions for its development.

\begin{figure*}
 \includegraphics[width=0.6\columnwidth]{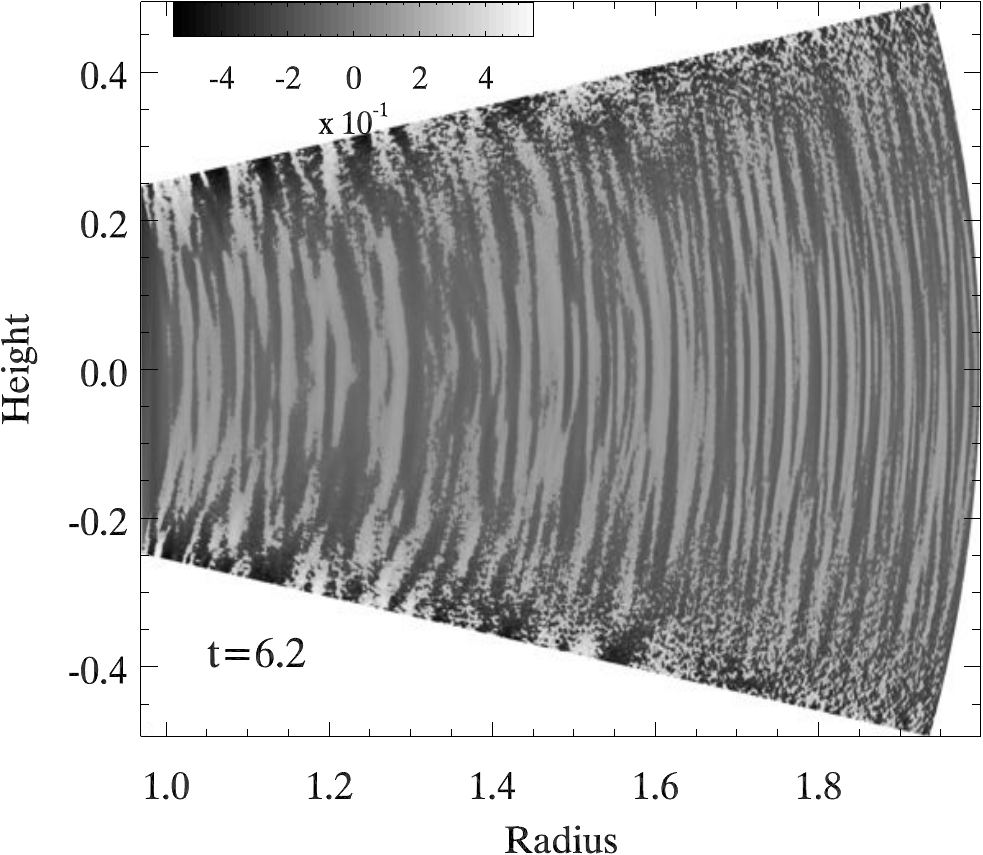}\hfill%
 \includegraphics[width=0.6\columnwidth]{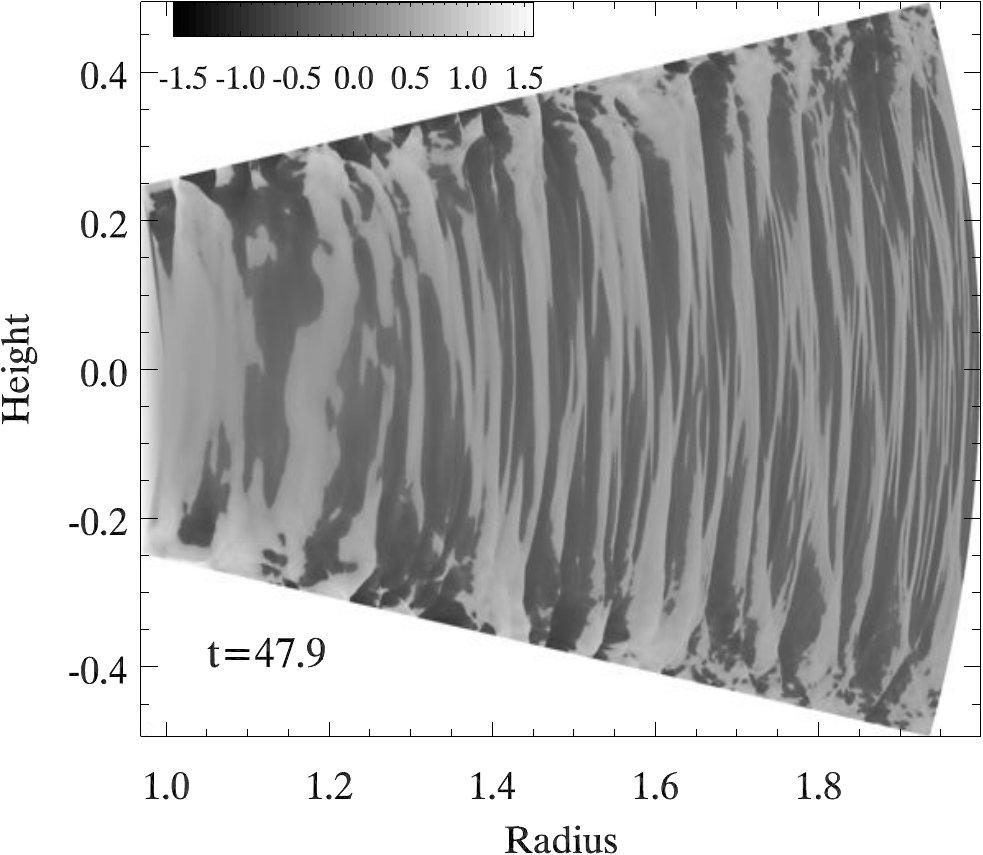}\hfill%
 \includegraphics[width=0.6\columnwidth]{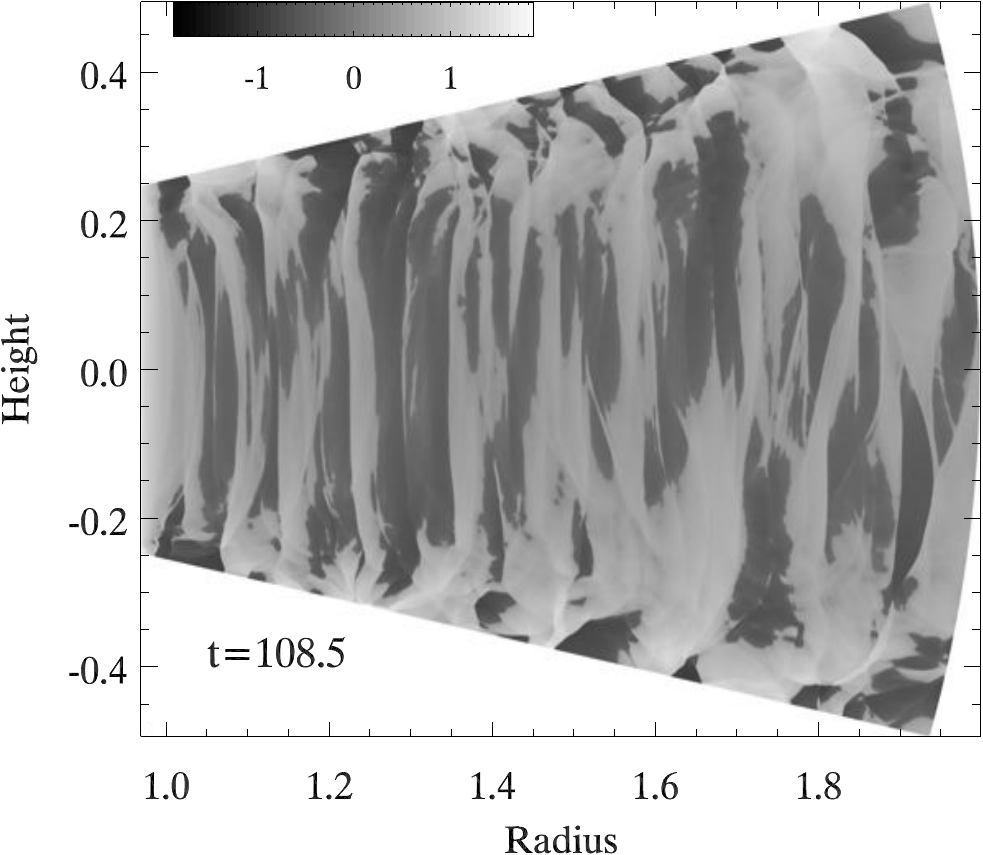}\hfill
 \includegraphics[width=0.6\columnwidth]{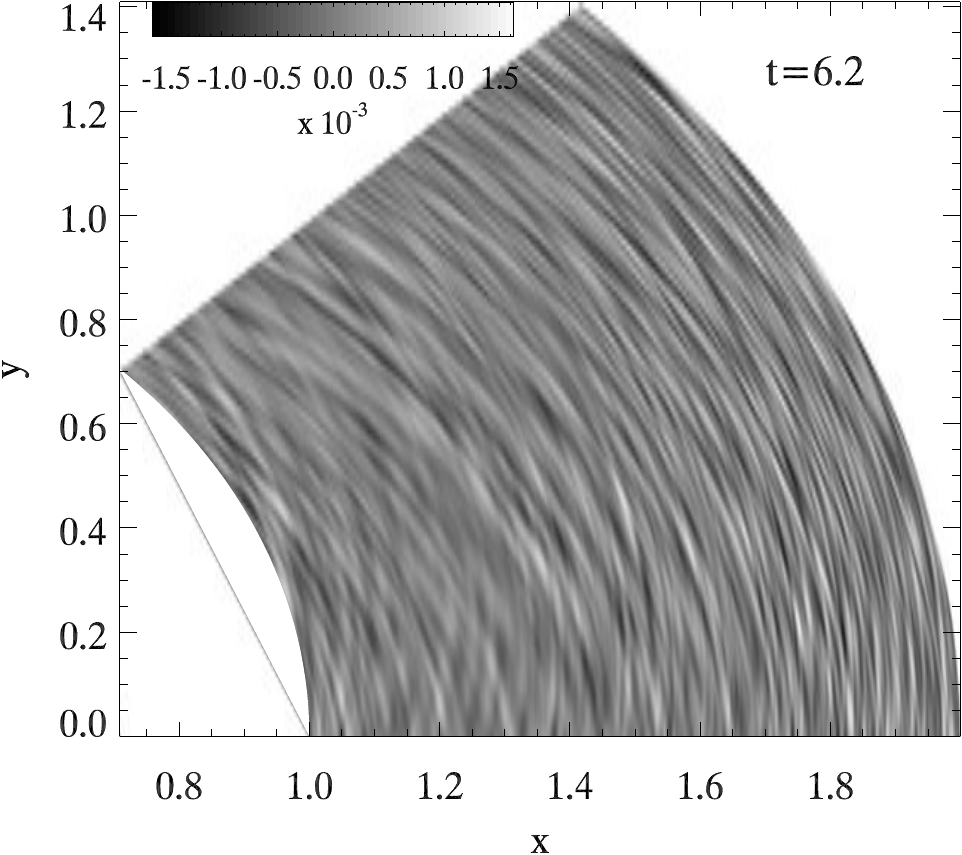}\hfill%
 \includegraphics[width=0.6\columnwidth]{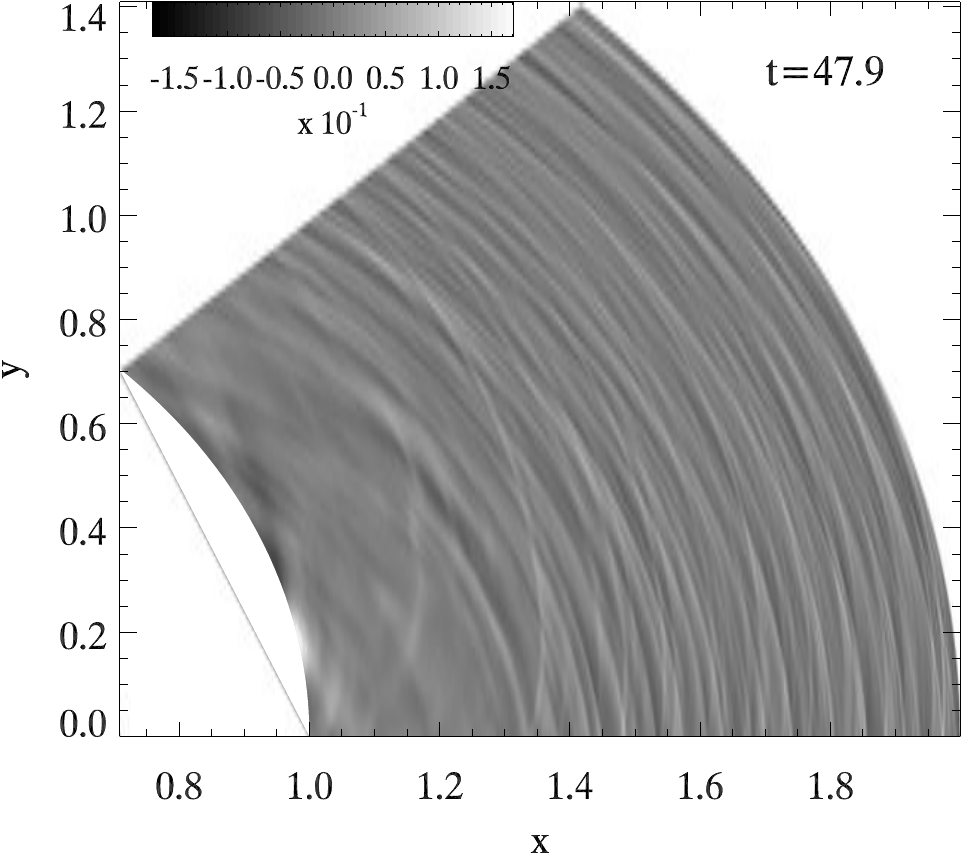}\hfill%
 \includegraphics[width=0.6\columnwidth]{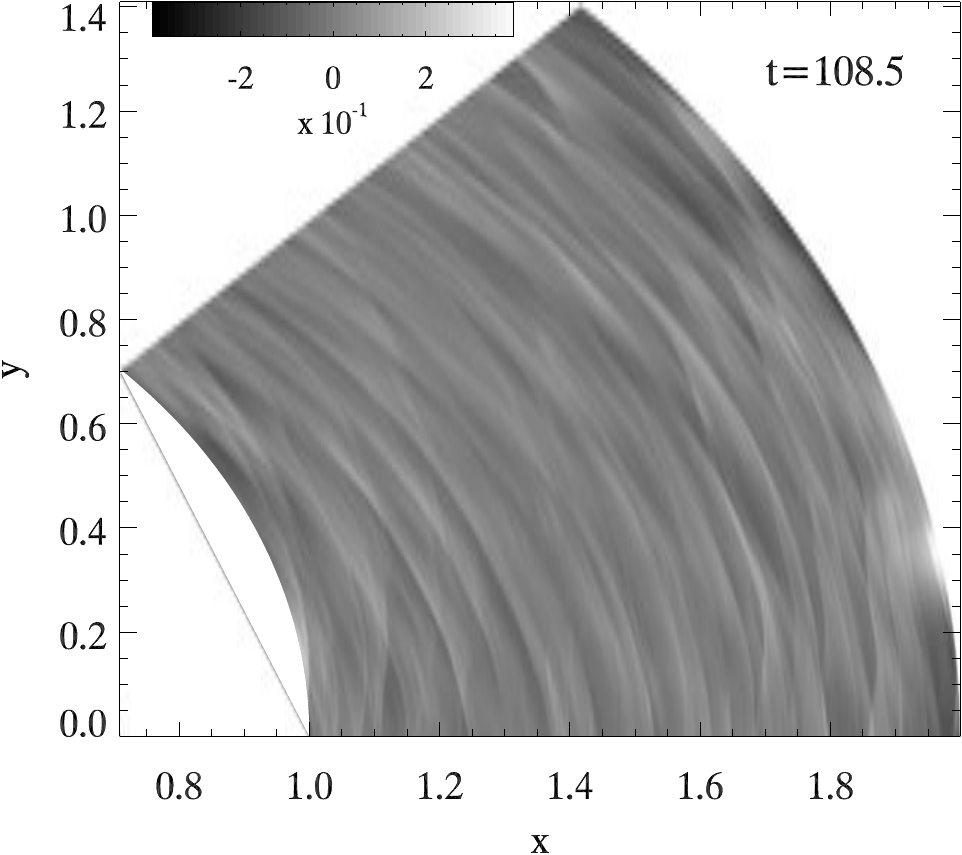}\hfill
 \caption{Perturbed density, $\delta \rho/{\overline \rho}$ in the merdional plane
 at $\phi=\pi/8$ (upper three panels) for the 3D simulation T1R--0--3D.
 Note that we have effectively stretched the grey-scale by plotting 
 the quantity ${\rm sign}(\delta \rho) \times |\delta \rho/ \rho|^{1/4}$
 in the upper panels.
 The lower panels show the relative density perturbations  
 $\delta \rho/{\overline \rho}$ at the disc midplane. No grey-scale stretching 
 has been applied to these lower panels.} \label{fig:NA1R}
\end{figure*}

The upper panels of Fig.~\ref{fig:NA1R} show contours of the perturbed density,
$\delta \rho/ \rho_0$ in a slice parallel the meridional plane at three different
times during the simulation, showing similar features to those presented
for the 2D-axisymmetric simulation in Fig.~\ref{fig:density-image-p1-5q1}. Perhaps more interesting are the lower panels of Fig.~\ref{fig:NA1R} which show the actual density $\rho$ in the ($R$, $\phi$) plane located at the disc midplane. Here the development of 
spiral density waves may be observed, similar in morphology to those that
arise in discs where turbulence is driven by the MRI \citep[e.g.][]{2003MNRAS.339..983P}. The 3D simulation presented here suggests that if the appropriate conditions prevail in astrophysical discs, the vertical shear instability may lead to a turbulent flow capable of supporting significant angular momentum transport.

\section{Theoretical Considerations}
\label{sec:analysis}
The GSF instability appears by rendering the inertial modes of
a rotating atmosphere unstable. The original analysis in Goldreich \& Schubert (1967, GS67 hereafter) demonstrated
the possibility of this instability by performing a point analysis
at a given location in a stellar radiative zone away from the equator,
equivalent to considering a location away from the midplane in a disc.
In this section we examine the mathematical structure of the instability by 
further extending previous analyses, including those of Urpin (2003) and 
Arlt \& Urpin (2004), by relaxing the point assumption.\par  
We shall focus on disturbances which are locally isothermal.  For the sake of 
completion of this important discussion we redo the original point analysis of 
GS67 in Appendix A, but without introducing the Boussineq approximation. 
Denoting $\sigma$ as the growth rate we find that the inertial mode response is
roughly given by
\beq
\sigma^2 = \frac{-\kappa_0^2 (c_0^2 k_z^2 + N_0^2) + 2\Omega_0 c_0^2 k_r k_z \frac{\partial \bar V}{\partial z}}
{c_0^2 (k_z^2 + k_r^2) + \kappa_0^2 + N_0^2},
\eeq
in which $c_0$ is the reference sound speed, $\kappa_0$ is the epicyclic frequency which, for
a keplerian disc, is given by $\Omega_0$, the local keplerian rotation rate at the point
in question.  $k_z$ and $k_r$ are the corresponding vertical and radial disturbance wavenumbers
respectively.  The local Brunt-Vaisaila frequency is $N_0$ and $\partial \bar V/\partial z$ is
the vertical gradient of the mean keplerian flow.  This quantity typically scales on the order
of magnitude of $(q/2)\Omega_0 (H_0/R_0)$ where $q$ is the same exponent of the radially
varying isothermal sound speed discussed in Section 2.  Supposing for this discussion that
$N_0$ is negligible it follows from this expression that if $H_0/R_0$ is small then instability
can only happen if the radial wavenumber conspires to be correspondingly large.  
In that limit the above expression implies
\beq
\sigma^2 \sim 2\Omega_0 \frac{k_z}{k_r} \frac{\partial \bar V}{\partial z}
-\kappa_0^2 \frac{k_z^2}{k_r^2},
\label{point_analysis_GSF_instability}
\eeq
indicating that instability is possible if $k_z/k_r \sim \order {q H_0/R_0}$.  The analysis
of Arlt \& Urpin (2004), for example, also similarly indicates that for the same rough conditions
the growth rate ought to scale as $\order {q \Omega_0 H_0/R_0}$.  The simulations
we have performed are consistent with this tendency where the radial length scales of 
the emerging structures are significantly shorter
than the vertical ones with growth rates of the instability on the order of 4 orbit times
for $H_0/R_0 \sim 1/20$ and $q = -1$.\par
Our goal is to develop a better physical understanding of the processes responsible for
this instability beyond invoking Solberg-H{\o}iland criteria. In this respect we notice
from Figure 1 that radial velocity fluctuations are considerably smaller in magnitude
than the corresponding meridional velocities.  \par
With these clues in mind, we show in the following how 
the processes involved in bringing about the instability 
is largely {\emph{anelastic}} and {\emph{radially geostrophic}} - by the latter expression
we mean to indicate dynamics which are in constant radial force balance between Coriolis
effects and pressure gradients.  Furthermore,
despite the varied simplifications we make to expose the essence of the 
physical process, the fundamental equations
describing the resulting linearised response remain inseparable in the radial
and vertical coordinates.  This means that the only recourse in establishing
any insight is through a further approximate solution of the resulting reduced equations. 
We find that the solution indicates that for given parameters describing disturbances
the instability {\emph{appears in pairs}}, as opposed to appearing individually
as indicated by (\ref{point_analysis_GSF_instability}).  Although we have not proved
it in this study, we conjecture the powerful driving of the instability
in the simulations may be, in part, caused by this feature.

\subsection{Equations of motion revisited and steady states rederived}
The equations of motion for axisymmetric inviscid dynamics in a cylindrical geometry 
are given by
\beqa
\left(\frac{\partial}{\partial t} + U\frac{\partial}{\partial R} + W\frac{\partial}{\partial Z}\right)
U -\frac{V^2}{R} &=& -\frac{1}{\rho}\frac{\partial c_s^2 \rho}{\partial R} 
- \frac{\partial \Phi}{\partial R},
\label{sec6:ueqn} \\
\left(\frac{\partial}{\partial t} + U\frac{\partial}{\partial R} + W\frac{\partial}{\partial Z}\right)V
+ \frac{UV}{R} &=& 0 , \label{sec6:veqn} \\
\left(\frac{\partial}{\partial t} + U\frac{\partial}{\partial R} + W\frac{\partial}{\partial Z}\right)
W &=& -\frac{1}{\rho}\frac{\partial c_s^2 \rho}{\partial Z} - \frac{\partial \Phi}{\partial Z}. 
\label{sec6:weqn}
\eeqa
Note that the ($R$, $\phi$, $Z$) velocity components are given
here by $(U,V,W)$.  We dispense with the subscripted scheme $(v_r,v_\phi,v_Z)$ used
in previous sections in order to simplify the notation.  The corresponding
equation of mass continuity is
\beq
\frac{\partial \rho}{\partial t} + \frac{1}{R}\frac{\partial R\rho U}{\partial r} + \frac{\partial \rho W}{\partial Z} = 0.
\label{sec6:rhoeqn}
\eeq
As mentioned above, we focus here on dynamics that are locally isothermal with an 
infinitely short cooling time ($\tau_{{\rm relax}} \rightarrow 0$).  This then is to be 
considered in the context of simulations T1R--0 to T4R--0 summarized in Table 1.  
Reciting therefore from Section 2: it means that the square of the sound speed is 
given by $c_s^2 = c_0^2 (R/R_0)^q$
where $R_0$ is the fiducial reference disc position and $c_0$ is the scaled
sound speed at that point.  The gravitational potential emanating from the
central object is $\Phi = -GM/(R^2+Z^2)^{1/2}$.
\par
The general equilibrium state solutions are found in eqns. (12)-(13) but, as we mentioned
earlier, perturbations superposed on this base state are difficult to analyse 
because the resulting equations are fundamentally inseparable so that
a typical normal-mode analysis is out of the question.
In order to facilitate some kind of tractable analysis  
we make the one and only approximation here: the radial and vertical
gradients of the potential $\Phi$ are expressed in terms of their corresponding first order 
Taylor Series expansions, i.e.
\beqa
& & \frac{\partial\Phi}{\partial R} \approx -\frac{GM}{R_0^2}\left(\frac{R_0}{R}\right)^2 
= -\Omega_0^2 R_0 \left(\frac{R_0}{R}\right)^2,
\nonumber \\
& & \frac{\partial\Phi}{\partial Z} \approx -\frac{GM}{R_0^3}\left(\frac{R_0}{R}\right)^3Z = 
-\Omega_0^2\left(\frac{R_0}{R}\right)^3Z,
\nonumber
\eeqa
in which $\Omega_0 = (GM/R_0^3)^{1/2}$ is the reference keplerian rotation rate at radius $R_0$.
The midplane density is chosen to be of the form $\rho_{mid} = \rho_0(R/R_0)^p$ where $\rho_0$
is the reference density and where $p$ is an arbitrary index (as referenced earlier).  
In the following analysis it will be convenient for our discussion to refer to the natural
logarithm of the density
instead of directly to the density itself, thus, we define $\Pi \equiv \ln \rho$.
Because we shall be concerned with perturbations around the steady states implied by the above
equations we shall represent these states by overbars.  As such, we have that
\beqa
& & \overline\Pi = \ln\rho_{mid}  - \frac{1}{2} \frac{Z^2}{H_0^2} \left(\frac{1}{{\cal H}^2}\right),
\label{basic_Pi} \\
& & \overline V = \Omega_0 R_0 \left(\frac{R}{R_0}\right)^{-3/2}
\left[1 + \varepsilon(R) + \frac{q}{4} \left(\frac{H_0}{R_0}\right)^2\left(\frac{Z}{H_0}\right)^2
\left(\frac{R}{R_0}\right)^{-2}
\right], \label{basic_V} 
\eeqa
where $H_0 \equiv c_0/\Omega_0$ is the local vertical scale height as referenced near the
end of Section 2.  The non-dimensionalised scale height ${\cal H}$ is accordingly 
given via the relationship ${\cal H}^2 \equiv (R/R_0)^{3+q}$ also as found in Sect.~2.  
The nondimensional quantity $\varepsilon$ is given by
\beqa
\varepsilon &\equiv& 
\left(\frac{H_0}{R_0}\right)^2{\cal C}^2 \left(\frac{R}{R_0}\right)
\left[R_0\frac{\partial \ln \rho_{mid}}{\partial R}+R_0\frac{\partial \ln {\cal C}^2}{\partial R}
\right] \nonumber \\
&=& (p+q)\left(\frac{H_0}{R_0}\right)^2\left(\frac{R}{R_0}\right)^{1+q}.
\nonumber
\eeqa
where ${\cal C}^2 \equiv (R/R_0)^q$.

\subsection{Linearised perturbations and non-dimensionalisation}
We introduce perturbations by writing for each dependent quantity
\[
U \rightarrow u', \qquad V \rightarrow \overline V + v', \qquad W \rightarrow w', \qquad
\Pi \rightarrow \overline \Pi + \Pi',
\]
and inserting these into the governing equations (\ref{sec6:ueqn})-(\ref{sec6:rhoeqn}). 
Linearising results in the expressions
\beqa
\frac{\partial u'}{\partial t} - 2\frac{\overline V}{R} v' &=& -c_s^2 \frac{\partial \Pi'}{\partial R},
\nonumber \\
\frac{\partial v'}{\partial t} + u'\frac{1}{R}\frac{\partial R\overline V}{\partial R}
+ w' \frac{\partial \overline V}{\partial Z} &=& 0, \nonumber \\
\frac{\partial w'}{\partial t} &=& -c_s^2\frac{\partial \Pi'}{\partial Z},
\eeqa
and
\beq
\frac{\partial \Pi'}{\partial t} = 
-\frac{1}{R}\frac{\partial Ru'}{\partial R} + \frac{\partial w'}{\partial Z}
-u'\frac{\partial \overline \Pi}{\partial R} - w' \frac{\partial \overline\Pi}{\partial Z}.
\eeq
It will now be made more transparent if we non-dimensionalise the above equations according 
to the quantities appearing.  We see that a natural time unit is given by the keplerian
rotation time $\Omega_0^{-1}$.  The radial and vertical length scales are naturally scaled
by $R_0$ and $H_0$ respectively.  Thus we write for the these quantities
\beq
t \rightarrow \Omega_0^{-1} t, \qquad Z \rightarrow H_0 z,\qquad R\rightarrow R_0 r,
\eeq
where $t,r,z$ are the corresponding non-dimensionalisations of the 
independent variables representing respectively time, radius and
height.  It is very important to note that the radial and vertical length scales are
disparate with respect to each other by a factor of $H_0/R_0$.  Because in all of our
simulations this ratio is quite small ($\sim 0.05$) we
shall treat this ratio as one of our ``small parameters" and formally represent it 
by $\epsilon \equiv H_0/R_0$ (not to be confused with $\varepsilon(R)$ 
defined earlier).  This disparity
must be kept in mind when the scalings invoked to recover the GSF instaiblity 
are formally made in the next section.
\par
Judging from the dynamics observed in the simulations the structures appearing
tend to be radially and vertically constrained. These spatial constraints 
(especially the radial confinement) indicate that perturbation velocities ought 
not to exceed the sound speed (at least initially). This is typical of the scalings frequently used to derive equations
appropriate to the dynamics in a small box of a disk 
\citep{1965MNRAS.130..125G, 2004A&A...427..855U} although we are not, technically, considering the dynamics
on such small scales yet.  In sum, therefore, we scale the dependent perturbation
velocities by
\[
u' \rightarrow c_0 u, \qquad v' \rightarrow c_0 v, \qquad w' \rightarrow c_0 w,
\]
where, as before, $u,v,w$ represent the corresponding non-dimensionalised component
velocities in the radial, azimuthal and vertical directions.  Therefore, the perturbation
equations now take on the following more transparent appearance 
\beqa
\frac{\partial u}{\partial t} &=&   2\frac{\bar v}{r} v -\epsilon {r}^q\frac{\partial \Pi'}{\partial r}, 
\label{perturbation_u_eqn:sec6}
\\
\frac{\partial v}{\partial t} &=&  -u\frac{1}{r}\frac{\partial r\overline v}{\partial r}
- \epsilon w \left(\frac{1}{2} qz r^{-7/2} \right)  
\label{perturbation_v_eqn:sec6}
\\
\frac{\partial w}{\partial t} &=& -r^q \frac{\partial \Pi'}{\partial z},
\label{perturbation_w_eqn:sec6}
\eeqa
and
\beq
\frac{\partial \Pi'}{\partial t} = -\epsilon\left(\frac{u}{r} + \frac{\partial u}{\partial r}\right)
-\epsilon\frac{\partial \overline\Pi}{\partial r} u 
-\frac{\partial w}{\partial z} - w \frac{\partial \overline\Pi}{\partial z},
\label{perturbation_Pi_eqn:sec6}
\eeq
where the non-dimensionalisation of the mean azimuthal flow $\overline V$ is given in terms
of the other redefined variables
\beqa
& & \overline\Pi \equiv \frac{p}{r} - z^2 \frac{1}{2{\cal H}^2}, \nonumber \\
& & \overline v \equiv \frac{\overline V}{\Omega_0 R_0} = 
r^{-3/2}
\left(1 + \varepsilon(r) + \frac{q}{4} \epsilon^2 z^2
r^{-2}
\right),
\eeqa
with $\varepsilon(r) = \epsilon^2(p+q)r^{1+q}$ and ${\cal H}^2(r) = r^{3+q}$.
We have written (\ref{perturbation_v_eqn:sec6}) in a seemingly curious way: 
the last term on the RHS of that equation is the product of the
vertical gradient of the mean azimuthal flow term, i.e. 
$-w \partial{\overline v}/{\partial z}$.  We have chosen to write it out
explicitly 
in order to bring to the fore the leading order scaling that sits in front of it
as it will effect how we proceed toward the reduced model
(see the next section).  

\subsection{Asymptotic scalings and resulting reduced equations}
The linearised equations of motion
(\ref{perturbation_u_eqn:sec6})-(\ref{perturbation_Pi_eqn:sec6}) are,
despite our efforts to simplify, still inseparable between the $r$ and $z$
variables.  In order to proceed asymptotically we must make further scaling choices.
These are guided by both the results of the numerical solutions as well
as by the discussion at the beginning of Section 6. 
At this stage we shall list them once again:
\begin{enumerate}
\item As estabilished by GS67, Urpin (2003) and Arlt \& Urpin (2004), 
growth rates scale as $\sim q \Omega_0 H_0/R_0$ which is, in our non-dimensionalised
time units, $\sim \epsilon q$,
\item For growth rates slow
on the timescale of the local disc rotation,
the emerging structures have radial dimensions ($\ell_r$)
considerably smaller than the corresponding
vertical dimensions ($\ell_z$).  That is to say, for $\epsilon q \ll 1$ the scaling analysis
of both GSF67 and Urpin (2003) indicate that $\ell_r/\ell_z \ll 1$ 
\item The numerical solutions also clearly indicate that during the
growth of the instability the radial velocity fluctuations
are significantly smaller than the corresponding meridional velocity fluctuations 
(see Figure 1).
\end{enumerate}
\par
In the following we describe scalings of (\ref{perturbation_u_eqn:sec6})-(\ref{perturbation_Pi_eqn:sec6})
that simplify them
into a set which is both more transparent and more amenable to further analysis
while retaining the essential physical processes involved in the instability.
We assume that $\epsilon \ll 1$ and treat $q$ as an order 1 quantity (although
a more general analysis can be done without this {\emph{a priori}} assumption achieving, in the end,
much of the same results discussed hereafter).  Furthermore we 
consider the analysis around the fiducial radius $r=1$.  
Since interest is in radial scales that are much smaller than the 
vertical scales and recalling that the $r$ scales dimensionally represent
physical length scales that are {\emph{longer}} than the dimensional
vertical scales $z$ by a factor of $\epsilon^{-1} (= R_0/H_0 \sim 20)$ 
we consider radial disturbances
\[
r-1 = \epsilon^2 x,
\]
where $x$ is order 1.  We leave the $z$ scales untouched as these are the {\emph{de-facto}}
reference scales of the analysis.  Because the growth rates are long by a factor of $\epsilon^{-1}$
we introduce a new long-time variable $\tau$ given by
\[
t = \tau \epsilon^{-1}.
\]
With this long-time scale assumed we find that in order to bring about non-trivial pressure balancing
with the inertial term 
in the vertical momentum equation it must follow that the pressure fluctuations must relatively scale
by $\epsilon$ as well.  This can be easily surmised by examining (\ref{perturbation_w_eqn:sec6})
and noticing that for $w$ order 1 and the time derivative scaling as order $\epsilon$, 
that the only way balance occurs is if the
pressure is correspondingly small by a factor of $\epsilon$.
This means introducing a new pressure fluctuation reflecting this scaling
through
\[
\Pi' = \epsilon \tilde\Pi.
\]
where $\tilde\Pi$ is the scaled pressure.\par
Finally as we have just intimated, in addition to the vertical velocity being order 1
we assume that the azimuthal velocity fluctuations are also unscaled (i.e. remaining order 1) in accordance with our
numerical observations.  We note here that scaling $v$ to be order 1 is 
also consistent with the pressure scalings assumed because it leads to a 
balance between the radial pressure gradient and the
Coriolis term in (\ref{perturbation_u_eqn:sec6}).
\par
However, we suppose that the {\emph{radial velocities are small}} in comparison
to the other velocity components and we propose that its relative smallness is similar to
the pressure field's scaling, i.e.
\[
u = \epsilon \tilde u,
\]
where $\tilde u$ is the correspondingly scaled radial velocity. Applying
these scalings assumptions to 
eqns.~(\ref{perturbation_u_eqn:sec6})-(\ref{perturbation_Pi_eqn:sec6}) results 
in the following equations at lowest order,
\beqa
0 &=& 2 v - \frac{\partial \tilde \Pi}{\partial x},\label{reduced_radial_geostrophy:sec6} \\
\frac{\partial v}{\partial \tau} &=& - \frac{1}{2} \tilde u - \frac{1}{2}q z w ,
\label{reduced_v_eqn:sec6} \\
\frac{\partial w}{\partial \tau} &=& - \frac{\partial \tilde \Pi}{\partial z},
\label{reduced_w_eqn:sec6} \\
0 &=& \frac{\partial \tilde u}{\partial x} + \frac{\partial w}{\partial z} - z w,
\label{reduced_anelastic_eqn:sec6}
\eeqa
with corrections to the above equations appearing at order $\epsilon^2$.  In this form
these reduced equations contain insight with regards to two very important physical
implications.  The first of these follows from the interpretation of
eqn.~(\ref{reduced_radial_geostrophy:sec6}) which says that the dynamics of the instability
occur under {\emph{radially geostrophic conditions}}, that is to say, that the processes
develop under conditions in which radial Coriolis effects balance radial pressure gradients. 
The second observation
is that the linear dynamics are anelastic
rather than incompressible in character.  By this we mean to say the following: since
on these radial/vertical length scales the mean (scaled) density profile has the form 
$\bar \rho = e^{-z^2/2}$,
eqn.~(\ref{reduced_anelastic_eqn:sec6}) may be equivalently written as
\beq
\frac{\partial \bar\rho\tilde u}{\partial x} + \frac{\partial \bar \rho w}{\partial z} = 0.
\label{reduced_anelastic_eqn_ver2:sec6}
\eeq
The fact that the dynamics here are not incompressible in the usual sense
is perhaps less surprising given that vertical stratification is non-negligible
under these spatial constraints - had we been interested in vertical scales that
were equally as short as the radial scales then stratification would not figure
prominently.
\par
  The other two equations describing the
vertical and azimuthal momentum balances retain their inertial terms and are largely
unaffected (directly) by these scalings.
\par
Before analyzing the solutions to these equations it is important to keep in mind that
the essential effect giving rise to the instability is present in the guise of 
the final term on the RHS of eqn.~(\ref{reduced_v_eqn:sec6}).  Additionally, in reflecting
upon these equations, it should be kept in mind that $\tilde u$ and $\tilde \Pi$ 
indicate real quantities that are intrinsically smaller (but not zero) compared
to the other terms.

\subsection{Approximate solutions and double instability}
The reduced system (\ref{reduced_radial_geostrophy:sec6}-\ref{reduced_anelastic_eqn:sec6})
may be combined into a single equation for the pressure perturbation $\tilde\Pi$:
\beq
\frac{\partial^2}{\partial \tau^2}\frac{\partial^2\tilde\Pi}{\partial x^2}
= -\frac{\partial^2\tilde\Pi}{\partial z^2} +
\left(1 + q\frac{\partial}{\partial x}\right)z\frac{\partial\tilde\Pi}{\partial z}.
\label{single_eqn_tildePi}
\eeq
We note here that a point analysis of (\ref{single_eqn_tildePi}), i.e., 
assuming $z = z_0$ is fixed and making a wave ansatz and proceeding similarly
to Appendix A recovers the content of the asymptotic growth rates contained
in eqn.~(\ref{point_analysis_GSF_instability}), indicating the consistency of
the scaling arguments we have exploited to get to this point.
\par
Nonetheless, examination of this equation, although appearing quite simple in many respects,
shows that it is also fundamentally inseparable owing to the $q\partial/\partial x$ term
on the RHS of the expression.  A more concerted future analysis is necessary to develop
proper solutions of these equations subject to proper boundary conditions on both
the vertical and radial boundaries of the system.  However, we may proceed analytically
by the following rationale.
\par
Keeping in mind that that equation (\ref{basic_Pi}) says 
that the basic state density profile $\overline\rho$ has the form $\sim e^{-z^2/2}$
we can develop solutions of these equations in which we require
that
\begin{enumerate} 
\item The pressure fluctuations $\tilde\Pi$
are zero on the radial boundaries located at $x=\pm\Delta$ and,
\item
The kinetic energy in the fluctuations decay as $z\rightarrow \pm \infty$.  Since
the kinetic energy involves terms that are appear as $\overline\rho \tilde u^2, \overline\rho v^2$
and $\overline\rho w^2$ then this boundary condition is satisfied even
if the velocity fields show {\emph{algebraic growth}} as $z\rightarrow\pm\infty$.
\end{enumerate}
The first boundary condition is not the same as the no-normal flow boundary conditions
of the simulations but we have checked and found that the resulting growth rates and
qualitative results are unchanged when implementing those conditions instead (see footnote
later).
This second condition is not outright unphysical but it, nevertheless, does not mirror
the conditions present in the simulations and we must keep this in mind when we
analyse the results later on. 
We assume all fields have normal-mode form in time
\[
\tilde\Pi(x,z,t) = \hat\Pi(x,z)e^{s\tau} + {\rm c.c.}
\]
where $s$ is the growth rate.  This form is also assumed for the other variables $\tilde u, v, w$.  
As a side note we point out
that with this solution ansatz inserted into eqn.~(\ref{reduced_w_eqn:sec6}) it follows that
$s\hat w = -\partial_z \hat\Pi$.  The equation for $\hat\Pi$ becomes
\beq
s^2\frac{\partial^2\hat\Pi}{\partial x^2}
= -\frac{\partial^2\hat\Pi}{\partial z^2} +
\left(1 + q\frac{\partial}{\partial x}\right)z\frac{\partial\hat\Pi}{\partial z}.
\label{sec6:partialy_decomposed_Pi_tosolve}
\eeq
We proceed to develop solutions of (\ref{sec6:partialy_decomposed_Pi_tosolve}) 
guided by the methods used to develop solutions to Hermite and Gegenbauer 
differential equations \citep{morse1953methods,2012ApJ...754...21L}.
With $m$ a positive integer we assume a tractable non-separable solution of the 
following form
\beq
\hat\Pi(x,z) = {\displaystyle \sum_{j=0}^m {\hat\Pi_{jm}(x) z^m}},
\label{Pi_Solution_Ansatz:sec6}
\eeq
where the radial functions $\hat\Pi_{jm}(x)$ are yet to be determined.
\footnote{For example, when developing solutions to the Hermite differential equation of some integer order, the constant coefficients in front of each power of $z$ are are determined through a standard recursive procedure in descending powers of $z$.  The same is done here except that the recursive procedure generates coupled ordinary differential equations in $x$.}
Inserting this solution ansatz into (\ref{sec6:partialy_decomposed_Pi_tosolve})
reveals that for a given $m$ and for each index $j = \cdots, m-4, m-2$ the following equations must be
satisfied
\beq
s^2\frac{\partial^2\hat\Pi_{jm}}{\partial x^2} = 
-(j+2)(j+1) \hat\Pi_{j+2,m} + j\hat\Pi_{jm} + q j\frac{\partial\hat\Pi_{jm}}{\partial x}.
\eeq
together with the corresponding ``top" equation, i.e. the ordinary differential equation for $j=m$,
\beq
s^2\frac{\partial^2\hat\Pi_{mm}}{\partial x^2} = 
  m\hat\Pi_{mm} + q m\frac{\partial\hat\Pi_{mm}}{\partial x}.
 \label{top_equation}
\eeq
The index $j$ starts from 0 (1) depending if $m$ is even (odd). 
In relation to the modes observed in our numerical solutions: breathing modes 
correspond to even values of $m$ while corrugation modes correspond to odd values 
of $m$. We also note that
each function $\hat\Pi_{jm}(x)$ must individually satisfy the boundary condition $\hat\Pi_{jm}(\pm\Delta) = 0$
for each $j$.
This is a rather cumbersome task and it is not our intention here to fully develop detailed
solutions.  Rather, we seek to determine the value of the growth rate $s$ which
may be calculated by solving the top equation (\ref{top_equation}) subject
to the boundary condition $\hat\Pi_{mm}(\pm\Delta) = 0$.  Elementary analysis of this
equation indicates solutions are of the form
\beq
\hat\Pi_{mm} = A_{m}{ e^{{mx}/{s^2}}}\cos kx, \qquad k = n\frac{\pi}{2}, \ \  {\rm where} \ \
 \ n=1,2,\cdots
 \label{Pi_mm_solution:sec6}
\eeq
where $A_{m}$ is an arbitrary constant.  The requirement for the consistency of this
solution follows from inserting (\ref{Pi_mm_solution:sec6}) directly into
(\ref{top_equation}) which, after factoring out $\hat\Pi_{mm}$
reveals the growth-rate relationship for $s$
\beq
s^4 + \frac{m}{k^2} s^2 + \frac{q^2 m^2}{4k^4} = 0,
\eeq
which has solution
\beq
s^2 = \frac{m}{k^2}\left(-1 \pm \sqrt{1-q^2 k^2}\right).
\eeq
It follows from an examination of the above equation that for $0 < qk < 1$
the two roots possible for $s^2$ are both negative which implies that
the four modes are all oscillatory.  However,
instability is possible once $qk>1$ 
since it implies that $s^2$ is now a complex number with a non-zero real
part.  Moreover, because of the sign convention, for $qk>1$ there are always
a pair of unstable modes and a pair of stable modes.  In other words
we always have in that case the four possible combinations: $s = \pm (|s_r| \pm i |s_i|)$
for some real values of $s_r,s_i > 0$.  This quality holds also for
no-normal boundary conditions at $x=\pm\Delta$.
\footnote{In this case, 
after inserting the solution ansatz into the top equation (\ref{reduced_v_eqn:sec6})
indicates that the no-normal boundary condition is enforced by
requiring $s^2\partial_x\hat\Pi_{mm} - qm\hat\Pi_{mm} = 0$
at $x=\pm\Delta$.  What emerges is a more complicated relationship between
$s$ and $k,m$ but resulting in the {\emph{same}} qualitative features discussed
for fixed pressure conditions. A more thorough exposition awaits a subsequent study.}
For unstable values ($|kq| > 1$) the expressions $|s_r|$ and $|s_i|$ are given by
\beq
|s_r| = \frac{\sqrt {m/2}}{|k|}\sqrt{|kq| - 1}, \qquad
|s_i| = \frac{\sqrt {m/2}}{|k|}\sqrt{|kq| + 1}.
\eeq
An elementary examination of the growth rate $s_r$ shows that the wavenumber corresponding to maximal
growth occurs for $|kq| = 2$.  If we let $\lambda_{{\rm max}}$ denote the corresponding wavelength
in {\emph{dimensional units}} then this wavelength of maximal growth is
\[
\lambda_{{\rm max}} = \pi R_0 \left(\frac{H_0}{R_0}\right)^2,
\]
where, for our expreriments corresponding to $q = -1, H_0/R_0 = 0.05$, implies
radial scales on the order of $\lambda_{{\rm max}} \sim 0.008 R_0$.  These scales are approximately the
length scales observed at the early stages of growth in both Figures 2 and 3, especially near
the $R_0 = 1$ boundary.  Similarly, the growth rate of the fastest growing mode denoted by
${\mathfrak Re}(s_{{\rm max}}) = |q| \sqrt{m/2}/2$ and, in dimensional units, this implies a maximum
growth rate $\sigma_{{\rm max}}$
\[
\sigma_{{\rm max}} = \frac{|q| \sqrt{m}}{2\sqrt 2} \left(\frac{H_0}{R_0}\right) \Omega_0 = 
\sqrt m |q|
\frac{\pi}{\sqrt 2} \left(\frac{H_0}{R_0}\right) {\rm {orbit}}^{-1}.
\]
For the simulations T1R--0 to T4R--0 (Table 1) this means a maximum growth
rate given by $\sigma_{{\rm max}} \sim 0.11 |q|\sqrt m$ orbit$^{-1}$.  It should be
kept in mind that these growth rates are for the individual velocity fields.
However, the growth rates
reported in Sect.~5 are for the perturbation kinetic
energies which depend upon the square of the velocity fields.  As such, the linear theory
therefore predicts the maximum growth rates {\emph{for the kinetic energy perturbations}} to $\sim 2 \sigma_{{\rm max}}$ which, if applied to our numerical experiments, would be
\[
\sim 0.22 |q|\sqrt m \ {\rm{orbit}}^{-1}.
\]
This is significant since we reported there that the growth times were measured to be
$\sim 0.24 $ orbit$^{-1}$.  The theory predicts: for the fundamental corrugation mode $m=1$
a growth rate of $\sim 0.22 $ orbit$^{-1}$, while for the first breathing mode $m = 2$ a growth rate of $\sim 0.31 $ orbit$^{-1}$. This suggests that the growth rates observed in the early stages of growth of simulations T1R--0 to T4R--0 is a convolution of these two fundamental modes of the system.
\par
The theory, as it stands, does not select for a preferred value of $m$ as the growth rates
all scale as $\sqrt m$ suggesting curiously that for larger values of $m$ one should expect 
faster growth rates.
This appears to be in conflict with the original point analysis of GS67 and others.
In particular, since $m$ roughly indicates the number
of nodes in the vertical direction one can treat $m$ as a proxy for
the vertical wavenumber $k_z$.  But according to the criterion 
obtained via the point analysis (\ref{point_analysis_GSF_instability}), 
increasing the vertical wavenumber ought to 
promote stability.  This conflict of interpretation needs to be
resolved in follow up work, however, we do offer two hypotheses regarding
this matter: 
\begin{enumerate}
\item Since the instability
appears to be global in character (as evinced by the numerical solutions)
that the classical point analyses serve only as a useful guide toward indicating
the possibility of instability only.
Additionally, since the reduced equations
derived here are more global in reach, it would be
logically more appropriate to be guided by their content in better
describing the emergence of the instability and/or,
\item The 
implementation of the decaying kinetic energy boundary condition for
$z\rightarrow\pm\infty$ may be somehow responsible for the feature 
regarding $m$ and the mode's stability character.  It could be
that high $m$ modes may become stable if one, instead, imposes
no normal flow conditions at some finite height away from the disc midplane.  Of course,
to do this requires non-trivial solution forms that are far more
complicated than the solution ansatz assumed in (\ref{Pi_Solution_Ansatz:sec6}).
\end{enumerate}

\section{Discussion and conclusion}
\label{sec:conclusion}
We have presented the results from a large suite of 2D-axisymmetric and 3D 
simulations, using two independent hydrodynamic codes, that examine
the stability and dynamics of accretion disc models in which
the temperature or entropy are strict functions of cylindrical radius.
Such thermal profiles lead to equilibrium angular velocity profiles that
depend on radius and height.
We find that these disc models are unstable to the growth of modes with
$|k_R/k_Z| \gg 1$ when the thermal evolution time is comparable to or
shorter than the dynamical time. This mode pattern is expected for systems
undergoing the Goldreich-Schubert-Fricke instability. The potential for this
vertical shear instability to apply to accretion discs has been investigated 
previously by \citet{1998MNRAS.294..399U}, \citet{2003A&A...404..397U}
and \citet{2004A&A...426..755A} using the Boussinesq approximation, and we have 
confirmed this with our own analysis that applies to fully compressible flows. 
The nonlinear simulations indicate that evolution of the instability leads to the 
dominant radial wavelength increasing from a fraction of the scale height
during early times to being comparable to the scale height at later times.
Instability also appears to generate disc vertical motions that correspond to
vertical breathing modes at early times, but at later times the dominant
motion corresponds to corrugation of the disc causing the disc midplane
to oscillate around its initial location at all radii (at least in axisymmetric flows). 
A full 3D simulation computed using a locally isothermal equation of state indicates 
that the instability generates a turbulent flow in its nonlinear saturated state,
leading to non-negligible transport of angular momentum through a Reynolds
stress with corresponding viscous alpha value $\alpha \sim 10^{-3}$. 

The requirement for quite rapid thermal evolution of the disc for strong 
instability to ensue suggests that the outer regions of protoplanetary 
discs may be places where this instability operates because of the low density
and optical depth there. We caution that the thermal evolution time
required for instability may depend on thermal gradients, such
that instability can operate for longer cooling times in discs with steeper
temperature/entropy profiles, but it seems unlikely that instability can
be sustained if the local thermal time scale greatly exceeds a few orbital periods.
We have shown that the instability is damped in viscous flows with dimensionless
kinematic viscosity $\nu \ge 10^{-6}$, so these outer disc regions would need to be 
stable against the MRI for the instability to operate (the instability
is not observed in global simulations of discs supporting fully developed MHD turbulence).
It is likely that the MRI is quenched by ambipolar diffusion in the outermost regions of protostellar 
discs \citep{2011ARA&A..49..195A}, so these may provide low density, magnetically 
inactive regions where the vertical shear instability can operate efficiently.

The vertical shear instability is an example of a baroclinic instability 
because the required vertical shear in a disc only occurs if the pressure 
is a function of both density and pressure, $P(\rho, T)$. An important question
that we have not addressed in this work is how a disc would evolve that is
subject to both the vertical shear instability and the subcritical baroclinic 
instability (SBI) studied by \citet{2007ApJ...658.1252P} and \citet{2010A&A...513A..60L}. 
The conditions required for the SBI to operate and be sustained strongly are a radial 
entropy gradient and a fairly rapid thermal relaxation time, and we have shown
that the vertical shear instability operates under these conditions also. Both 
\citet{2007ApJ...658.1252P} and \citet{2010A&A...513A..60L} report that a strongly 
sustained instability is obtained for thermal time scales close to the local orbital
period, leading to a highly complex flow in which long-lived vortices are formed,
and an effective viscous stress of $\alpha \sim$ few $\times 10^{-3}$ is maintained
by the Reynolds stress in compressible flows. The 3D simulation we presented in
Sect.~\ref{sec:nonaxi} used a locally isothermal equation of state, and so is not 
subject to the SBI,
but it seems likely that the combined action of the two instabilities in a disc with longer
thermal evolution time and appropriate entropy gradient will
generate a complex flow containing long-lived vortices and vertical motions that correspond
to corrugation of the disc, accompanied by a Reynolds stress that leads to efficient 
outward angular momentum transfer. It is worth noting that the vertical shear instability
is linear, whereas the SBI is a finite-amplitude instability, so it is possible that the
SBI may be stimulated by perturbations generted during the development of the 
vertical shear instability. We will present a study of these 
two instabilities operating in tandem in a future publication to explore these hypotheses.
Given the role that vortices may play in the trapping of solids during planet formation
\citep[e.g.][]{1995A&A...295L...1B,2003ApJ...582..869K}, this is clearly an
important issue for further investigation.

\section*{Acknowledgements}
Part of this work used the \NIII code developed by Udo Ziegler at the Leibniz 
Institute for Astrophysics (AIP). All computations were performed on the QMUL HPC
facility, purchased under the SRIF and CIF initiatives. We 
acknowledge the hospitality of the Isaac Newton Institute for
Mathematical Sciences, where much of the work presented in this paper
was completed during the `Dynamics of Discs and Planets' research
programme. RPN acknowledges useful and informative discussions with 
Steve Lubow on the subject of corrugation modes in discs during an early 
phase of this project. We acknowledge comments provided by the referee of
an earlier version of this work that have led to substantial improvements
in this paper.

\bibliographystyle{bib/mn2e}
\bibliography{bib/refs}


\appendix

\section{Recital of GS67 with local isothermal assumption}
Owing to the importance of the original study presented by GS67,
we repeat here the calculation contained in that work wherein we assume
the equation of state of the gas is locally isothermal with a temperature profile
varying in the nominal radial direction.  We also extend that analysis
to include the effects of compressibility and we do {\emph{not}} assume outright
that the hydrodynamic flow is incompressible.\par
The original analysis was performed in a local rotating reference
frame (the original intention being to examine the possibility of such instabilities
in the interiors of rotating stars).  The equations of motion considered in that work are akin to the local
shearing sheet approximation familiar to accretion disc theory \citep{1965MNRAS.130..125G}. Since this is our concern here we shall use that as our starting model set.
The equations for axisymmetric dynamics in that environment are therefore
\beqa
\frac{du}{dt} - 2\Omega_0 v &=& - c_s^2\frac{\partial \Pi}{\partial x} -
\frac{\partial c_s^2}{\partial x} + 3\Omega_0^2 x, \nonumber
\\
\frac{dv}{dt} + 2\Omega_0 u &=& 0,
\nonumber \\
\frac{dw}{dt}  &=& - c_s^2\frac{\partial \Pi}{\partial z} - g, \nonumber \\
\frac{d\Pi}{dt} + \frac{\partial u}{\partial x} + \frac{\partial w}{\partial z} &=& 0,
\eeqa
where the expressions appearing are consistent with those developed in the body of the text
recalling especially that $\Pi = \ln \rho$ and $\Omega_0 =$ constant (see Sect.~6). 
The remaining expressions are defined again for convenience below.
The only difference is that we use lower case letters for the radial, aziumthal and vertical
velocities (i.e. $u,v,w$) in order to distinguish them from the ones used to describe dynamics in a cylindrical geometry examined in the main text of our study.  
Typically speaking, the vertical component of gravity is vertically varying and
is given by $g=\Omega_0^2 z$ - but {\emph{for this analysis it is treated as a constant}}.
\par
As done in GS67, one can do a point expansion (see the discussion of GS67 right prior to eq. 17 of that work) around any nominal level $z=z_0$.   We start by considering the mean states which we represent with overbars.
Thus,
\beqa
-2\Omega_0 \overline V &=& -c_s^2 \frac{\partial \overline\Pi}{\partial x} - \frac{\partial c_s^2}{\partial x}, \nonumber \\
0 &=& -c_s^2  \frac{\partial \overline\Pi}{\partial z} - g. \nonumber
\eeqa
Note that the mean azimuthal flow state $\bar v$ has been decomposed
into a
keplerian piece (the term $-(3/2)\Omega_0 x$) plus a deviation about that state $\overline V$,
i.e.  $\bar v = -q\Omega_0 x + \overline V$.
As before, $c_s^2$ is the sound speed (implicitly a function of $x$ because of the radial dependence of the vertically isothermal temperature profile).
In steady state we find that
\[
\frac{\partial\overline \Pi}{\partial z} = -\frac{g}{c_s^2},
\]
and, most importantly, the mean gradient of the azimuthal flow is
\beq
\bar V_z \equiv \frac{\partial \bar V}{\partial z} = \frac{g}{2\Omega_0} \frac{\partial \ln c_s^2}{\partial x}.
\eeq
Perturbations of $\Pi$ and all other variables around their reference means states are 
introduced with prime notation, e.g.
$
\Pi \rightarrow \bar \Pi + \Pi',
$  
etc.
We assume an isothermal equation of state for
the perturbations as well. Linearised perturbations of the equations of motion
reveal
\beqa
\frac{\partial u'}{\partial t} - 2\Omega_0 v' &=& -c_s^2 \frac{\partial \Pi'}{\partial x} \nonumber \\
\frac{\partial v'}{\partial t} + \Omega_0\left(1/2 +  \overline V_x\right) u' + \overline V_z w' &=& 0 \nonumber \\
\frac{\partial w'}{\partial t} &=& -c_s^2 \frac{\partial \Pi'}{\partial z} \nonumber \\
\frac{\partial \Pi'}{\partial t} + u'\frac{\partial \overline\Pi}{\partial x} + w'\frac{\partial \overline\Pi}{\partial z}
+\frac{\partial u'}{\partial x} + \frac{\partial w'}{\partial z} &=& 0.
\eeqa
Note we have utilised the shortened notation 
$\overline V_x \equiv \partial \overline V/\partial x$.
To emphasize: the analysis we carry out here pointedly departs from that
done in GS67 in two respects: (i) we assume an isothermal equation of state for
the disturbances and (ii) we allow for compressibility (cf. eq. 29 of GS67).
Since $g$ is constant, the above equations are easily combined into a single one for 
$\Pi'$ yielding,
\beqa
& & \frac{\partial^4 \Pi'}{\partial t^4} -\left[\left(\frac{\partial\overline\Pi}{\partial z}
+ \frac{\partial}{\partial z}\right)c_s^2 \frac{\partial}{\partial z}
+\left(\frac{\partial\overline\Pi}{\partial x} + \frac{\partial}{\partial x}\right)c_s^2 \frac{\partial}{\partial z}
-\kappa_0^2\right]\frac{\partial^2 \Pi'}{\partial t^2} \nonumber \\
& & \hskip 0.0cm
+ \left[
2\Omega_0 \bar V_z \left(\frac{\partial\overline\Pi}{\partial x} + \frac{\partial}{\partial x}\right)c^2
- \kappa_0^2\left(\frac{\partial\overline\Pi}{\partial z} +\frac{\partial}{\partial z}\right)c_s^2
\right]\frac{\partial \Pi'}{\partial z} =0
\eeqa
where the epicyclic frequency is represented by $\kappa_0$ and is
related to the steady state quantities by
\[
\kappa_0^2  = 2\Omega_0^2 (1/2+ \bar V_x).
\]
Paraphrasing directly from GS67 (right prior to eq. 17 of that work): The next step is
to expand the unperturbed variables and their derivatives in
Taylor series about some point $x=x_0$ and $z=z_0$.  The latter restriction
is not as arbitrary as it might seem - it just makes the calculation more
transparent.  Discarding terms of order $(x-x_0)/x_0$ and $(z-z_0)/z_0$
the perturbation variables may be expanded in
plane waves of the form
$
\sim e^{i\left(\omega t + k_x x + k_z z\right)},
$
revealing the dispersion relationship
\[
\omega^4 - \left[ c_0^2(k_x^2 + k_z^2) + igk_z + \kappa_0^2\right]\omega^2
-2\Omega_0 \bar V_z c_0^2 k_x k_z + \kappa_0^2 (c_0^2 k_z^2 + ik_z g) = 0,
\]
in which $c_0^2$ is the sounds speed at the point in question and where
$
\kappa_0 \rightarrow \Omega_0.
$
As is standard for problems of atmospheres, we can recast the vertical
wave dependence to have a complex character (indicating a basic wavey pattern) by defining
$k_z \rightarrow k_z -  i g/(2c_0^2)$ which renders the
dispersion relation into the form
\beq
\omega^4 - \left[ c_0^2(k_x^2 + k_z^2) + \kappa_0^2 + N_0^2\right]\omega^2
-2\Omega_0 \bar V_z c_0^2 k_x k_z + \kappa_0^2 (c_0^2 k_z^2 + N_0^2) = 0,
\label{dispersion_relation}
\eeq
where we have defined the Brunt-Vaisaila frequency
\[
N_0^2 = \frac{g^2}{4c_0^2}.
\]
Solutions for $\omega$ are easy to obtain and write down.  It is more instructive, however, to assess the
stability characteristics straight from an analysis of (\ref{dispersion_relation}) itself.
\par
The classical GSF instability is one in which the modes pass through zero frequency
before becoming unstable and they describe the influence that the
vertical shear gradient has upon {\it inertial modes}.  In our dispersion relation this amounts to
the instability condition
\beq
-2\Omega_0 \bar V_z c_0^2 k_xk_z + \kappa_0^2 (c_0^2 k_z^2 + N_0^2) < 0.
\label{GF_criterion}
\eeq
When $c_0^2 k_z^2 \gg N_0^2$ the condition reduces to
\[
-2\Omega_0 \bar V_z + \kappa_0^2  \frac{k_z}{k_x} < 0,
\]
which is essentially the first term appearing in eq.~33 of GS67 (misprinted). Also, this condition is
identical to that found in Urpin (2003), eq.~20.  These latter correspondences
follow from realizing that
\[
\bar V_z \leftrightarrow \frac{\partial \Omega}{\partial z},\qquad
\frac{\kappa_0^2}{2\Omega_0} \leftrightarrow
\frac{1}{R_0^2}\left(\frac{\partial R^2 \Omega}{\partial R}\right)_{R=R_0}
\]
and, thus, recovering the classical Goldreich-Schubert condition.
The dispersion relation (\ref{dispersion_relation}) has the general form
\[
\omega^4 - B\omega^2 + C = 0
\]
where $B$ and $C$ are obviously identified with
the terms in (\ref{dispersion_relation}).  This equation has imaginary
solutions for $\omega$ if (i) $C<0$ with $B>0$ or (ii)
$B^2 - 4C < 0$.
The GSF criterion (\ref{GF_criterion}) is essentially the condition that $C<0$.
Condition (ii) is for acoustic-inertial modes to be unstable.
This amounts to
\beq
\left[ c_0^2(k_x^2 + k_z^2) + \kappa_0^2 + N_0^2\right]^2
- 4\left[-2\Omega_0 \bar V_z c_0^2 k_xk_z + \kappa_0^2 (c_0^2 k_z^2 + N_0^2)
\right] < 0.
\eeq
A detailed examination of the behaviour of this condition shows that it does not occur 
for reasonable values of the parameters leading us to conclude that
the acoustic modes (per se) do not become unstable as well.

\label{lastpage}

\end{document}